\documentclass[12pt]{article}
\pdfoutput=1

\usepackage[utf8]{inputenc}
\usepackage{amsmath,amssymb,amscd}
\usepackage{listings}
\usepackage{caption}
\usepackage{dsfont}
\usepackage{slashed}
\usepackage{color}
\usepackage[caption=false]{subfig}
\usepackage[pdftex]{graphicx}
\usepackage{epstopdf}
\usepackage{listings}
\usepackage{epsfig}
\usepackage{listings}
\usepackage{caption}
\usepackage{cite}
\usepackage{hyperref}

\usepackage{multirow}

\usepackage{xcolor}
\hypersetup{
    colorlinks,
    linkcolor={red!0!black},
    citecolor={blue!50!black},
    urlcolor={blue!80!black}
}

\newcommand{\bea}{\begin{align}}
\newcommand{\eea}{\end{align}}
\newcommand{\beq}{\begin{equation}}
\newcommand{\eeq}{\end{equation}}
\newcommand{\nbea}{\begin{align*}}
\newcommand{\neea}{\end{align*}}
\newcommand{\nbeq}{\begin{equation*}}
\newcommand{\neeq}{\end{equation*}}

\newcommand{\ket}[1]{\left| #1 \right>} 
\newcommand{\bra}[1]{\left< #1 \right|} 

 \newcommand{\aem}{\alpha_{\textrm em}}

\newcommand{\br}[1]{\mathcal{BR}_{#1}}
\newcommand{\avgam}{\langle \gamma \rangle}
 \newcommand{\gvu}{g_{u}}
 \newcommand{\gvd}{g_{d}}
  \newcommand{\gvs}{g_{s}}
   \newcommand{\gau}{\tilde{g}_{u}}
    \newcommand{\gad}{\tilde{g}_{d}}
    \newcommand{\gas}{\tilde{g}_{s}}

      \newcommand{\amc}{{\sc MadGraph5}\_a{\sc MC@NLO}}
      \newcommand{\fr}{{\sc Feyn\-Rules}}

 \def\mev   {\textrm{ MeV}}   
  \def\nn   {\nonumber}  
\def \Tr {\textrm{Tr}}     
\def \grpp {g_{\rho \pi \pi}}
\def \grop {g_{\rho \omega \pi}}
\def \etap{\eta^\prime{}\!}     
\def \Azero{\mathcal{A}_0^\mu}
\def \Aoct{\mathcal{A}_8^\mu}

\def \neff{N_\mathrm{eff}}

\def\eff{\mathcal{E}}

\numberwithin{equation}{section}

\begin{document}

\begin{titlepage}

\baselineskip=21pt
 {\small
 \rightline{CERN-TH-2020-003}
 }
\vskip 1in

\begin{center}

{\large {\bf Light Dark Sectors through the Fermion Portal}}

\vskip 0.75in

{\bf Luc~Darm\'e},$^{1}$
{\bf Sebastian~A.~R.~Ellis}$^{2}$
and {\bf Tevong~You}${^{3,4}}$

\vskip 0.2in

{\small {\it
$^1${INFN - Laboratori Nazionali di Frascati, via E. Fermi 40,\\ 00044 Frascati, Italy}\\
$^2${SLAC National Accelerator Laboratory, 2575 Sand Hill Road, \\ Menlo Park,  CA 94025, USA} \\
$^3${CERN, Theoretical Physics Department, Geneva, Switzerland}\\
$^4${DAMTP, University of Cambridge, Wilberforce Road, Cambridge, CB3 0WA, UK; \\
Cavendish Laboratory, University of Cambridge, J.J. Thomson Avenue, \\
\vspace{-0.01cm}
Cambridge, CB3 0HE, UK}

}}

\vskip 0.2in

{\bf Abstract}

\end{center}

\baselineskip=18pt \noindent


{\small
Pairs of Standard Model fermions form dimension-3 singlet operators that can couple to new dark sector states. This ``fermion portal" is to be contrasted with the lower-dimensional Higgs, vector and neutrino singlet portals. We characterise its distinct phenomenology and place effective field theory bounds on this framework, focusing on the case of fermion portals to a pair of light dark sector fermions. We obtain current and projected limits on the dimension-6 effective operator scale from a variety of meson decay experiments, missing energy and long-lived particle searches at colliders, as well as astrophysical and cosmological bounds. The \texttt{DarkEFT} public code is made available for recasting these limits, which we illustrate with various examples including an integrated-out heavy dark photon. 
}


\end{titlepage}
\newpage

\tableofcontents

\section{Introduction}

In recent years, the search for dark matter has broadened beyond weakly-interacting massive particles in the GeV-TeV range. There has been a resurgence of interest in improving experimental sensitivity to sub-GeV dark matter, while much progress has also been made in neutrino experiments (see e.g. Refs.~\cite{Battaglieri:2017aum,Beacham:2019nyx, Buonocore:2019esg} for recent overviews). In addition to uncovering the nature of dark matter, these searches could open a window onto a rich dark sector that must often accompany it. The dark sector's experimental signatures often share many similarities with those of dark matter and neutrinos. Moreover, dedicated experiments have been proposed to look specifically for the spectacular signal of long-lived particles decaying back to the visible sector. The generality of such dark sectors requires an appropriate framework in which to characterise the sensitivity of these various searches. Many models, either simplified or top-down motivated, have been constructed and used as benchmarks. Here we propose a bottom-up effective field theory approach to characterise the phenomenology of light dark sector searches more generally. 

Dark sector fields are singlets under the Standard Model $SU(3)_c \times SU(2)_L \times U(1)_Y$ gauge groups. They are motivated by observations of neutrino masses and dark matter, which require additional particle content that may well involve an extended dark sector, and arise generically in many models of new physics addressing a variety of problems. Additionally, dark sectors with potentially rich phenomenology are a generic prediction of compactified string theory (see e.g.~\cite{Acharya:2016fge,Acharya:2017kfi,Halverson:2018xge,Acharya:2018deu} for recent studies and reviews). Indeed, there is nothing exotic about singlet charge assignments under the Standard Model gauge groups; we already know of particles that are $SU(3)_c \times SU(2)_L$ singlets, so it is reasonable to suppose others may exist that go a step further in being uncharged under $U(1)_Y$ as well.

In full generality, the visible sector of the Standard Model can be described as communicating with dark sectors through so-called ``portal'' operators, $\mathcal{O}_\text{SM}^{(d)}$. They are formed by singlet combinations of Standard Model fields. For a portal operator of (mass) dimension $d$, we may write its Lagrangian interaction term with a dark sector operator  $\mathcal{O}_{\rm DS}^{(d^\prime)}$, with dimension $d^\prime$, as
\begin{equation}
\mathcal{L} \supset \frac{c_{ij} \mathcal{O}_{\rm SM}^{(d) i}\mathcal{O}_{\rm DS}^{(d^\prime) j}}{\Lambda_{ij}^{d+d^\prime - 4}} \, .
\end{equation}
The quantity $\Lambda_{ij}$ is a dimensionful scale and $c_{ij}$ is a dimensionless coefficient; if $d+d^\prime > 4$ then the interaction is associated with an effective non-renormalisable Lagrangian where $\Lambda_{ij}$ is related to the heavy mediator mass and $c_{ij}$ is a Wilson coefficient. If the light dark sector is only connected to the Standard Model through heavy mediators, as for example in hidden valley models~\cite{Strassler:2006im}, the resulting scale suppression would provide a natural explanation for the weakness of the dark sector's interactions with the visible sector.

The first few portals ordered by dimensionality are listed in Table~\ref{tab:portals}. The lowest-dimensional portal operators are the well-known Higgs, vector, and neutrino portals (see e.g. Refs.~\cite{Holdom:1985ag,Schabinger:2005ei, Patt:2006fw, Batell:2009di, Falkowski:2009yz} and references therein):    
\begin{align}
&|H|^2 \quad (d=2) \, , \\
&F_{\mu\nu} \quad (d=2) \, , \\
&LH \quad (d=5/2) \, ,
\end{align}
where $d$ is the mass dimension, $F_{\mu\nu}$ is the electromagnetic (or hypercharge) gauge field strength and the Higgs and lepton doublet fields are denoted by $H$ and $L$. The low operator dimensionality of these portals allows them to form renormalisable interactions ($d+d^\prime \leq 4$) with the hidden sector. For example, the Higgs portal may couple to another scalar; the vector portal can kinetically mix with a hidden sector gauge field strength; and the neutrino portal can be responsible for neutrino masses through a right-handed neutrino coupling.

The next lowest-dimensional singlet operators involving Standard Model fermions are those that we dub ``fermion portals".\footnote{The fermion portal terminology has also been used for the unrelated model of Ref.~\cite{Bai:2013iqa}.} These are singlet combinations of Standard Model fermion fields, of which the lowest-dimensional takes the form
\begin{align}
&\psi_i \psi_j \quad (d=3) \, , \label{eq:fermionportal1}  
\end{align}
where $\psi$ represent the Standard Model fermions and the different types of contractions are left implicit. At the level of unbroken electroweak symmetry, these portals can only form higher-dimensional operators of dimensions 5, 6 or greater with the dark sector,\footnote{The portal of Eq.~\ref{eq:fermionportal1} can make a dimension 4 interaction with a hidden sector vector boson, but a light gauge boson is often included as part of the vector portal phenomenology. Depending on the nature of the new vector boson, and whether it mixes with hypercharge or a conserved current of the Standard Model, the phenomenology can be very different. Here we shall be concerned with heavy mediators.} suppressed by some effective field theory cut-off scale. The heavier the cut-off, the weaker the interaction between the visible and dark sectors. Searches for light, weakly-coupled dark sectors in fermion portals could then also yield complementary information about the scale of heavier new physics mediating the interaction. 


\begin{table}[t]
	\begin{center}
		\begin{tabular}{|c|c|c|}
				\hline
				Portal & Operator & Dimension \\
				\hline
				Higgs & $|H|^2$ & 2 \\
				Vector & $F_{\mu\nu}$ & 2 \\
				\hline
				Neutrino & $LH$ & $5/2$ \\
				\hline
				Fermion &
				$\psi\psi$ & 3 \\
				\hline
			\end{tabular}
	\end{center}
	\caption{Portal operators in the Standard Model, ordered by their mass dimensionality. }
		\label{tab:portals} 
\end{table}


For example, the neutral current dimension-3 fermion portal of Eq.~\ref{eq:fermionportal1}, $\mathcal{O}^{(3)}_{SM}$, can form a dimension-5 operator with a derivatively coupled scalar, 
\begin{equation}
\left(\frac{\partial_\mu \phi}{\Lambda}\right) \mathcal{O}^{(3)}_{SM} \, .
\end{equation}
The prototypical example of this is the familiar axion, where the scale suppression is related to the axion's symmetry-breaking scale. For some recent phenomenological studies of constraints on the scale of the axion decay constant, see for example Refs.~\cite{Baumann:2016wac, Bauer:2017ris, Brivio:2017ije, Bjorkeroth:2018dzu,Chang:2018rso}.

In this paper we characterise the sensitivity of dark sector searches focusing on the case of neutral current fermion portals to a pair of light dark sector fermions.\footnote{We note that there are also 3-fermion singlet combinations, $ \psi_i \psi_j \psi_k \quad (d=9/2)$, such as $d_i u_j  d_j$. Since they carry baryon number $B=1$ they must couple to a hidden sector fermion that carries the opposite baryon number. Such a singlet fermion with baryon number and no lepton number has been discussed e.g. in Refs. \cite{Ma:2017dbl, Elor:2018twp}. Ref. \cite{Ma:2017dbl} also categorises the various singlet fermion lepton number assignments that would forbid the renormalisable neutrino portal operator at tree level. Indeed, there exists a rich set of possibilities for dark sector fermions beyond the familiar right-handed neutrino with lepton number $L=1$, though we will not consider them any further here.} In this case the dark sector fermion $\chi$ forms a dimension-6 four-fermion operator with the Standard Model fermion pair of Eq.~\ref{eq:fermionportal1},
\begin{equation}
\frac{1}{\Lambda^2}(\bar\chi \Gamma \chi) \mathcal{O}^{(3)}_{SM} \, ,
\end{equation}
where we focus on the tensor structures $\Gamma \subset \{\gamma_\mu, \gamma_\mu \gamma_5\}$. We consider in this work four-fermion operators coupling the dark sector fermions to quarks (with effective couplings $g_u^{ij}$ and $g_d^{ij}$  where the indices $i,j$ refer to the generation of SM fermions) and charged leptons (with effective couplings $g_l^{ij}$). We will typically choose particular ratios $g_u : g_d : g_l$ and present the limits in terms of $\Lambda / \sqrt{g}$, where $\Lambda$ is a mass scale and $g$ represents the overall coefficient (that we also refer to as the effective coupling).

Light dark sectors are typically probed by an extremely wide range of experiments usually referred to as the \textit{intensity frontier} of particle physics. They share a relatively low center-of-mass energy, high available statistics and very good background rejection. We will use three general type of accelerator-based experiments (a complete list of the searches we have implemented can be found in Table~\ref{tab:explist} in Appendix~\ref{app:Code}): 

\begin{itemize}

\item First, there are the dedicated flavour experiments, which can be either based at $e^+ e^-$ colliders such as the B-factory experiments BaBar~\cite{Lees:2017lec}, Belle-II~\cite{Kou:2018nap}, or at beam dump facilities such as the Kaon factories NA62~\cite{NA62:2017rwk} or E949~\cite{Artamonov:2009sz}. These experiments are typically used for missing energy (mono-photon) dark sector searches or for indirect limits using invisible $B/K$ meson decays. 

\item Second, we have neutrino-focused experiments that are typically based on proton beam dumps. These include the past experiments LSND~\cite{Athanassopoulos:1997er}, CHARM~\cite{Bergsma:1985is} or the current near-detector experiments MiniBooNE~\cite{Aguilar-Arevalo:2018wea}, and NO$\nu$A~\cite{Ayres:2004js}. Here dark sector particles, abundantly produced at the beam dump, can travel alongside neutrinos and either  scatter or decay in the detector if they are sufficiently long-lived.

\item The third class of experiments are dark sector-oriented ones. Past experiments were typically searching for axion-like particles or dark photons and were based on electron beam dump experiments. Their sensitivity relies on the electron's bremsstrahlung into the dark sector, whose large cross-section typically compensated for their somewhat lower statistics. Additionally, a significant fraction of proposed new experiments will be either LHC-based (such as FASER~\cite{Feng:2017uoz}, CODEX-b~\cite{Aielli:2019ivi}, and MATHUSLA~\cite{Lubatti:2019vkf}) or based on a proton beam dump (such as SHiP~\cite{Anelli:2015pba} or a possible extension of SeaQuest~\cite{Berlin:2018pwi}). While we aim for a comprehensive coverage of existing experimental limits, we do not attempt to provide projections for all upcoming intensity frontier experiments and will instead focus on a few representative examples. A more complete list can be found in, e.g. Refs.~\cite{Battaglieri:2017aum,Beacham:2019nyx}.

\end{itemize}

We also discuss astrophysical and cosmological constraints on dark sector fields linked to the SM by a fermion portal. While such particles can constitute all or a fraction of the observed dark matter relic density and must not overclose the universe,\footnote{For some recent works involving four-fermion operators with dark matter at the LHC, see e.g.~\cite{Bertuzzo:2017lwt,Belyaev:2018pqr,Choudhury:2019tss,Choudhury:2019sxt}).} we will not derive limits based on this criteria here, as they depend strongly on the inner dynamics of the dark sector states. Indeed, we do not require that the dark sector states make up any of the dark matter, or be cosmologically stable. On the other hand limits from the observed cooling rates of supernovas and the cosmic microwave background can provide relatively model-independent constraints on fermion portal dark sectors with masses below the tens of MeV range. These indirect observational constraints can be complementary dark sector probes to the direct experimental searches listed above.  

The paper is structured as follows. In Section~\ref{sec:production}, we summarise the various production mechanisms relevant for the fermion portal operators, from light to heavy meson decay processes and direct parton-level production. Section~\ref{sec:acceleratorsearches} focuses then on the various relevant search channels at accelerator and beam-dump based experiments, including decay and scattering signatures of dark sector particles, invisible meson decay limits and mono-X searches. Section~\ref{sec:astrophyics} presents a selection of the relevant astrophysical limits, from SN1987A cooling to an estimate of early universe cosmological limits. Finally, Section~\ref{sec:numerics} is dedicated to a presentation of the results obtained using our public code \texttt{DarkEFT}. We show different representative choices of models and in particular illustrate the effectiveness of the formalism through the example of a relatively heavy dark photon (in the tens of GeV range). The appendices contain more detail about the meson decay production mechanisms as well as a brief presentation of the \texttt{DarkEFT} companion code, released alongside this paper and dedicated to the recasting of existing dark sector limits into constraints on the fermion portal operators.

\section{Dark sector production through the fermion portal}
\label{sec:production}

Dark sector fermions can be produced via the fermion portal through various processes involving quark and lepton bilinears. In this work we will be agnostic as to the flavour pattern of the Wilson coefficients of the effective operators. These may in principle involve both flavour-diagonal and off-diagonal couplings. Our results will be as general as possible, so that they may be applied to any model of dark sector fermions coupling to the SM via a mediator far off-shell. 

The production channels of dark fermions we consider here are:
\begin{itemize}
	\item Light meson decays: this channel depends on the nature of the meson and the operator -- axial vector or vector.  It typically proceeds through an associated decay, e.g. $\pi^0 \rightarrow \gamma \chi \chi$, or via a fully dark decay, e.g. $\pi^0 \rightarrow\chi \chi$. This channel requires non-zero couplings $g_u$ or $g_d$ to up or down quarks respectively.
	\item Heavy meson decays: this channel will be important for vector charmed quarks, e.g. $J/\Psi$, or for flavour-violating rare meson decays. This channel depends on couplings to heavy quarks such as $g_c$ to charm or $g_b$ to bottom. The latter in particular will require non-flavour-diagonal couplings.
	\item Direct production either at the parton level, via $p p \rightarrow \chi \chi$ or  $ p p \rightarrow \textrm{jet} + \chi \chi$, or in electron colliders, via $e e  \rightarrow \chi \chi$ or mono-photon $e e  \rightarrow \gamma \chi \chi$. 
\end{itemize}

The production of light dark sector fermions through bremsstrahlung, $p(e) N \to p(e) N \chi \chi$, is also possible, although often with smaller cross-sections than some of the processes listed above. This production mechanism, when paired with a search for missing momentum/energy as an experimental signature, can be quite powerful to search for light dark sector particles~\cite{Izaguirre:2014bca, Izaguirre:2015yja, Berlin:2018bsc}, as it does not require a visible decay, and therefore a further $g/\Lambda^2$ suppression. We defer study of this production mechanism and detection approach to future work.

Interestingly, parton-level production is usually irrelevant in standard portal searches for low mass dark sectors, since QCD perturbativity breaks down before the mediator scale. In our fermion portal model, this is no longer the case as long as the mediator mass is larger than the QCD scale, so that direct production becomes essentially constant for low dark sector masses. 

In the following, we will consider the most generic case in which there are two dark sector states, $\chi_1$ and $\chi_2$, with masses $M_1$ and $M_2$ respectively (we will generally take $M_2 > M_1$). All our results can be trivially applied to a one state scenario by setting $M_1 = M_2$. We now consider in turn each of these production mechanisms.

\subsection{Light meson decays}
\label{sec:lightmesdecays}
Light mesons are abundantly produced in proton-based colliders and beam dump experiments. Furthermore, their decays typically proceed with a relatively long lifetime which tempers the strong suppression from the fermion portal's high scale. For a meson $M$, the final production number $N_\text{prod}$ of the dark sector state $\chi_2$ (in association with $\chi_1$ and possibly another SM particle $X$) is given by 
\begin{align}
	N_{\rm prod} = \mathcal{N}_M \times \mathrm{BR} \left(M \rightarrow \bar{\chi}_1 \chi_2 (+X) \right) \propto  \frac{M_m^4}{\Lambda^4} \ ,
\end{align}
where $M_m$ is the meson mass and $\Lambda$ is the scale of the fermion portal operator. Heavy mesons are then expected to interact more with the dark sector than lighter fermions and can dominate the production rate. This is somewhat reminiscent of Higgs portal phenomenology, where the dark sector also couples more strongly to heavier fermions, though the effect is even more pronounced in the fermion portal case as it has a quartic dependence on the meson mass compared to a quadratic dependence for the Higgs portal. Note that this effect is balanced in part by the fact that lighter mesons have a smaller decay width, as summarised in Table~\ref{tab:effcoup}, so that their branching ratio into dark sector fields is enhanced.

Depending on the pseudo-scalar or vector nature of the light unflavoured mesons, the type of operator (vector or axial vector) will be critical in determining the possible channels for dark sector production in meson decay. We shall consider effective vector-vector (V-V) operators of the form 
\begin{equation}
\mathcal{L} \supset \sum_{q \in u,d} \frac{g_q}{\Lambda^2}(\bar\chi_1 \gamma_\mu \chi_2) (\bar{q} \gamma^{\mu} q) \, ,
\label{eq:opV}
\end{equation}
and axial-vector couplings (AV-AV) corresponding to operators of the form 
\begin{equation}
\mathcal{L} \supset \sum_{q \in u,d} \frac{\tilde{g}_q}{\Lambda^2}(\bar\chi_1 \gamma_\mu \gamma^5 \chi_2) (\bar{q} \gamma^{\mu} \gamma^5  q) \, ,
\label{eq:opAV}
\end{equation}
where we have included the possibility of having two states in the dark sector, with a mass splitting $\Delta_\chi \equiv |M_2| - |M_1|$. Depending on which one is the most relevant, we will also use the normalised splitting
\beq
\delta_\chi \equiv \frac{|M_2| - |M_1|}{|M_1|} \ .
\label{eq:dChi}
\eeq
The splitting can be taken to zero to recover the single state case.

We will consider Dirac fermion dark sector states $\chi_1$ and $\chi_2$. In the following we shall summarise the results for the various amplitudes; details of the calculation can be found in Appendix~\ref{app:mesdecaycalc}. For most decays, we present both the small splitting limit, $\delta_\chi \ll 1$, and a saturated splitting where $M_1 \ll M_2 $ ($\delta_\chi \gg 1$). For all numerical results we use the full amplitude which are straightforwardly derived once the effective couplings relevant to the corresponding decaying meson are known. In all the rest of this section, and unless explicitly specified otherwise, one can obtain the decay rates for mixed operators AV-V or V-AV by replacing $M_1$ by $-M_1$ in the V-V or AV-AV expressions respectively. The result from the V-V operator can also be used for the so-called ``pseudo-Dirac case''~\cite{TuckerSmith:2001hy} when $\chi_1$ and $\chi_2$ are Majorana states originating from a single Dirac field.\footnote{The inclusion of a small Majorana mass term triggers the splitting of the Dirac fermion into two Majorana fermions. An important subtlety is the fact that if one insists on keeping positive masses from both Majorana fermions, the mixing matrices become complex. In the case of an initial vector coupling for the Dirac fermion coupling, the interactions term then contains both a leading vector $\bar{\chi}_1 \gamma^\mu \chi_2$ interaction and a sub-leading axial-vector one $\bar{\chi}_1 \gamma^\mu \gamma^5 \chi_2$. For larger splitting, both interactions become relevant and it is preferable to use instead a negative mass $M_1$ with a purely axial-vector interaction $\bar{\chi}_1 \gamma^\mu \gamma^5 \chi_2$. Note that this limit also extends naturally to the case where the Majorana component dominates, in which case both the masses are positive.} Interestingly, most of the phenomenology (for instance the type of meson decays, the flavour-violation effects, etc.) depends only on the Standard Model part of the operators so that the following results can also give a rough estimate of the constraints one could expect for other dark sectors not considered here, for instance with scalar fields instead of fermions.

\paragraph{Vector coupling}

Due to the vector nature  of the effective operator, the decay of light pseudo-scalar mesons have to proceed through the axial anomaly with an associated photon production. The dominant production mechanism is then $\pi^0, \eta, \eta^\prime \rightarrow \gamma \chi \chi$ with a decay amplitude of the form
\begin{align}
\nonumber \Gamma_{P, V} = \frac{2 g^2_{P}}{ \pi f_\pi^2  \Lambda^4} &\times \frac{\aem}{3 (4  \pi)^5} \int_{(|M_1|+|M_2|)^2}^{M_P^2} ds \ \frac{s (M_P^2 -s )^3}{M_P^3} \\
&\times \begin{cases}
\sqrt{1-\frac{4 M_1^2}{s}}\left(1+\frac{2 M_1^2}{s}\right) \quad \hfill \textrm{(small splitting, V)} \\[0.3em]
	 2 \left(1-\frac{4 M_1^2}{s}\right)^{3/2} \quad  \hfill \textrm{(small splitting, AV)} \\[0.3em]
	  \left( 2 + \frac{ M_2^2}{s} \right)   \left( 1 - \frac{ M_2^2}{s} \right)^2 \quad \hfill \textrm{(non-degenerate)}
	\end{cases}
\end{align}
and we have used the effective couplings $g_{P}$ as defined in Table~\ref{tab:effcoup}, with $P \equiv \pi^0, \eta, \eta^\prime$ and the pion decay constant $f_\pi = 130.7 $ MeV. The strong numerical suppression factor arises in part from the loop-induced axial anomaly and in part from the phase space suppression for this 3-body decay. Furthermore, we recover as expected the scale suppression by $M_P^4/\Lambda^4$.

\begin{figure}[t]
	\centering
	\subfloat[]{%
		\label{fig:BRemV}%
		\includegraphics[width=0.47\textwidth]{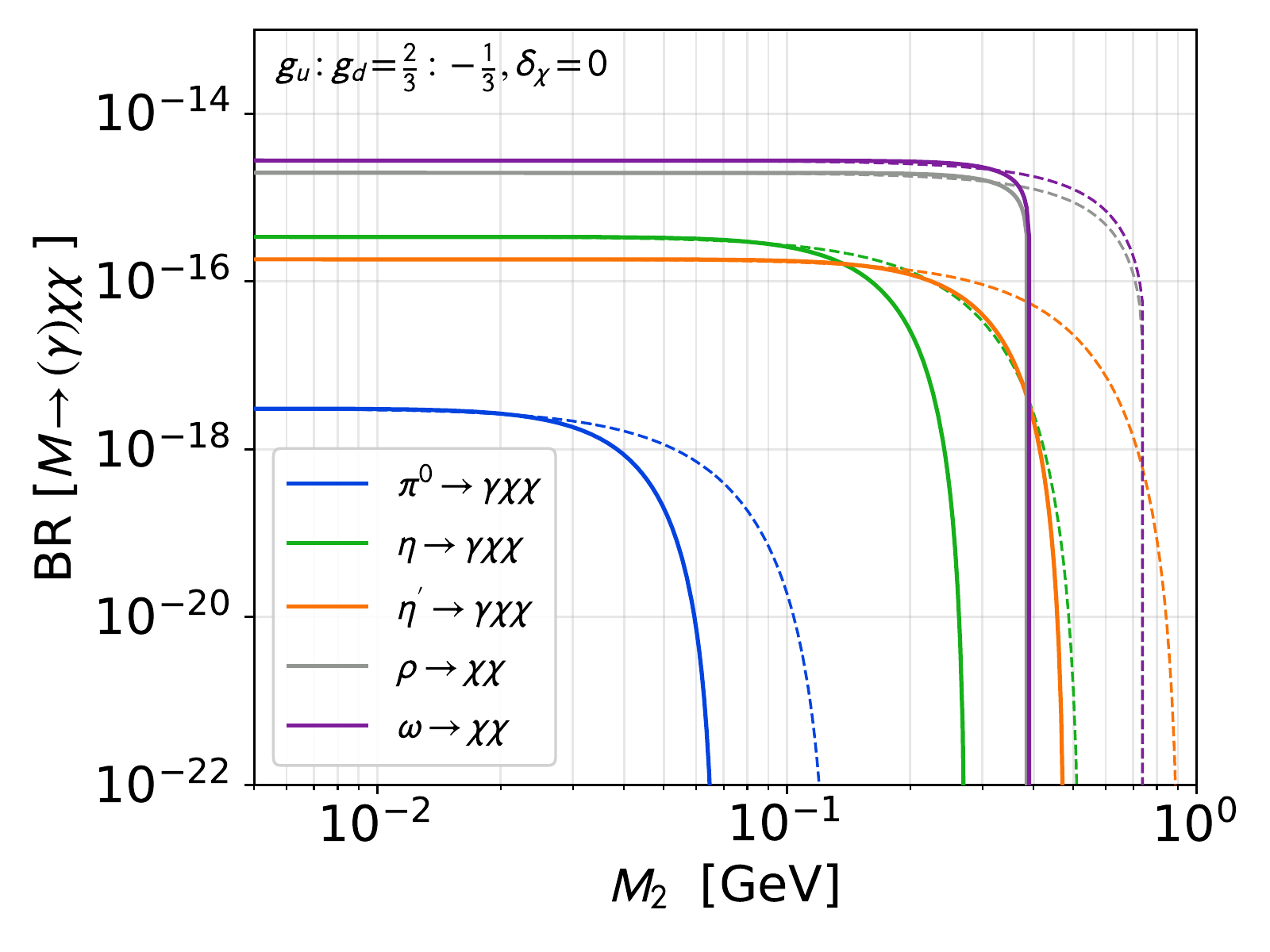}
	}%
	\hspace{0.0\textwidth}
	\subfloat[]{%
		\label{fig:Zal}%
		\includegraphics[width=0.47\textwidth]{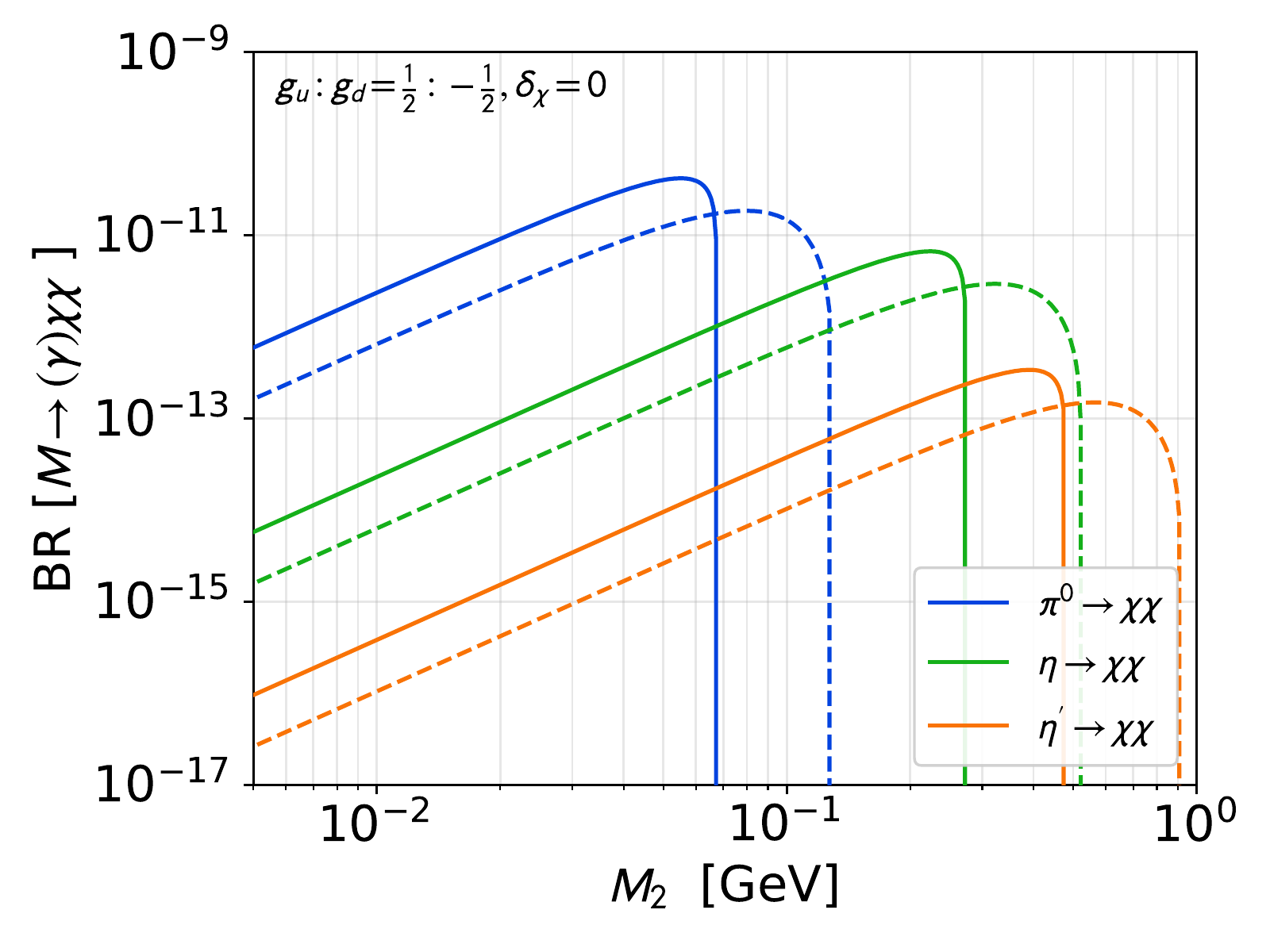}
	}%
	\hspace{0.0\textwidth}
		\subfloat[]{%
		\label{fig:BRbmlV}%
		\includegraphics[width=0.47\textwidth]{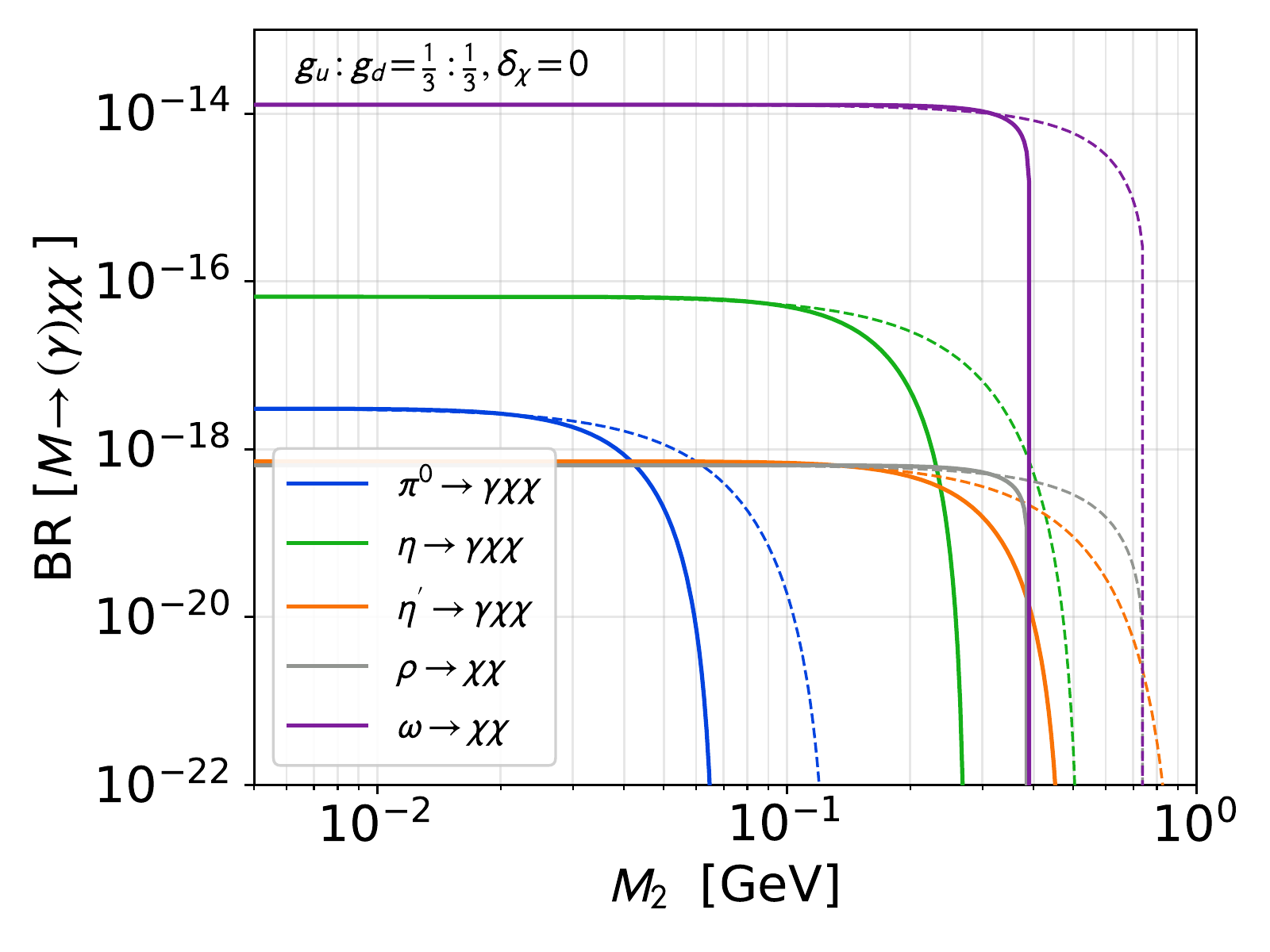}
	}%
	\hspace{0.0\textwidth}
	\subfloat[]{%
		\label{fig:uni}%
		\includegraphics[width=0.47\textwidth]{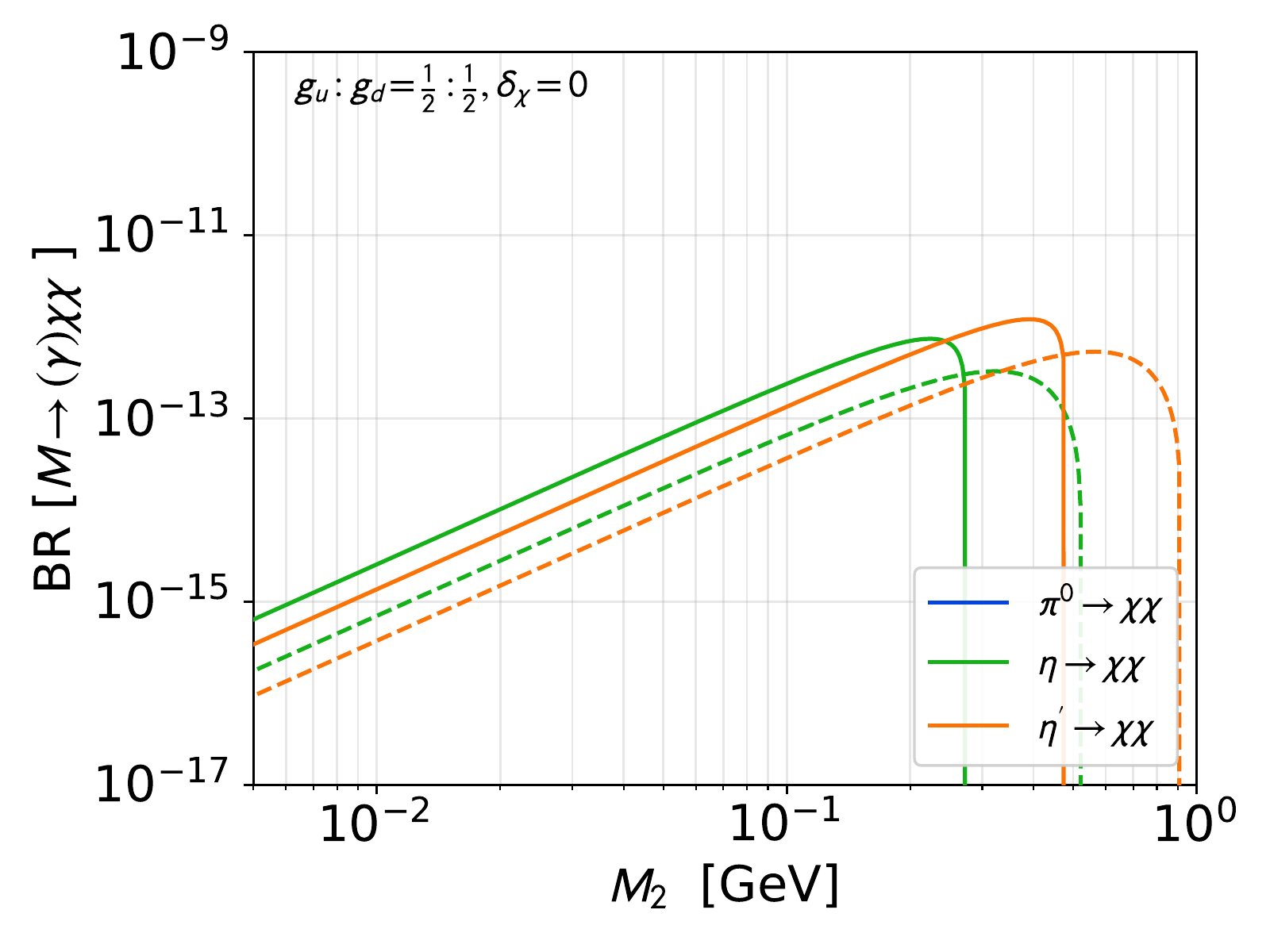}
	}%
	\caption{Branching ratios as a function of dark sector fermion mass for various meson decays for \text{\textbf{(a)}}, \textbf{(c)} the vector-current operator and \textbf{(b)}, \textbf{(d)} the axial-vector operator, with $\delta_\chi = 0$ ($ \delta_\chi = 20$) between the dark sector states in solid (dashed) lines. The upper and lower figures show the effect of changing the relative couplings to up- and down-quarks. Couplings to strange-quarks have been assumed to align with the down-quark coupling.}
	\label{fig:BR}
\end{figure}

Both effects combined imply that the dominant production mechanism in most cases will in fact be vector meson decays. Indeed, for dark sectors coupling to the SM through a vector current, vector mesons can decay directly into dark sector particles, e.g. $\rho, \omega \rightarrow \chi \chi$, with a decay width given by
\begin{align}
\label{eq:GammaVectMesondecay}
\Gamma_{U} = \frac{g^2_{U} f_\pi^2}{ 24 \pi }  \times  \frac{M_U^3}{\Lambda^4} \left( 1 - \frac{(M_2-M_1)^2}{M_U^2} \right)^{3/2} \left( 1 - \frac{(M_2+M_1)^2}{M_U^2} \right)^{1/2} \left( 2 + \frac{(M_2 + M_1)^2}{M_U^2} \right) \, ,
\end{align}
where $U \equiv \rho, \omega$ with the effective couplings $g^2_{U}$ defined in Table~\ref{tab:effcoup}. While the $\Lambda$ suppression remains, the two-body nature of the decay and the absence of $\aem$ suppression strongly enhance this decay compared to the previous one. Altogether, as can be seen in Fig.~\ref{fig:BRemV} and Fig.~\ref{fig:BRbmlV}, the production for the vector coupling case is strongly dominated by the decay of vector mesons.


\begin{table}[t]
	\begin{center}
	\resizebox{1.\textwidth}{!}{\begin{minipage}{1.11\textwidth}
		\begin{tabular}{|c|c|c|c|c|}
				\hline
				\rule{0pt}{14pt}Meson decay & Vector current  & Axial-vector current & Dark photon & $\Gamma_M$  (GeV) \\
				\hline
				\hline
				\rule{0pt}{14pt}
			 $\pi \rightarrow \gamma X X $         & $g_{\pi^0} = 2 \gvu + \gvd $  & /  & $e \varepsilon$  &  $7.7 \cdot 10^{-7}$ \\
			  $\eta \rightarrow \gamma X X $ & $g_{\eta} =  1.5 \gvu -0.7 \gvd +0.6 \gvs $&  / & $e \varepsilon$   &  $1.3 \cdot 10^{-6}$ \\
			  	$\eta^\prime \rightarrow \gamma X X $ & $g_{\eta^\prime} =  1.2 \gvu -0.6 \gvd -0.9 \gvs $&  / & $1.3 ~ e \varepsilon$   &  $2.0 \cdot 10^{-4}$ \\
				\hline
				\hline
				$\rho \rightarrow  X X $         & $ g_{\rho} =1.3 \gvu - 1.3 \gvd $  & /  & resonant    &  $0.15$   \\
				$\omega \rightarrow  X X $ & $ g_{\omega} = 1.2\gvu + 1.2 \gvd $&  / & resonant  &  $8.5 \cdot 10^{-3}$  \\
				\hline
				\hline
			 $\pi \rightarrow  X X $      &/   & $ \tilde{g}_{\pi^0} = (\gau - \gad)/\sqrt{2} $ & /   &  $7.7 \cdot 10^{-7}$   \\
		$\eta \rightarrow X X $      &/   & $\tilde{g}_{\eta} = 0.6 \gau+0.6 \gad-0.9 \gas $ & /    &  $1.3 \cdot 10^{-6}$  \\
		$\eta^\prime \rightarrow  X X $      &/   & $\tilde{g}_{\eta^\prime} = 0.5 \gau+0.5 \gad+ 1.1\gas $ &/   &  $2.0 \cdot 10^{-4}$    \\		
			\hline
			\end{tabular}
			\end{minipage}}
		\caption{ Approximate scaling of the effective couplings for the various meson decays through vector and axial-vector currents, presented in Section~\ref{sec:production}. The dark photon case is also listed for comparison. We refer the reader to Appendix~\ref{app:mesdecaycalc} for details. The last column lists the Standard Model width of the meson, $\Gamma_M$.}
		\label{tab:effcoup} 
	\end{center}
\end{table}


In Fig.~\ref{fig:BRemV} and Fig.~\ref{fig:BRbmlV} we summarise the corresponding branching ratios for the vector case as a function of the dark sector fermion mass. The couplings are chosen to be either aligned to the electromagnetic ones, $g_u=2/3$, $g_d=g_s=-1/3$, for Fig.~\ref{fig:BRemV}, or following a ``baryonic'' coupling $g_u=g_d=g_s=1/3$ for Fig.~\ref{fig:BRbmlV}. We have set the splitting at $\delta_\chi=0$ and $20$ for the solid and dashed lines respectively. In particular, following the scaling presented in Table~\ref{tab:effcoup}, we see that the production from $\rho$ meson decay is strongly suppressed in the baryonic regime. Note that the definition of the effective coupling $g_\rho$ in Table~\ref{tab:effcoup} is presented to one-decimal precision, and therefore should not lead to an exact cancellation in the baryonic regime when $g_u = g_d$. This is reflected in Fig.~\ref{fig:BRbmlV}, where we have used the full numerical precision included in the \texttt{DarkEFT} code.

\paragraph{Axial Vector coupling}

Here the dominant contributions from meson decay arises from the direct decay  $\pi^0, \eta, \eta^\prime \rightarrow \chi \chi$, with a decay width given by

\begin{align}
\label{eq:GammaPAV}
\Gamma_{P,AV} = \frac{\tilde{g}^2_{P} f_\pi^2}{ 8 \pi }  \times  \frac{M_P}{\Lambda^4} \left( M_1 + M_2 \right)^2 \times \left( 1 - \frac{(M_2-M_1)^2}{M_P^2} \right)^{3/2} \left( 1 - \frac{(M_2+M_1)^2}{M_P^2} \right)^{1/2}  \, ,
\end{align}
where we have used the effective couplings $\tilde{g}^2_{P}$ defined in Table~\ref{tab:effcoup} and taken $P \equiv \pi^0, \eta, \eta^\prime$. Notice that, similarly to the standard calculation of the decay $\pi^0 \rightarrow \nu \nu $, the decay amplitude depends quadratically on the dark sector mass $M_1$ due to the helicity suppression of the decay amplitudes~\cite{Herczeg:1981xa}.

In Fig.~\ref{fig:Zal} and Fig.~\ref{fig:uni} we summarise the corresponding branching ratios for the axial-vector case as a function of the dark sector fermion mass with a splitting $\delta_\chi = 0$ ($20$) in solid (dashed) lines. The couplings are chosen to be either $Z$-aligned, with $g_u=1/2$, $g_d=g_s=-1/2$, or uniform across the quarks as in the baryonic case. Notice that this latter case corresponds to a pion-phobic regime and strongly suppresses the production rate at small masses.

Finally, one can obtain the overall number of produced dark sector particles at a given beam dump point by factoring in the typical ratio of mesons per Proton-on-Target (PoT) for the various beam dumps and beam energies. These ratios are listed in Table~\ref{tab:nbmeson}. We have checked using the code \texttt{EPOS-LHC}~\cite{Pierog:2013ria} as distributed within the package CRMC~\cite{CRMC} that the number of
mesons 
per proton-on-target 
were consistent with the ones used in the works referenced in Table~\ref{tab:nbmeson}. Where only partial information on meson production was available we completed the table using the above codes. The overall normalisation (typically given by the number of $\pi^0$ per PoT) tends however to vary strongly from analysis to analysis. In the best cases,  \texttt{GEANT4} simulations of the full hadronic and electromagnetic cascade, supplemented by experimental data are used (as for MiniBooNE in~\cite{Aguilar-Arevalo:2018wea}). Several other studies use either \texttt{PYTHIA8} simulations of a $pp$ process with similar center-of-mass energy, or experimental data from $pp$ collision to extract the meson multiplicity, which tend to underestimate the production (hence leading to conservative limits). Intermediate approaches, such as those presented above using \texttt{EPOS-LHC}, simulate the $p N$ process, but do not include the subsequent showers. In our case, we typically do not choose between the various methods since the overall normalisation depends on the analysis that we will use later for  recasting limits; certain studies choose only to keep primary mesons while others use the full hadronic shower components. In particular, this is the case for the projection for NO$\nu$A from Ref.~\cite{deNiverville:2018dbu} which chooses to keep only one $\pi^0$ meson per proton on target. For the case of SHiP, where we provide estimates for a $10$ events reach, we use the overall normalisation to be around $\sim 10 \pi^0 / $PoT, based on the \texttt{EPOS-LHC} results (though we note that a \texttt{PYTHIA8} approach in Ref.~\cite{Buonocore:2018xjk} found around  $\sim 6 \pi^0 / $PoT).


\begin{table}[t]
	\centering
	\resizebox{1\textwidth}{!}{\begin{minipage}{1.07\textwidth}
		\begin{tabular}{|c|c|c|c| c c c c c|}
		\hline
				\rule{0pt}{14pt}Experiment & $E_{\rm beam}$&  Target & PoT & $N_{\pi^0}$ & $N_{\eta}$ & $N_{\eta^\prime}$ & $N_{\rho}$ & $N_{\omega}$ \\
				\hline
				\hline
				\rule{0pt}{14pt}
			 CHARM~\cite{Bergsma:1985is}& $400$ GeV& Cu & $2.4 \cdot 10^{18}$ &$2.4$ & $0.3$ & $0.03$ & $0.3$ & $0.25$ \\
			 LSND~\cite{Athanassopoulos:1997er} & $0.8$ GeV& Water & $0.92 \cdot 10^{23}$ &$0.14$ & $0.$ & $0$ & $0$ & $0$ \\
			 MiniBooNE~\cite{Aguilar-Arevalo:2018wea} & $8$ GeV& Fe & $1.86 \cdot 10^{20}$ &$2.4$ & $0.1$ & $0$ & $0.1$ & $0.1$ \\
				\hline
				\rule{0pt}{14pt}
		SHiP~\cite{Anelli:2015pba} & $400$ GeV& W / Pb* & $2 \cdot 10^{20}$ & $10$ & $1$ & $0.08$ & $1.1$ & $1$ \\
			NO$\nu$A~\cite{deNiverville:2018dbu}& $120$ GeV& C & $3 \cdot 10^{20}$ &$1$ & $ 1/30$ & $ 1/300$ & $1/30$ & $1/30$ \\
			SeaQuest~\cite{Berlin:2018pwi} & $120$ GeV& Fe & $1.44 \cdot 10^{18}$ &$3.5$ & $0.4$ & $0.04$ & $0.4$ & $0.4$ \\
				\hline
				\rule{0pt}{14pt}
			HL-LHC (barn)~\cite{Feng:2017uoz} & $14$ TeV& pp & $\mathcal{L}=3$ ab${}^{-1}$ &$4.3$ & $0.5$ & $0.05$ & $0.5$ & $0.5$ \\		
						\hline
			\end{tabular}
			\end{minipage}}
		\caption{Beam and target characteristics for various experiments, along with the total number of protons on target (PoT) and the average number of a given meson per proton. $E_{\rm beam}$ is the beam kinetic energy. The SHiP design is not final. References point either to an analysis paper in the case of existing constraints, or to prospective bounds in the case of future experiments. The ratios quoted for NO$\nu$A account only for primary mesons (see~\cite{deNiverville:2018dbu}). For MiniBooNE, we use the ratios and normalisation from the recent off-target analysis~\cite{Aguilar-Arevalo:2018wea}. }
		\label{tab:nbmeson} 
\end{table}


\subsection{Heavy meson decays}
\label{sec:heavymes}

Due to their larger masses, heavy meson decays into dark sectors are significantly less suppressed than the light ones. Despite their low production rates in most intensity frontier experiments, they can still be important production mechanisms for new light states.\footnote{For example in the case of Higgs portal scenarios, where the presence of a Yukawa interaction in the coupling between the dark sector fields and the quarks typically favour heavy mesons. We leave for future work the recasting of existing studies such as~\cite{Feng:2017vli}.} Furthermore, since their kinematic distributions differ significantly from the ones of the lighter mesons, the final experimental efficiencies for dark sector particles originating from their decay cannot be inferred from existing decay and scattering searches which do not include them. Consequently, we will follow two complementary directions: We calculate directly the limits from their invisible decays, for which one need not assume any particular detection efficiencies, and we present naive 10-event projections at SHiP as an order-of-magnitude estimate for the reach of searches based on heavy meson decays.

We shall focus on the decays mediated by the vector effective operator, which can be separated into flavour diagonal ones, $\phi, J/\Psi, \Upsilon \to \chi \chi$, and flavour-violating ones such as $B\to K \chi \chi, \pi \chi \chi$, etc. The expression for the decay width of the former has been already estimated in Eq.~\eqref{eq:GammaVectMesondecay}, with the corresponding decay constants given in Table~\ref{tab:effcoupheavymes} of Appendix~\ref{app:heavymesdecaycalc}. 

\begin{figure}[t]
	\centering
		\includegraphics[width=0.7\textwidth]{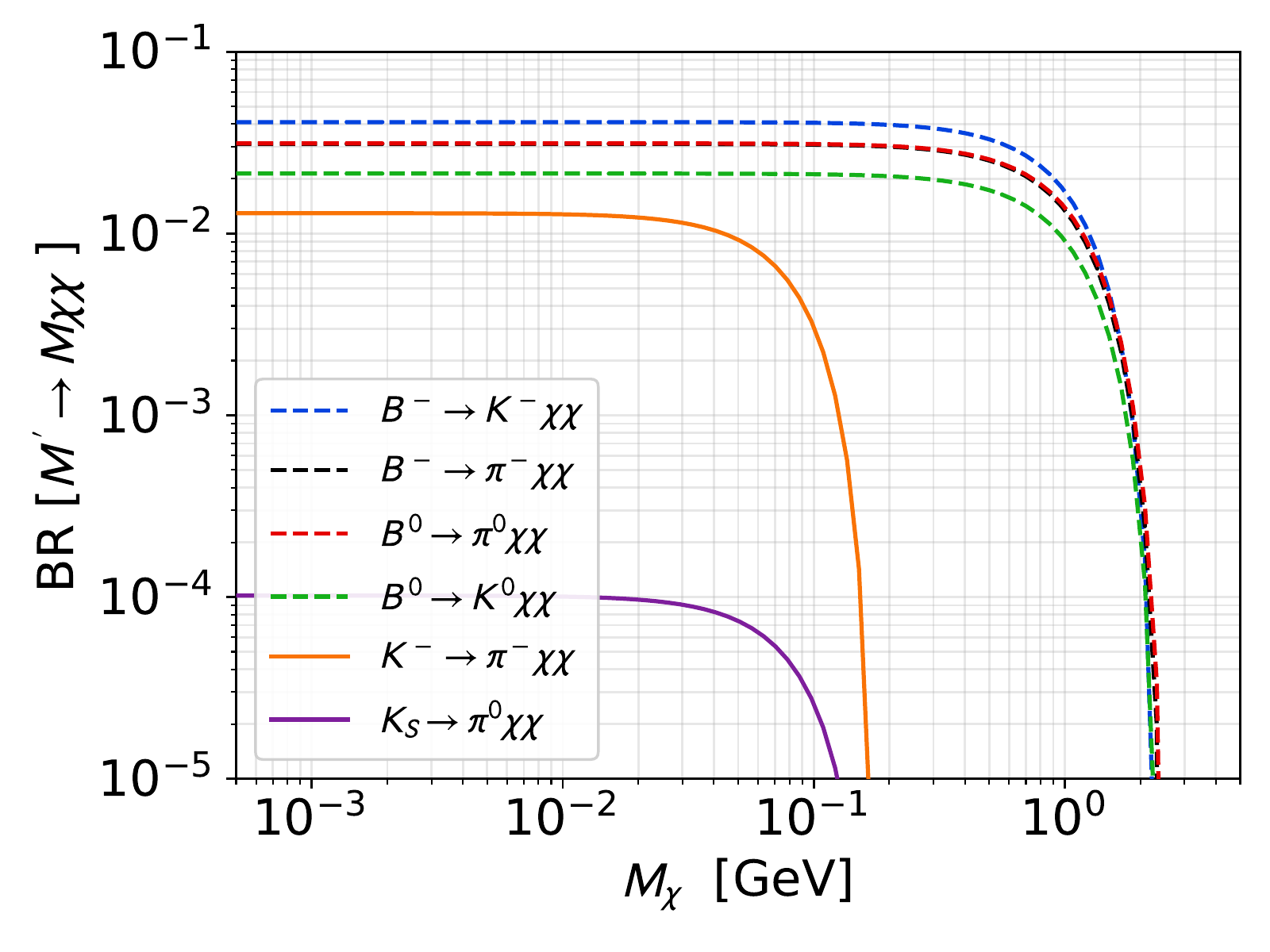}
	\hspace{0.02\textwidth}
	\caption{Branching ratios as function of the light dark sector mass for $B$ and $K$ heavy mesons, with $M_1 = M_2 = M_\chi$. We have set the relevant couplings $g_{ij}=1$ and the effective scale $\Lambda = 1$ TeV. }
	\label{fig:heavymesBR}
\end{figure}

For the three-body decay of heavy pseudo-scalar mesons, we extend the study of Ref.~\cite{Bjorkeroth:2018dzu} for the case of axions to our four-fermion portal operator generated by an off-shell mediator (see also Ref.~\cite{Kamenik:2011vy}). We consider the $B$ and $K$ mesons, whose sensitivities are enhanced by their heavier mass and flavour-violating decays through off-diagonal vector couplings: 
\begin{equation}
\mathcal{L} \supset \sum_{ij} \frac{g_{ij}}{\Lambda^2}(\bar\chi_1 \gamma_\mu \chi_2) (\bar{q}_i \gamma^{\mu} q_j) \, .
\end{equation}
The three-body decay of a heavy meson $P$ (mass $M$) into a lighter meson $P^\prime$ (mass $M^\prime$) and two dark sector fermions $\chi$ through a contact interaction is derived in Appendix~\ref{app:heavymesdecaycalc}. In the massless $\chi$ limit, denoting $g_{P P^\prime}$ as the relevant effective coupling (defined in Table~\ref{tab:effcoupheavymes} in terms of effective quark couplings $g_{ij}$ where $i,j$ denote the quark flavour), the decay width reduces to
\begin{equation}
\Gamma_{P \to P^\prime \chi \chi} =\frac{g_{P P^\prime}^2|f_+(0)|^2}{\Lambda^4} \frac{\left(M^8 -8M^6{M^\prime}^2 + 8M^2{M^\prime}^6 - {M^\prime}^8 + 24M^4{M^\prime}^4\log{\frac{M}{M^\prime}}\right)}{768\pi^3 M^3} \, .
\end{equation}
The hadronic form factor $f_+(q^2)$ is obtained from lattice results in Ref.~\cite{AlHaydari:2009zr}; for a momentum transfer $q^2 \equiv (p_P - p_{P^\prime})^2 \to 0$ its value is between $0.23$ and $1$ depending on the decay process (see Table~\ref{tab:effcoupheavymes}). Since the $q^2$ dependence of the form factor does not vary significantly for the purpose of setting limits, we take these constant values as a good approximation. The branching ratios are plotted for various $B$ and $K$ meson decays in Fig.~\ref{fig:heavymesBR} as a function of the light dark sector mass with no splitting between the two dark sector masses.  


\begin{table}[t]
	\begin{center}
	\resizebox{1.\textwidth}{!}{\begin{minipage}{1.14\textwidth}
		\begin{tabular}{|c|c|c|c|c|c|c|}
				\hline
				\rule{0pt}{14pt}Meson decay & $\phi \to \chi \chi$ & $J/\Psi \to \chi \chi$ & $\Upsilon \to \chi \chi$ & $B \to K \chi \chi$  & $B \to \pi \chi \chi$ & $K \to \pi \chi \chi$  \\
				\hline
				\hline
				\rule{0pt}{14pt}
			$f_+ (0)$ / Decay const. & $241$ MeV & $418$ MeV & $649$ MeV & $0.32$ & $0.27$ & $1.$  \\[0.4em]
			\hline
			Effective coupling $(g_{PP'})$ & $g_{ss}$ & $g_{cc}$& $g_{bb}$& $g_{bs}$& $g_{bd}$& $g_{sd}$ \\[0.3em]
			\hline
			\end{tabular}
			\end{minipage}}
		\caption{Form factors $f_+ (0)$ in the case of pseudo-scalar mesons decay from~\cite{AlHaydari:2009zr,Bjorkeroth:2018dzu} of the form $P' \to P \chi \chi$ and effective decay constant for the flavoured vector meson decay from~\cite{Hazard:2016fnc}, as well as the relevant effective couplings $(g_{PP'})$ for the various heavy meson decays through vector currents, presented in Section~\ref{sec:heavymes}.}
		\label{tab:effcoupheavymes} 
	\end{center}
\end{table}


In order to get an estimate for the number of $B$ mesons produced at SHiP, we multiply the number of protons on target, $N_\text{prot} = 2 \times 10^{20}$, by the ratio of the $B$ meson production cross-section per nucleon, $\sigma_B \simeq 3.6$ nb, to the total proton-nucleon cross-section, $\sigma_{pN} \simeq 40$ mb. For $K_S^0$, $K^-$ and $K^+$ we find that the multiplicities of 0.232, 0.224 and 0.331, respectively, from Ref.~\cite{Choi:2019pos} agree well with our \texttt{EPOS} simulations for proton-proton collisions at $\sqrt{s} = 27.4$ GeV. However, the target for SHiP is tungsten where we find instead a factor of $\sim 2.5$ enhancement in multiplicities. These numbers are used to generate a naive projection of limits assuming 10 events, based on the routine presented in the next sections.\footnote{The study of such signatures in the case of a new light decaying pseudo-scalar was done for example in Refs.~\cite{Kamenik:2011vy,Dobrich:2018jyi}.}

\subsection{Parton-level dark sector production}
\label{sec:directprod}
When the centre-of-mass energy $E_{\rm cm}$ of the relevant beam-dump or collider experiment is larger than $\mathcal{O} (1)$ GeV, direct parton-level production can become relevant. In the case of dark sector production at beam dump experiments for a target of atomic number $Z$ and mass number $A$, the final direct cross-section production is given by
\begin{align}
\sigma_{p + A \rightarrow \chi \bar{\chi}} = Z  \sigma_{p + p \rightarrow \chi \bar{\chi}}  + (A-Z) \sigma_{p + n \rightarrow \chi \bar{\chi}} \ . 
\end{align}
The total number of produced pairs of dark sector particles  can then be deduced as
\begin{align}
    N_{\rm direct} = \frac{\sigma_{p + A \rightarrow \chi \bar{\chi}} }{\sigma_{p + A}} N_{\rm PoT} \simeq  \frac{ Z/A ~\sigma_{p+p \rightarrow \chi \bar{\chi}} +  (1-Z/A) ~\sigma_{p + n \rightarrow \chi \bar{\chi}} }{\sigma_{pp}} N_{\rm PoT} \ ,
\end{align}
where $N_{\rm PoT}$ is the number of protons on target and $\sigma_{p + A}$ is the total scattering cross-section on the material. The second equality makes the (strong) assumption that ``screening'' effects, which make the typical proton-nuclei cross-sections scale proportionally to $A^m$ rather than $A$ (typically with $m \sim 0.7$ for the energy of intensity frontier experiments), apply similarly to new physics processes as to the Standard Model~\footnote{Note furthermore that this assumes that the experiment is designed such that all protons of the beam interacts with the target. This approximation also does not account for the hadronic shower development.} (see e.g.~\cite{Glauber:1955qq,AlHaydari:2009zr,Gustafson:2015ada}).

Assuming for now that both final dark sector states have similar mass, the cross-section can be written for a process with a center-of-mass energy $\sqrt{s}$ as 
\begin{align}
    \sigma(p (P_1) + N (P_2) \rightarrow \chi \bar{\chi}) = \int_0^1\int_0^1 dx_1 dx_2 \sum_{q,\bar{q}} f_q^N (x_1) f_{\bar{q}}^N (x_2) \times \sigma ( ~q (x_1 P_1) \ \bar{q}(x_2 P_2) \rightarrow \chi \bar{\chi}) \ ,
\end{align}
where we have introduced the parton distribution function (PDF) $f_q^N$ for the parton $N$ (proton $p$ or neutron $n$), the sum runs over all quarks and antiquarks and the initial momentum of the quarks in the cross-section $\sigma ( q \bar{q} \rightarrow \chi \bar{\chi})$ depends on the momentum fractions $x_1$ and $x_2$.

\begin{figure}[t]
	\centering
		\subfloat[]{%
		\label{fig:proddirecta}%
		\includegraphics[width=0.47\textwidth]{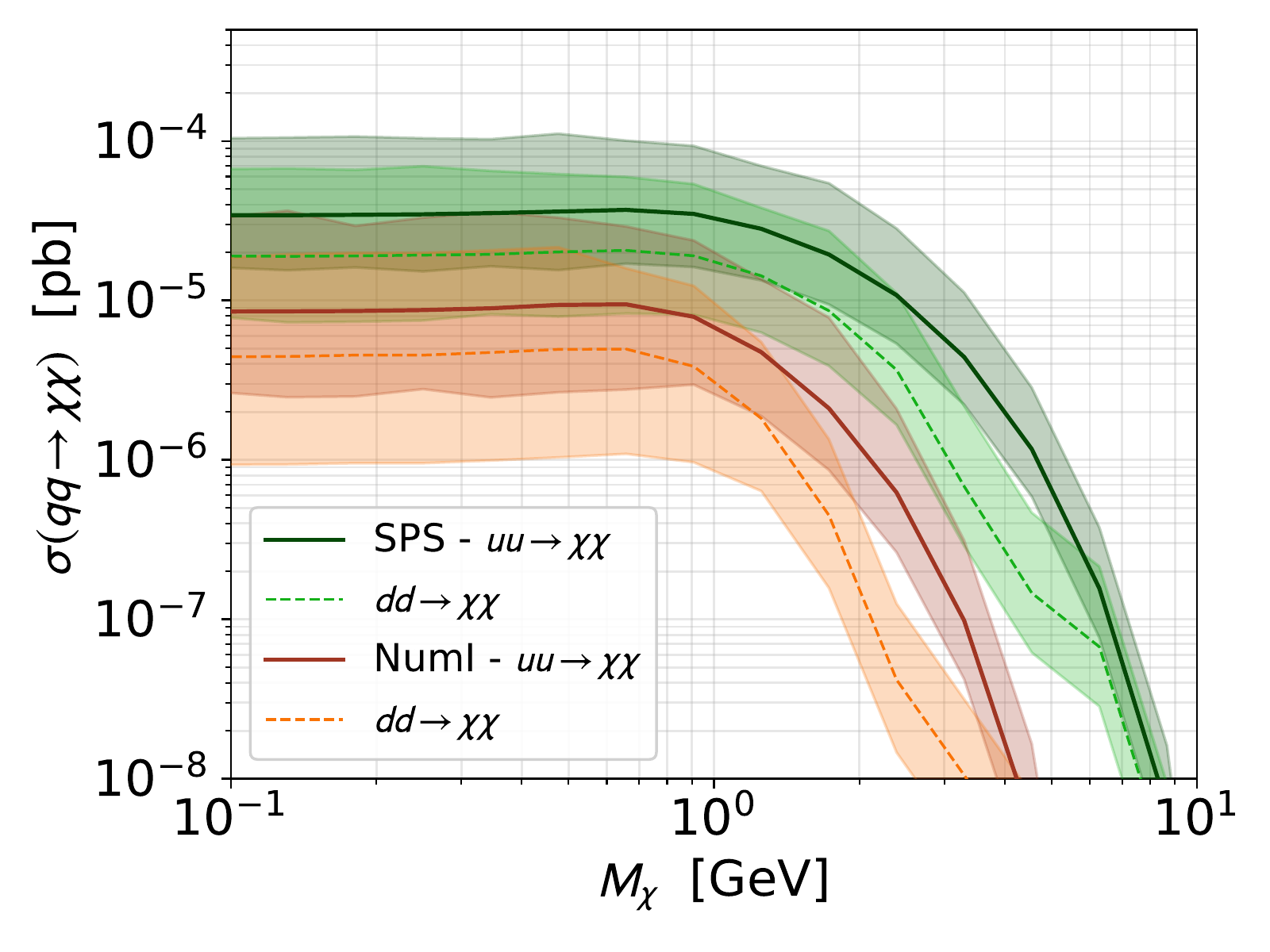}
	}%
			\subfloat[]{%
		\label{fig:proddirectb}%
	\includegraphics[width=0.47\textwidth]{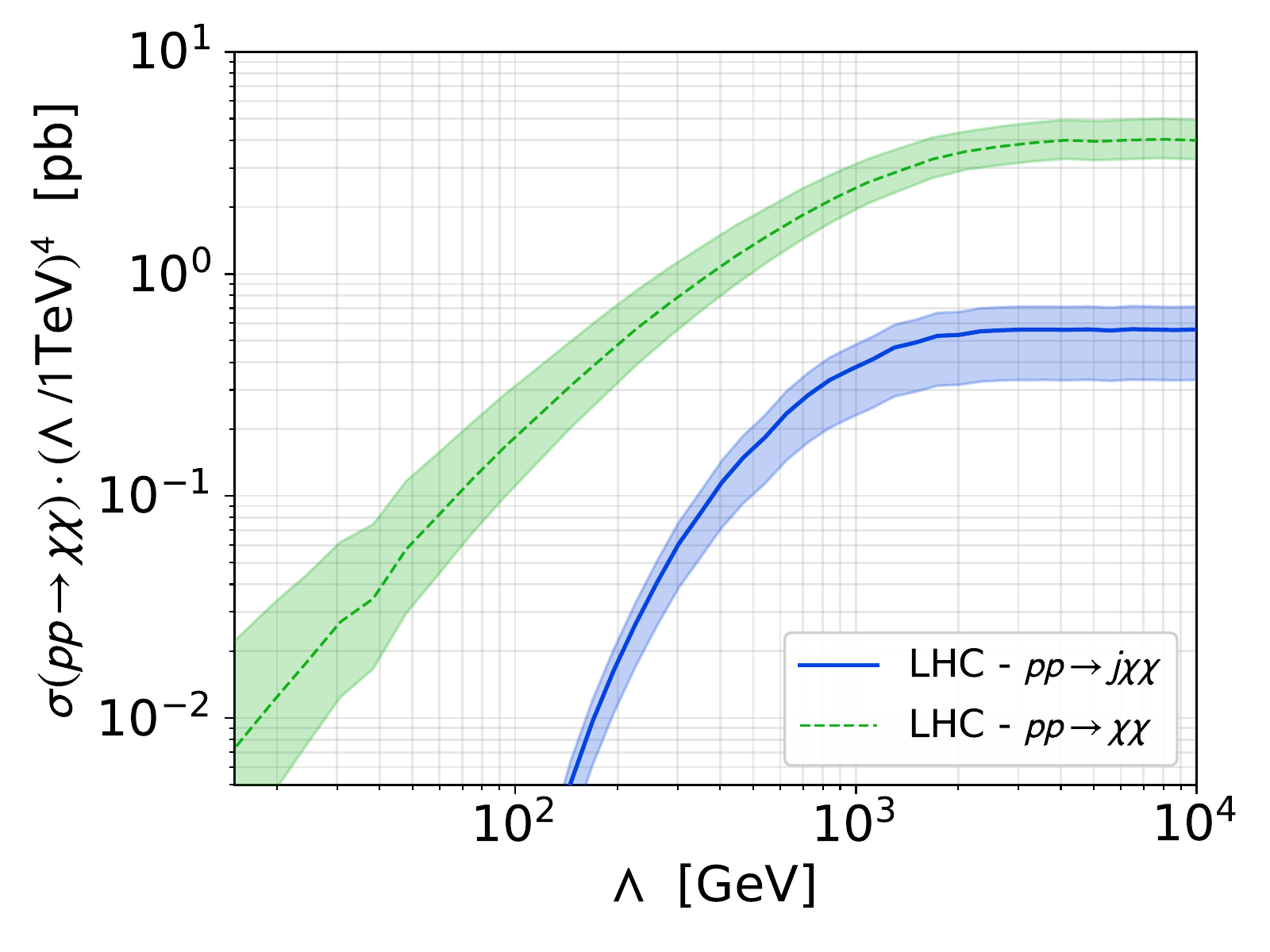}
	}%

	\caption{(a) Cross-sections for the proton partonic production processes at $\Lambda = 1$ TeV for $u \bar{u}, d \bar{d} \rightarrow \chi \bar{\chi} $ for the SPS ($400$ GeV) beam and for the NuMI ($120$ GeV) beam
	(b) Cross-section times $\Lambda^4$ at LHC for $p p \rightarrow \chi \bar{\chi} + \rm{jets}$ as a function of the new physics scale $\Lambda$. In both cases, the theoretical errors on the cross-section 
	are obtained by varying the renormalisation scale by a factor of $2$ or $1/2$ with \texttt{MadGraph}.
}
\end{figure}

In practice the neutron PDFs can be determined from those for protons using isospin symmetry. Factorising the effective couplings between the dark sector fermions and the quarks, $g_u$ and $g_s$, we obtain
\begin{align}
    \sigma_{p + p \rightarrow \chi \bar{\chi}} & = 2 \int_0^1\int_0^1 dx_1 dx_2 \left[ f_u^p f_{\bar{u}}^p ~g_u^2 +  f_d^p f_{\bar{d}}^p ~g_d^2 \right] \times \sigma_{q \bar{q} \rightarrow \chi \bar{\chi}} \ , \\
        \sigma_{p + n \rightarrow \chi \bar{\chi}} & =  \int_0^1\int_0^1 dx_1 dx_2 \left[ f_u^p f_{\bar{d}}^p +  f_d^p f_{\bar{u}}^p \right] (g_u^2 + g_d^2) \times \sigma_{q \bar{q} \rightarrow \chi \bar{\chi}} \ ,
\end{align}
where we neglected the quark masses and factored out the couplings in the last cross-section to obtain~\footnote{Similar results can be applied in the case of the dark photon of mass $M_V$, which we will use to recast existing searches with $M_1 \sim M_2 \sim M_\chi$:
\begin{align}
\sigma_{q \bar{q} \rightarrow \chi \bar{\chi}} = \frac{1}{36 \pi} \frac{s}{(s-M_V^2)^2+M_V^2 \Gamma_V^2}\sqrt{1-\frac{4 M_\chi^2}{s}} \left(1 + 2 \frac{M_\chi^2}{s}  \right) \ ,
\end{align}
and the couplings are given as $g_q^2 = g_D \epsilon e Q_q$ with $Q_q$ the quark electric charge, $g_D$ the dark gauge coupling and $\epsilon$ the kinetic mixing. }
\begin{align}
\sigma_{q \bar{q} \rightarrow \chi \bar{\chi}} = \frac{1}{36 \pi} \frac{s}{\Lambda^4}  \left( 1 - \frac{(M_2-M_1)^2}{s} \right)^{3/2} \left( 1 - \frac{(M_2+M_1)^2}{s} \right)^{1/2} \left( 2 + \frac{(M_2 + M_1)^2}{s} \right) \ .
\end{align}
It turns out that in the regime of interest, the PDF ratios follow the scaling 
\begin{align}
\label{eq:sigmapnpp}
    \sigma_{p + p \rightarrow \chi \bar{\chi}}  & \sim ( 2 g_u^2 + 1.2 g_d^2 )\sigma_{p + p \rightarrow \chi \bar{\chi}}^0 \\ 
     \sigma_{p + n \rightarrow \chi \bar{\chi}}  &\sim \sigma_{p + p \rightarrow \chi \bar{\chi}}  / 2 \sim  (g_u^2 + 0.5 g_d^2) \sigma_{p + p \rightarrow \chi \bar{\chi}}^0 
\end{align}
where $ \sigma_{p + p \rightarrow \chi \bar{\chi}}^0  $ is estimated in the electromagnetic alignment $g_u = 2/3, g_d = -1/3 $ and shown in Table~\ref{tab:CSdirect_limit} in the limit $M_1, M_2 \ll E_{\rm cm} $. This scaling is accurate at $20\%$ in the region of low $M_1 + M_2$ mass compared to the center-of-mass energy $E_{\rm cm}$. We have estimated the cross-section using CTEQ6.6M pdfs both directly from the above formula and through \amc ~platform~\cite{Alwall:2014hca} with the effective theory model implemented in \fr~\cite{Alloul:2013bka} to create the UFO module~\cite{Degrande:2011ua}. The renormalisation scale choice was left dynamical, chosen as the sum of the transverse mass of the outgoing dark sector fields divided by $2$.\footnote{ When estimated directly, we chose $Q=500$ GeV for the LHC though we note that Eq.~\eqref{eq:sigmapnpp} was not significantly modified by varying it from $250$ GeV to $1$ TeV.} We show in Fig.~\ref{fig:proddirecta} the cross-sections for the $u \bar{u}, d \bar{d} \rightarrow \chi \bar{\chi} $ (for a proton-proton process), which corresponds to choosing $g_u=1, g_d=0$ and $g_d=1,g_u=0$ respectively in the equations above. Note that in the numerical process, we combine both curves using the relations from Eq.~\eqref{eq:sigmapnpp} and account for a slight dependence of the coefficient on the energy of the initial beam.

We implemented the dark photon direct production with the same approach in order to recast the limits later on. In this case the renormalisation scale is set to the dark photon mass corresponding to the resonant Drell-Yan production~\cite{Batell:2009di}. Since the dark photon is typically quite light, we probe a different renormalisation scale range in this case, but one still has typically
\begin{align}
    \sigma^{\rm DP}_{p + p \rightarrow V \rightarrow \chi \bar{\chi}} \sim 2 \sigma^{\rm DP}_{p + n \rightarrow V \rightarrow \chi \bar{\chi}} \ .
\end{align}


\begin{table}[t]
	\begin{center}
		\begin{tabular}{|c|c|c|c|c|}
			\hline
			 &\rule{0pt}{15pt} LHC (14 TeV)  & LHC (13 TeV)  & SPS ($400$ GeV) & FNAL ($120$ GeV) \\
			\hline
			\hline
			\rule{0pt}{15pt}
		$ \sigma^{0}$ ($p p \rightarrow \chi_1 \chi_2$) & $4.6$ pb & $4.0$ pb & $0.016$ fb&   $0.004$ fb \\[0.2em]
			\hline
		\end{tabular}
		\caption{Cross-section for direct production in various high-energy proton beam experiments for $M_1 + M_2 \ll E_{\rm cm}$ for the effective couplings (as defined in Eq.~\eqref{eq:opV} and Eq.~\eqref{eq:opAV}) chosen as $g_u = 2/3, g_d=-1/3$, with the operator scale of $\Lambda = 1$ TeV.}
		\label{tab:CSdirect_limit} 
	\end{center}
\end{table}

 As long as the total mass of the dark sector fermion pair $M_1 + M_2 \ll E_{\rm cm} \ll \Lambda$, the cross-section is roughly constant and independent of the actual dark sector masses. But an additional difficulty arises when the effective scale becomes of the order of the centre-of-mass energy of the process considered; the effective theory starts becoming unreliable since direct mediator production should dominate. This issue is mostly relevant for LHC searches and has been intensely studied in recent years following searches for dark matter at the LHC through effective operators. While various approaches have been suggested (see e.g.~\cite{Racco:2015dxa} for a brief summary), the general strategy is to restrict on an event-by-event basis at the Monte-Carlo generator level the typical energy of the process to be below the effective field theory scale.\footnote{Depending on the model, one can choose the most generic scale: $E_{cm}$, or, if the process is derived for example from a dark photon model, the maximal virtuality $Q_{tr} = \sqrt{|p_{\chi_1}+p_{\chi_2}|^2 }$ of the mediator (as used recently in e.g.~\cite{Belyaev:2018pqr}).}

While we will later consider briefly for completeness some of the mono-X limits from the LHC, our dominant interest and strongest bound will come from direct ``dark'' production of a dark sector pair that is subsequently detected by a dedicated experiment such as FASER or MATHUSLA. Removing the requirement for an additional high-$p_T$ particle significantly increases the potential reach of the EFT. We illustrate this in Fig.~\ref{fig:proddirectb} for the associated  $pp \rightarrow \chi \chi X$ dark sector production (relevant for mono-X searches) where we present the typical cross-section as a function of the effective operator scale, following the procedure described in Ref.~\cite{Racco:2015dxa}.

\section{Hunting for the fermion portal's dark sectors}
\label{sec:acceleratorsearches}

\subsection{Long-lived dark sectors}
\label{sec:decay}

Searches for hidden particles in both beam-dump and accelerator-based experiments can typically be decomposed between a production stage and detection stage. The former was described in the previous section; it usually takes place at an interaction point (either the beam dump target or collision point for accelerators), while the latter occurs in a shielded detector farther away. Furthermore, most of the searches follow simple cut-and-count strategies (up to rare exceptions, for instance missing energy searches~\cite{Lees:2017lec,Banerjee:2017hhz}); we can therefore decompose the expected number of signal events as
\begin{align}
\mathcal{N} \simeq 	\mathcal{N}_{\rm prod} \times \mathcal{E} \times \mathcal{P}_{\rm sig} \, ,
\label{eq:Nevents}
\end{align}
where $\mathcal{N}_{\rm prod}$ is the number of produced dark sector states, $\mathcal{E}$ is a detection efficiency which contains all the details about the search channel efficiency of the experiment considered, and $\mathcal{P}_{\rm sig}$ is the probability that a dark sector state leads to a signal event in the detector (for instance a decay or a scattering). In most of this section, we will place limits on our fermion portal framework by reinterpreting existing dark sector searches focused on dark matter, dark photons or dark Higgs bosons. We neglect to a first approximation the difference in kinematics.

For a relativistic long-lived particle decaying through the operators~\eqref{eq:opV} or~\eqref{eq:opAV}, the detection probability depends directly on the probability of observing a decay in the decay volume of the detector.
\begin{align}
 \mathcal{P}_{\rm sig} = e^{-\Gamma_2 \frac{D}{\hbar c \gamma}} \left( 1 - e^{-\Gamma_2 \frac{L}{\hbar c \gamma}} \right) \ ,
 \label{eq:DetProb}
\end{align}
where we have defined $\Gamma_2$ to be the decay width of the heavier unstable state $\chi_2$, $D$ the distance it travels before entering the decay volume, $L$ the distance travelled in the decay volume and $\gamma$ its boost factor.
We present in Table~\ref{tab:paramExp} the values of these parameter for various experiments. In particular, the value of the boost factor is critical in determining the lower reach of the experiments (which corresponds to the short lifetime limit for $\chi_2$). We obtained the average boost factor from direct simulations of meson decays using \texttt{BdNMC}~\cite{deNiverville:2016rqh} modified to handle the decay of a dark photon mediator into two dark sector states $\chi_1$ and $\chi_2$ of different masses~\cite{Darme:2017glc,Darme:2018jmx}.\footnote{We obtain the average boost factor for the particles which intersect the detector, so it is slightly higher than the average boost factor at the interaction point.} The boost factor in the case of parton-level production (relevant for dark sector masses around the GeV) is then obtained by rescaling the average energy of the dark sector pair according to its invariant mass $M_{12}^2 = (p_1 + p_2)^2$ as $\sqrt{E_{\rm mes}^2 - M_{12}^2}$, with $E_{\rm mes}$ the average meson kinetic energy. An important difference with respect to the fermion portal's phenomenology is that the typical fermion portal boost factor from parton level events is significantly lower than the one typically obtained from dark bremsstrahlung of dark photons (since in the latter case the dark photon carries off most of the beam energy). 


\begin{table}[t]
	\begin{center}
		\begin{tabular}{|c|c|c|c|c|}
			\hline
			\hline
			\rule{0pt}{16pt}
			Experiment & Distance   & Decay length & Average $|\vec{p}|$ [GeV] & PoT / Lumi. \\
			 &  $D$ [m] & $L$ [m]  &  Meson &\\
			\hline
			\hline
			\rule{0pt}{16pt}
			LSND~\cite{Athanassopoulos:1997er}       & $34$    & $8.3$  & $0.1$   & $0.92 \times 10^{23}$\\
			MiniBooNE~\cite{Aguilar-Arevalo:2018wea}      & $488$    & $R= 6$  &  $1$   & $1.86 \times 10^{20}$ \\
		    SBND~\cite{Antonello:2015lea}        & $110$    & $5$  & $1$    & $6.6 \times 10^{20}$ \\
		    \hline
		    \rule{0pt}{16pt}
		     NO$\nu$A~\cite{Ayres:2004js}        & $990$    & $14.3$  & $8$   & $3 \times 10^{20}$  \\
		    SeaQuest~\cite{Aidala:2017ofy,Berlin:2018pwi}       & $5$    & $5$  & $8$   & $1 \times 10^{20}$  \\
		    CHARM~\cite{Bergsma:1985is}        & $480$    & $35$  & $14$    & $2.4 \times 10^{18}$ \\
			SHiP~\cite{Anelli:2015pba}        & $60$    & $65$  & $14$   & $2 \times 10^{20}$ \\
			\hline	
			\rule{0pt}{16pt}
			MATHUSLA~\cite{Lubatti:2019vkf}        & $100$    & $35$  & $1000.$  &  $3 \times 10^{3}$ fb${}^{-1}$ \\						
			FASER~\cite{Feng:2017uoz}        & $480$    & $10$  & $1000.$  & $3 \times 10^{3}$ fb${}^{-1}$\\				
			\hline	
		\end{tabular}
		\caption{Experimental data for various relevant high-intensity frontier experiment. Note that MiniBooNE is a spherical detector of radius $ R\sim 6 $m. For standard beam-dump experiments, the average boost factor has been determined from direct simulations using a modified \texttt{BdNMC}~\cite{deNiverville:2016rqh,Darme:2017glc,Darme:2018jmx} from a dark photon mediator. For LHC-based experiments, we used Ref.~\cite{Berlin:2018jbm} for FASER and MATHUSLA. Energy quantities are in GeV while distances are in meters.}
		\label{tab:paramExp} 
	\end{center}
\end{table}


Existing limits in the literature usually focus either on the case of a decaying dark photon or on the inelastic dark matter scenario, where the decay of the heaviest state is mediated by an off-shell dark photon. To estimate the sensitivity to the fermion portal we will recast these results for our effective theory in three steps:
\begin{enumerate}
\item We simulate the typical number of produced dark sector particles, both for the existing inelastic dark matter limits, $\mathcal{N}_{\rm prod}^{\rm DP}$, (typically for $\delta_\chi \equiv |M_2|/|M_1| - 1 = 0.1 $) and for our effective theory with the required $M_1$ and $M_2$, $\mathcal{N}_{\rm prod}^{\rm eff}$. Note that we use a kinetic mixing parameter $\varepsilon=0.001$ and dark sector coupling $\alpha_D = 0.1$ for the former,\footnote{We define the kinetic mixing as
\begin{align}
\mathcal{L} &\supset -\frac{1}{2}\frac{\varepsilon}{\cos\theta_w}B_{\mu\nu}F^{\prime\mu\nu} 
\end{align}
with $B_{\mu\nu}$ the hypercharge field strength, $F^{\prime\mu\nu}$ the dark photon field strength.} and $\Lambda / \sqrt{g} = 1$ TeV for the latter (with the effective coupling's ratio either electromagnetically-aligned or Z-aligned). This choice has no consequences on the results since the scaling with respect to these parameters is trivial.
\item Focusing on the very long-lived case, where $c \tau \gamma \gg D, L $, the parameters of the experiments can be cancelled out of the ratio
\begin{align}
\frac{ \mathcal{P}^{\rm DP}_{\rm sig}}{ \mathcal{P}^{\rm eff}_{\rm sig}} = \frac{\Gamma_{\rm DP}}{\Gamma_{\rm eff}} \simeq 700 \left(\frac{M_1}{\Lambda/\sqrt{g_e}} \right)^4  \left(\frac{1}{\varepsilon_{\rm lim}} \right)^2 \ ,
\end{align}
where $g_e$ is the effective coupling to electrons.
\item Finally, we assume that the efficiencies of the experiment between the inelastic dark matter and our effective theory case will be similar for equivalent invariant masses of the $\chi_1 \chi_2$ pair, so that using Eq.~\eqref{eq:Nevents} for both the inelastic dark matter and the fermion portal case leads to 
\begin{align}
\Lambda_{\rm lim} = 410 \textrm{ GeV} \times \sqrt{g_{e}}   \left(\frac{0.001}{\varepsilon_{\rm lim}} \right)^{1/2}  \left(\frac{\mathcal{N}_{\rm prod}^{\rm eff}}{\mathcal{N}_{\rm prod}^{\rm DP}} \right)^{1/8} \, .
\end{align} 
\end{enumerate}

This procedure can also be used to obtain the bounds for different splittings between the two hidden sector states. In this case we first proceed to estimate the efficiency around the limit by inverting Eq.~\eqref{eq:Nevents}. Assuming that  the detection efficiency $\eff$ depends dominantly on the invariant mass of the $\chi_1 \chi_2$ pair $M_{12}$, we need to match the efficiency for the production and detection of two states with splitting $\delta_\chi$ to the efficiency $\eff^\prime$ of two states with splitting $\delta_\chi^\prime $.  We have
\begin{align}
    \eff^\prime (M_2^\prime) \sim \eff \left(M_2^\prime \frac{2+\delta_\chi^\prime }{(1+\delta_\chi)(1+\delta_\chi^\prime )} \right) \ ,
\end{align}
where we have assumed that the first efficiency was given as a function of $M_1$, while we want the recasted one as a function of $M_2^\prime$ (which is the most relevant mass for large splitting). 

Finally, a last difficulty occurs due to the lower kinematic threshold $2 m_e$ of these decay searches based on $\chi_2 \to \chi_1 e^+ e^-$. Since most of the initial limits are estimated at small splitting, we cannot estimate the efficiency in the range $M_{12} \in [2 m_e , \frac{2 + \delta_\chi}{\delta_\chi} 2 m_e]$. We sidestep the issue by scaling $\eff$ from the available range of
\begin{equation*}
\left[\frac{2 + \delta_\chi}{\delta_\chi} 2 m_e , \frac{M_\pi^0}{3} \right] \, ,  
\end{equation*}
to 
\begin{equation*}
\left[\frac{2 + \delta_\chi^\prime}{\delta_\chi^\prime} 2 m_e , \frac{M_\pi^0}{3} \right] \, ,
\end{equation*}
where the upper limit is arbitrarily chosen to the be significantly larger than the lower threshold and nonetheless small enough so that the effect from the $\pi^0$ production threshold remains small (we have checked that the limits do not significantly depend on the precise value for the upper limit).

An important subtlety, however, lies with the appearance of the lower limit in $\Lambda$, arising typically when the long-lived particle decays mainly \textit{before} reaching the detector. In this case one needs to use all the geometric parameters of the experiment to estimate the decay probability. We typically use the upper limit to deduce the lower limit, using the fact that at fixed masses, the production rate scales simply as $1/\Lambda^4$ so that the only technically challenging quantity to estimate is the decay probability ratio. More precisely, we require
\begin{align}
  \mathcal{P}^{\rm up}_{\rm sig} =     \left( \frac{\Lambda_{\rm low}}{\Lambda_{\rm up}} \right)^4 \times \mathcal{P}^{\rm low}_{\rm sig} \ .
\end{align}
Assuming that the decay probability~\eqref{eq:DetProb} in the lower regime is dominated by the exponentially-suppressed first contribution, the above equation leads to a simple transcendental equation on $\Lambda_{\rm low}$,
\begin{align}
    \Lambda_{\rm low}^{-4} ~\exp \left(\frac{A}{\Lambda_{\rm low}^4}\right) = \Lambda_{\rm up}^{-4} \mathcal{P}^{\rm up}_{\rm sig} \ , \quad \textrm{ where} \quad A \equiv \frac{D \Lambda_{\rm up}^4 }{c \tau_{2} \avgam} \, ,
\end{align}
where $\avgam$ is the average boost factor of a $\chi_2$ particle, with the relevant values collected in Table~\ref{tab:paramExp}.
These type of equations can be solved using the Lambert $W$-function. An expansion in logarithms in the regime of interest to us then leads to the simple expression
\begin{align}
    \Lambda_{\rm low}^{-4} = \frac{1}{A} \left[ \ln\left(\frac{A \mathcal{P}^{\rm up}_{\rm sig} }{\Lambda_{\rm up}^4} \right) +  \ln\left( \ln\left( \frac{A \mathcal{P}^{\rm up}_{\rm sig} }{\Lambda_{\rm up}^4} \right)\right)\right] \ .
    \label{eq:lowelimit}
\end{align}
Estimating the parameter $A$ along with the decay probability for the upper limit $\mathcal{P}^{\rm up}_{\rm sig} $ implies knowing the geometric properties of the various detectors under consideration. We have collected all the relevant parameters in Table~\ref{tab:paramExp}.\footnote{Note that when the production of dark sector particles occurs at the LHC, the cross-section has a non-trivial dependence on the scale $\Lambda$ due to the cuts described in Sec.~\ref{sec:directprod} In this case we parametrise $\sigma (\Lambda) \sim a \times \Lambda^{b}$ and use the solution $X^c e^{A X} = B \implies X = -\frac{A}{c} W(-\frac{A B^{1/c}}{c})$ to find the full solution.} Notice that this leads to \emph{conservative} lower limits.  Indeed, we only include the average boost factor from the dominant production mechanism, but in the short lifetime limits it is actually the high energy tail of the distribution which dominates the population in the detector and fixes the limit. Full simulation of the production and decay of the heavy states is therefore likely to lead to significantly improved lower limits with more parameter space coverage.

 Most experimental limits are based on observing the creation of a positron-electron pair in the detector. The 3-body decay $\chi_2 \rightarrow \chi_1 e^+ e^-$ width can be straightforwardly estimated as 
\begin{align}
\label{eq:chi2Gamma}
\Gamma_2 &= \frac{g_l^2}{\Lambda^4} \frac{M_2^5}{ \pi^3}  \times \mathcal{G} (M_1,M_2) \, ,
\end{align}
where the function $\mathcal{G}$ depends on the type of effective coupling to leptons. For a vector effective coupling to leptons, in the saturation limit ($M_2 \gg M_1$) and limit of near-degenerate small splitting, we have
\begin{align}
\mathcal{G^V}  = \begin{cases}
\frac{1}{60 M_1^5}   \left( \Delta_\chi^2-M_l^2\right)^{5/2}  , \qquad \qquad \hfill \textrm{Near-degenerate,} \\[0.3em]
\frac{1}{384} \left( 1 - \frac{2 M_1}{M_2} \right) \ , \qquad \qquad  \hfill  \textrm{Saturation,} 
\end{cases} 
\end{align}
For an axial-vector coupling, we have instead
\begin{align}
\mathcal{G^{AV}}  = \begin{cases}
\frac{1}{60 M_1^5}   \left( \Delta_\chi^2-M_l^2\right)^{3/2} \left( \Delta_\chi^2+ 6M_l^2\right)   , \qquad \textrm{ Near-degenerate,} \\[0.3em]
\frac{1}{384} \left( 1 + \frac{2 M_1}{M_2} \right) \ , \hfill \textrm{Saturation,} 
\end{cases} \
\end{align}
where in both cases we neglected the lepton mass in the saturation limit. In both expressions, we have assumed both $M_1,M_2 >0$. Note that as for the meson decay amplitude, in the case of a pseudo-Dirac setup (with a $\bar{\chi}_2 \gamma^\mu \gamma^5 \chi_1$ dark fermions operator and negative mass $M_1$) one can directly use the vector coupling result with positive mass.

\begin{figure}[t]
	\centering
	\subfloat[]{%
		\label{fig:GammaChi2a}%
		\includegraphics[width=0.47\textwidth]{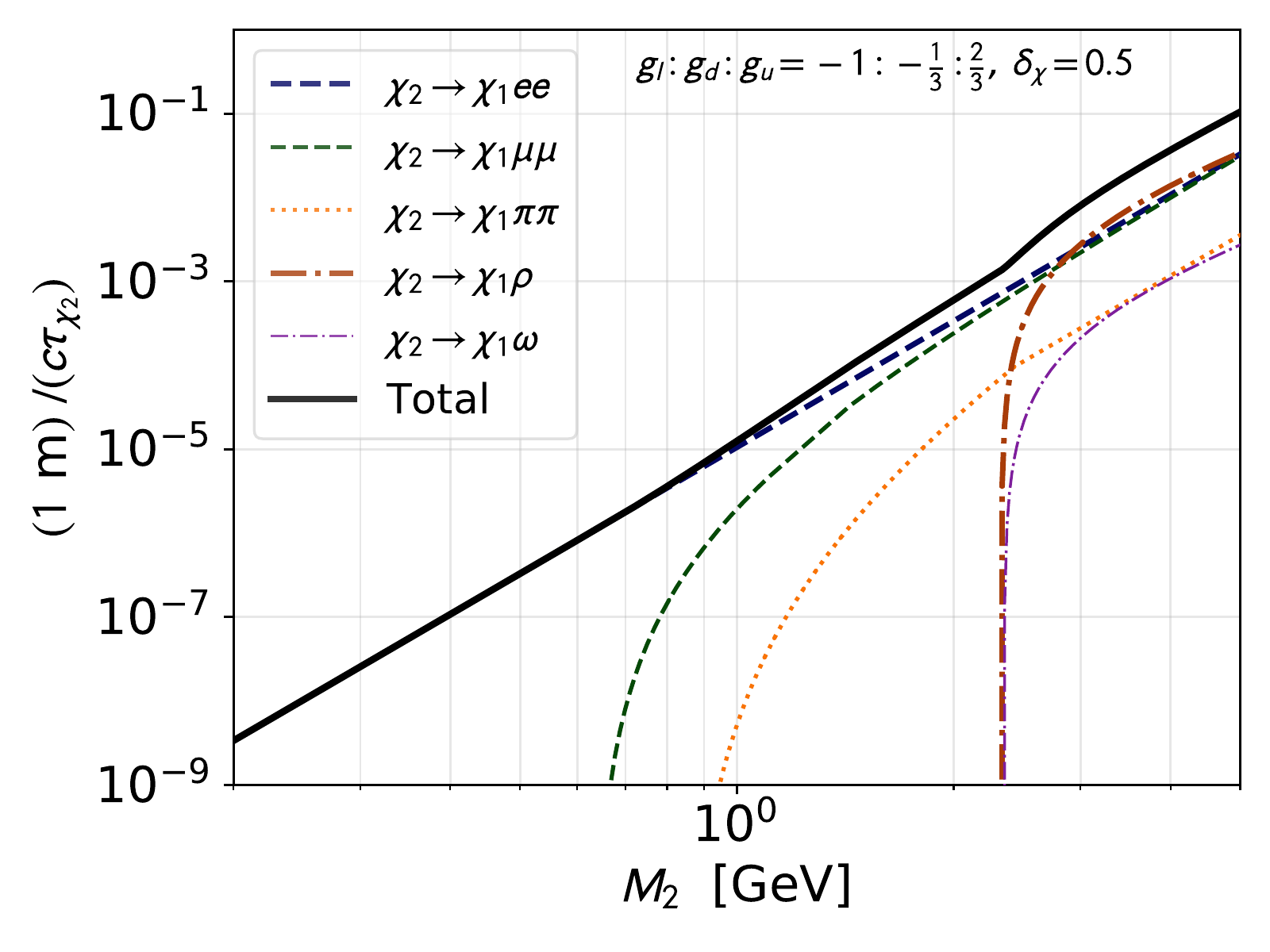}
	}%
	\hspace{0.02\textwidth}
	\subfloat[]{%
		\label{fig:GammaChi2b}%
		\includegraphics[width=0.47\textwidth]{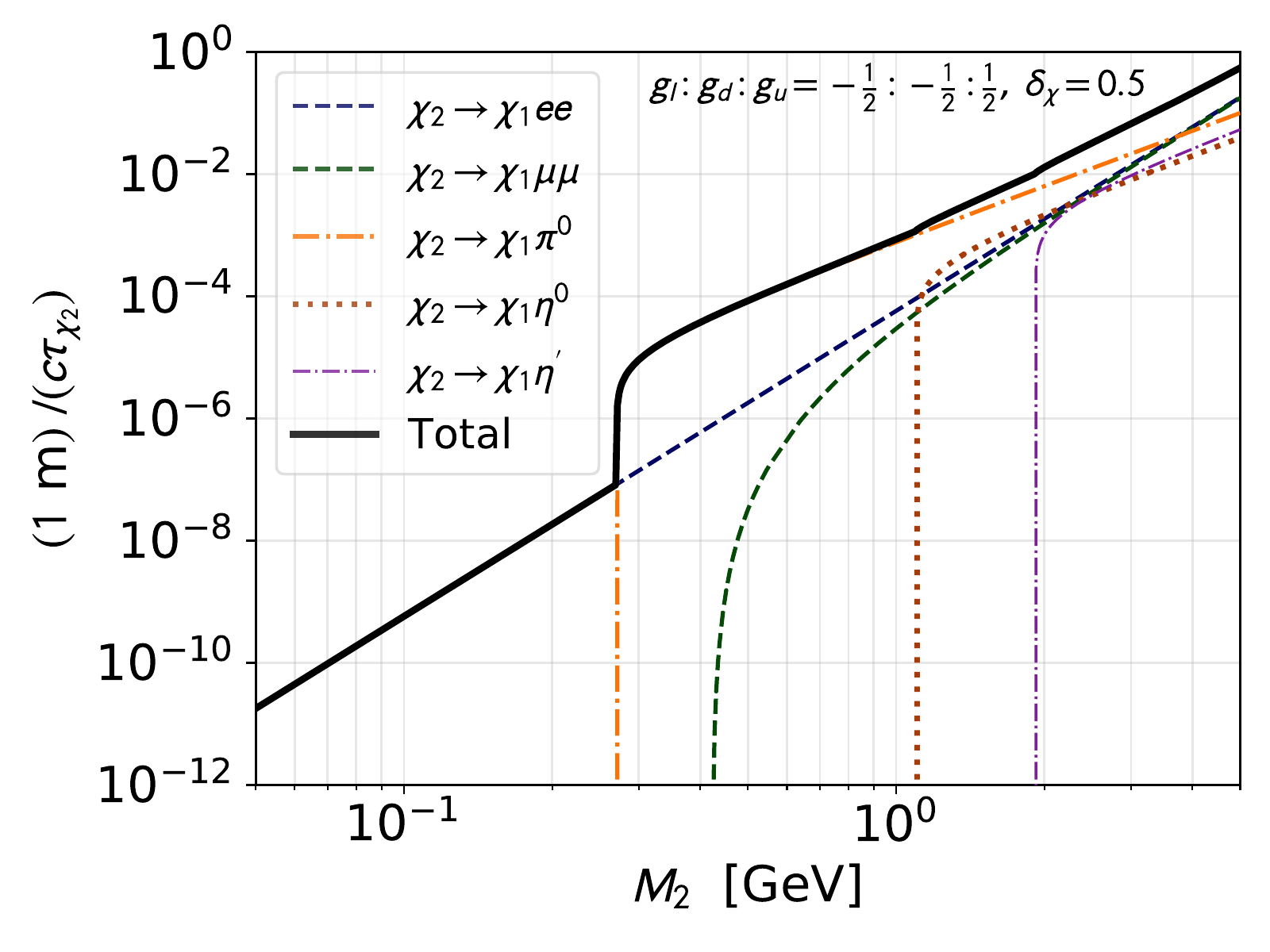}
	}%
	\hspace{0.02\textwidth}
	\caption{Decay width of $\chi_2$ for a vector-current operator with electromagnetically-aligned coupling (a) and an axial-vector current operator with Z-aligned coupling (b). The normalised splitting $\delta_\chi$ is defined in Eq.~\ref{eq:dChi}.}
\end{figure}

For larger mass splitting between the dark sector states more decay channels become available. While this does not modify the expected numbers of $e^+ e^-$ pairs in the long lifetime limits, it can alter significantly the lower bounds from the short-lived limits. The decay into a pair of muon-antimuon opens up once $\Delta_\chi > 2 m_\mu$, while the possible decay into hadronic states depends on the type of effective operator with quarks. The relevant channels are
\begin{itemize}
    \item Vector coupling --- The dominant hadronic processes are $\chi_2 \rightarrow \chi_1 \pi^+ \pi^-$, $\chi_2 \rightarrow \chi_1 \rho$ and $\chi_2 \rightarrow \chi_1 \omega$. The corresponding decay width can be estimated with the same techniques developed in the previous section and in Appendix~\ref{app:mesdecaycalc}, for instance with the vector meson dominance (VMD) formalism. We show the corresponding decay width in Fig.~\ref{fig:GammaChi2a}.
    \item Axial-vector coupling --- The dominant hadronic processes are the direct decay into pseudo-scalar meson $\chi_2 \rightarrow \chi_1 \pi^0$, $\chi_2 \rightarrow \chi_1 \eta$ and $\chi_2 \rightarrow \chi_1 \eta^\prime$. The former in particular strongly dominates over the leptonic decay for  $\Delta_\chi \sim m_{\pi^0}$.  We show the corresponding decay width in Fig.~\ref{fig:GammaChi2a}.
\end{itemize}

In most cases the leptonic channels contribute significantly to the decay width so that neglecting the hadronic decay channels will be a good approximation.\footnote{This is an important difference with respect to the usual dark photon scenario, for which the hadronic decay channels can be enhanced through resonant mixing with vector mesons; see e.g.~\cite{Ilten:2018crw} for a recent overview.} The notable exception is the two-body $\chi_2 \rightarrow \chi_1 \pi^0$ decay in the case of an axial-vector current. Altogether we will use the total decay width when estimating the lower limit as in Eq.~\ref{eq:lowelimit}.

Finally, in the case of a very long-lived particle (or a completely stable one), it is possible to search for a scattering signature in the detector. However, the strong suppression from the off-shell nature of the portal implies that such limits are hardly relevant compared to the mono-X searches presented in the next section. We based our limits on the standard dark photon portal searches, such as for instance the one from MiniBooNE~\cite{Aguilar-Arevalo:2018wea}. We included projections for upcoming experiments based on Refs.~\cite{deNiverville:2016rqh,deNiverville:2018dbu}.\footnote{Note that more recent limits have been derived very recently in Ref.~\cite{Buonocore:2019esg}.} Since the processes involved are very similar, we use the same techniques as presented above for the very long-lived regime to recast the existing limits. Note that the coupling dependence of the limit depends on the nature of the targets and on the precise form factor for scattering off a proton or a neutron. Altogether, we approximate the scaling as being $\sim |g_d| + |g_u|$ since this gives already a qualitative understanding of the typical size of the bounds.\footnote{Since the reference limits are for a dark photon, this scaling is exact in this case. Furthermore, for nuclei with similar number of protons and neutrons, isospin symmetry should ensure that $g_u$ and $g_d$ are treated on equal footing. A more precise implementation of the form factors' effect is left for future work.}

\subsection{Collider and mono-X searches}
\label{sec:monoX}

Constraints from mono-photon searches at $e^+e^-$ colliders are standard bounds on most models of dark photons. For the fermion portal, the off-shell nature of the mediator degrades the signal quality since one can no longer search directly for a bump in the data. A thorough analysis of this case was conducted in Ref.~\cite{Essig:2013vha}, where a limit corresponding to 
\begin{align}
\frac{\Lambda}{\sqrt{g_e}} ~\lesssim~ 50 \textrm{ GeV} \ , 
\end{align}
was found, valid when the dark sector mass is below a few GeV (at higher masses, the limited energy of the BaBar beam starts impacting the production cross-section~\cite{Essig:2013vha}). We note that the actual limit could be enhanced with a dedicated analysis, since it was derived by Ref.~\cite{Essig:2013vha} in a signal-only setup and with a cut-and-count approach for each bin of the reconstructed missing $m_{\chi \bar{\chi}}$ mass in the original analysis~\cite{Aubert:2008as}. This implies that the most recent BaBar limits~\cite{Lees:2017lec} using instead the final $53$ fb$^{-1}$ dataset do not improve the bounds straightforwardly.

For Belle II, the projected limits from Ref.~\cite{Essig:2013vha} is  
\begin{align}
\frac{\Lambda}{\sqrt{g_e}} ~\lesssim~ 100 \textrm{ GeV} \, ,
\end{align}
assuming that the dominant radiative Bhabha scattering background can be reconstructed and subtracted with a systematic uncertainty of $5\%$. Once again, a proper analysis including background simulation and fitting of the distribution is likely to give a stronger limit. Note that a recent study has considered displaced vertices signatures at Belle-II~\cite{Duerr:2019dmv}; we leave its recasting for future work.

Limits from LEP on extra-dimensions obtained at DELPHI~\cite{Abdallah:2008aa} have been shown in Ref.~\cite{Fox:2011fx} to lead to the limit 
\begin{align}
    \frac{\Lambda}{\sqrt{g_e}} \lesssim 500 ~\textrm{GeV} \ .
\end{align}
We will use directly this limit, and follow Ref.~\cite{Bertuzzo:2017lwt} in adding a lower limit $\Lambda \gtrsim 200$ GeV to account for the breakdown of the effective theory around the LEP centre-of-mass energy.

Let us now turn to the limits at the LHC. The strongest bounds on an invisible dark sector comes from mono-jet searches, and in particular the analysis from ATLAS~\cite{Aaboud:2017phn} with $36.1$ fb${}^{-1}$. This analysis was recasted in Refs.~\cite{Bertuzzo:2017lwt,Belyaev:2018pqr} for a variety of effective operators in the context of dark matter searches at LHC. We used the upper limits from these references for a vector-vector operator to extract the limit cross-section at $g_u=2/3,g_d=-1/3$:  $\sigma_{\rm lim} \sim 0.28$ pb. We can then find the upper and lower limits from MET searches, where the lower limit arises due to the limitations of the EFT approach (as discussed in Sec.~\ref{sec:directprod}),
by solving
\begin{align}
    \sigma_{\rm mono} (\Lambda) = \sigma_{\rm lim}~\frac{1}{2 g_u^2 + g_d^2} \ .
\end{align}
In practice, we solve once and for all the limits in $\Lambda$ as a function of $2 g_u^2 + g_d^2$, then substituted the exact value depending on the precise values of the coefficient. This typically leads to limits of $\Lambda >1.3$ TeV or $\Lambda < 0.3$ TeV for $2 g_u^2 + g_d^2 = 2$, with the upper limits reaching up to the tens of TeV for couplings near the perturbativity limit. Note that when considering associated production, the final limits typically turn out to grossly underestimate the reach compared to simplified models; moreover, additional effects such as multi-jet production also become relevant~\cite{Belwal:2017nkw}. Our limits may then be considered a conservative estimate.

\subsection{Invisible meson decays}
\label{sec:invmesdecay}
Let us first discuss the limits from invisible decay of $\pi^0$ meson. The current best bound is fixed by the NA62 collaboration (see e.g. Refs.~\cite{Moulson:2013oga , NA62:Kaon2019}) to be
\begin{align}
    \br{\pi^0 \rightarrow \nu \nu} \leqslant 4.4 \cdot 10^{-9} \ .
\end{align}
Using the expression for the $\pi^0 \rightarrow X X $ width for the axial-vector effective operator from Eq.~\eqref{eq:GammaPAV} one can readily deduce the limits for any given parameters. As we will see in the next sections, this limit can be quite stringent and leads to limits in the hundred of GeV on $\Lambda$.

For flavour-diagonal couplings to second and third generation quarks, the invisible decay of heavy vector mesons is constrained from BaBar and BES measurements~\cite{Aubert:2009ae,Ablikim:2007ek}:
\begin{align}
    \br{J/\Psi  \to \rm{inv.} }&< 7.2 \times 10^{-4} \quad \text{(BES)} \, , \\
    \br{\Upsilon (1S) \to \rm{inv.} }&< 4 \times 10^{-4} \quad \text{(BaBar)} \, .
\end{align}

More sensitive limits can be obtained from searches for invisible heavy meson decays with off-diagonal couplings. Belle has placed the strongest bounds on neutral $B$ meson invisible decays~\cite{Grygier:2017tzo}: 
\begin{align}
    \br{B^0 \to K^0 \nu \bar{\nu}} &< 1.3 \times 10^{-5} \quad \text{(Belle)} \, , \\
    \br{B^0 \to \pi^0 \nu \bar{\nu}} &< 0.9 \times 10^{-5} \quad \text{(Belle)} \, .
\end{align}
For charged $B$ meson invisible decays the best limits are from BaBar~\cite{Aubert:2004ws,delAmoSanchez:2010bk}, though we also include a projected future bound from Belle-II~\cite{Abe:2010gxa},  
\begin{align}
    \br{B^\pm \to K^\pm \nu \bar{\nu}} &< 1.3 \times 10^{-5} \quad \text{(BaBar)} \, , \\
    (\br{B^\pm \to K^\pm \nu \bar{\nu}} &< 1.5 \times 10^{-6}) \quad \text{(Belle-II projected)} \, , \\
    \br{B^\pm \to \pi^\pm \nu \bar{\nu}} &< 1.0 \times 10^{-4} \quad \text{(BaBar)} \, . 
\end{align}
Finally, we also include the following Kaon invisible decay bounds from the NA62~\cite{NA62:Kaon2019}, E949 and E787 experiments~\cite{Adler:2008zza}, as well as a future projection NA62~\cite{Fantechi:2014hqa}.
\begin{align}
    \br{K^0_L \to \pi^0 \nu \bar{\nu}} &< 0.46 \times 10^{-10} \quad \text{(NA62)} \, , \\
    \br{K^+ \to \pi^+ a} &< 0.73 \times 10^{-10} \quad \text{(E949+E787)} \, , \\
    (\br{K^+ \to \pi^+ a} &< 0.01 \times 10^{-10}) \quad \text{(NA62 projected)} \, .
\end{align}
Note that for the charged Kaon decays only an invisible axion $a$ as a final state was considered, not $\nu \bar{\nu}$ as in the other cases above, though we expect the constraints to be of similar order of magnitude. Similar limits can also be obtained for the axial-vector operator case from fully invisible decays of $B$ and $K$ meson (see e.g. Ref.~\cite{Barducci:2018rlx}).

\section{Astrophysical and Cosmological limits on the fermion portal}
\label{sec:astrophyics}
\subsection{Limits from Supernova 1987A and stellar cooling}

The cooling of stars and supernova 1987A are known to place strong bounds on light new physics, and as such there has been great effort to quantify these constraints (for some examples of these studies, see e.g. Refs.~\cite{An:2013yfc,Hardy:2016kme,Chang:2018rso,Tu:2017dhl,Zhang:2014wra,Dreiner:2013mua,Dreiner:2013tja,Fayet:2006sa,Dreiner:2003wh,Barbieri:1988av,Raffelt:1996wa,DeRocco:2019jti,DeRocco:2019njg,Chang:2016ntp,Keil:1996ju,Turner:1987by,Raffelt:1987yt,Ellis:1987pk,Mayle:1987as,Bertolami:2014wua}). Typically, the bound arises from the requirement that the energy loss from a star to new light states should not exceed the energy loss to neutrinos. The dominant production mechanisms of such light states are Bremsstrahlung and SM particle annihilation in the stellar interior. If new light particles exist and can be produced in a star or supernova, there exist two typical regimes bracketing the bounds. In one regime the interaction strength with the SM becomes too weak so that the production of light states is no longer an effective cooling mechanism, and the energy loss to new physics is dwarfed by the energy loss to neutrinos. In the other regime, the interaction with the SM becomes so strong that the light particles are produced in abundance, but interact so frequently in the stellar interior that they are unable to exit the star/supernova. In between these two limiting regimes is when cooling takes place (too) efficiently as compared with SM processes, and can thus be excluded.

\subsubsection*{Supernova 1987A}

\noindent Light particles can be produced in the proto-neutron star at the heart of a supernova. This will occur as long as the masses of the light particles are below the characteristic energy scale of the star. The core temperature of the supernova is $T_{\rm c} \sim 30$ MeV. This enables the placing of constraints on particles with masses as large as $m_\chi \sim \mathcal{O}(100)$ MeV, since the observed cooling of the supernova agrees within uncertainties with SM estimates.

In the case where couplings to quarks and leptons are electromagnetically aligned, assuming a thermal distribution of state near the core of the proto-neutron star, the dominant production mechanism of new light states in the supernova is through electron-positron annihilation~\cite{DeRocco:2019njg}.\footnote{If the number densities do not follow a thermal distribution, bremsstrahlung is the dominant production mechanism. If this is the case, the constraint is modified by $\mathcal{O}$(few)~\cite{Chang:2018rso}.} Many of the analyses in the literature consider the special case where dark fermions are coupled to a dark photon with a similar mass. Their exclusion results are therefore not straightforward to recast in the language of effective operators. However, we obtain a limit from the analysis in Ref.~\cite{DeRocco:2019njg} which was performed under the assumption that the dark photon was decoupled and therefore allows for a simple recasting. The limit is obtained on the invariant mass of the pair of light fermions. Therefore, when we consider a large splitting $M_2 \gg M_1$, the upper limit on $M_2$ tends to be greater than naively expected. This can be seen in e.g. Fig.~\ref{fig:LimitsEM}, where the splitting is large, and therefore the invariant mass of the light pair is almost entirely dominated by $M_2$.

When two light states of differing mass are produced, their mass difference can have an important effect on the lower boundary of the limit on $\Lambda/\sqrt{g}$, where particles become trapped instead of streaming out of the star. For example, if the mass splitting is much greater than $\sim 30$ MeV, the likelihood for the lighter $\chi_1$ to up-scatter into $\chi_2$ by interacting with SM particles is exponentially suppressed. If the $\chi_2$ decay rate into $\chi_1$ + SM is sufficiently large, this means that to a good approximation, there is no appreciable $\chi_2$ population in the star, and there can be no annihilation or scattering of $\chi_{1,2}$ to result in a trapping limit~\cite{Chang:2018rso}. Thus in this situation, there would be no lower limit on $\Lambda/\sqrt{g}$. If on the other hand, the decay rate of $\chi_2$ is very small, even though the mass splitting may be too large for up-scattering to occur, there is still a significant population of $\chi_2$. The dark sector particles may therefore annihilate back into SM fermions, resulting in trapping and a lower limit on $\Lambda/\sqrt{g}$. The precise location of this limit would depend on the mass splitting and the decay length of $\chi_2$. For this reason, in our figures we show the upper limit on $\Lambda/\sqrt{g}$ as a solid line, while the lower limit is dotted.

Finally, the case of an axial-vector operator is more intricate as the production is strongly modified with respect to the dark photon case. As a conservative upper limit,  we thus simply use the bound from invisible $\pi^0$ decay from SN1987A cooling inferred in Ref.~\cite{Natale:1990yx} for a core temperature of $50$ MeV:
\begin{align}
    \br{\pi^0 \rightarrow \nu \nu} \lesssim 1 \cdot 10^{-13} \ .
\end{align}
The upper limits on the effective operator are then obtained in the same way as in Sec.~\ref{sec:invmesdecay}. We treat the lower limit on the suppression scale in the same way as we do for the case of the vector operator above.

\subsubsection*{Stellar cooling}

\noindent The characteristic temperature at the cores of Horizontal Branch and Red Giant stars is $T\sim 10$ keV. This core temperature results in constraints that only apply to dark sector particles with masses as large as $m_\chi \sim \mathcal{O}(50)$ keV. The two classes of stars differ in their densities, chemical potentials and photon plasma masses. This leads to slightly different limits being obtained from the two types of stars (see e.g. Ref.~\cite{Hardy:2016kme}). However, due to their core temperature being significantly lower than that of a supernova, the limit that can be derived on $\Lambda/\sqrt{g}$ is also less strong. Indeed, in Ref.~\cite{Dreiner:2013mua}, it was found that the upper limit on the suppression scale of the fermion portal operator would be $\Lambda/\sqrt{g} \sim v$, where $v$ is the usual electroweak VEV. This can be understood intuitively from the fact that at stellar core temperatures, plasmon decay is the dominant process not only for dark sector particle production~\cite{Dreiner:2013mua}, but also for neutrino production. In the limit where both neutrinos and dark sector particles are massless, requiring that the luminosity in dark states not exceed the luminosity in neutrinos is equivalent to requiring that the suppression scales in their decay rates be similar in size. We do not show these bounds in our figures, since as discussed in the following section, below $m_\chi \sim 5\text{ MeV}$, much stronger constraints can be obtained from considerations of the early universe.

\subsection{Early universe and relic density bounds}

Four-fermion operators like the ones considered have previously been used to describe dark matter interacting with the Standard Model in a model-independent way~\cite{Fox:2011fx,Essig:2013vha,Bertuzzo:2017lwt,Belyaev:2018pqr}. This approach for dark matter typically faces several difficulties. The final relic density obtained from the freeze-out of dark matter annihilating through a fermion portal operator typically scales as
\begin{align}
    \Omega h^2 \sim 0.1 \left( \frac{\Lambda}{125 \textrm{ GeV}} \right)^4 \left( \frac{1 \textrm{ GeV}}{M_\chi} \right)^2 \ ,
\end{align}
where we have considered only one vector operator of the form $ \frac{1}{\Lambda^2}(\bar\chi \gamma_\mu \chi) (\bar{e} \gamma^{\mu} e)$. This illustrates that the effective interactions are typically too suppressed to lead to the proper relic density through the standard freeze-out mechanism. (Note that including more effective operators and treating properly the annihilation into mesons improves somewhat the picture, as can be seen in, e.g. Ref.~\cite{Choudhury:2019tss} for the case of scalar dark matter, but there is still an overabundance of thermal dark matter when $\Lambda \gtrsim 200$ GeV.) One can still obtain the correct relic density through the freeze-in mechanism, as pointed out in Ref.~\cite{Bertuzzo:2017lwt}, although this typically implies an extremely high effective scale and adds a dependence on the reheating temperature. More generally there have been many studies of additional dynamics in the dark sector beyond the fermion portal operator which may contribute to fixing the proper relic density which require either other operators or more dark sector particles. Some of the earlier examples of such setups include Secluded DM~\cite{Pospelov:2007mp} and related scenarios such as co-decaying DM (see e.g. Refs.~\cite{Dror:2016rxc,Kopp:2016yji}), or additionally Cannibal DM~\cite{Pappadopulo:2016pkp} or co-scattering~\cite{DAgnolo:2017dbv,DAgnolo:2019zkf}, along with many more recent examples. Additionally, exotic cosmological histories can also modify the relic density, for instance through a late phase transition (see e.g.~\cite{Dimopoulos:1990ai,Cohen:2008nb} and the subsequent literature).

The strongest limit on a possible dark matter candidate below $\sim 10$ GeV comes from the cosmic microwave background (CMB) constraints on the dark matter annihilation cross-section. As shown, in, e.g. Ref.~\cite{Slatyer:2015jla} unsuppressed s-wave annihilation in this case is excluded by the CMB spectral shape~\cite{Aghanim:2018eyx} (see also~\cite{Choudhury:2019tss} for a recent study of the dim-6 case for scalar dark matter).

Finally, while we do not focus explicitly on this case in this work, very strong additional limits from the CMB arise when one of the dark sector fermion is lighter than around $5$ MeV (see e.g.~\cite{Sabti:2019mhn} for an up-to-date estimate). For dark fermions light enough to behave as radiation at neutrino decoupling, the strongest limits come from the effective number of relativistic degrees of freedom $N_{\rm eff}$ in the early universe. As was studied in detail in Refs.~\cite{Brust:2013xpv,Baumann:2016wac,Baumann:2018muz} CMB-S4 observatories can in principle completely exclude any light relativistic relic up to arbitrarily high decoupling temperature, provided it was in thermal equilibrium with the Standard Model and that the reheating temperature was higher than its decoupling temperature. We will adapt the calculations of Refs.~\cite{Baumann:2016wac,Baumann:2018muz} to the fermion portal case in order to obtain order-of-magnitude limits. We focus on the case of a fermion portal involving electrons, $\frac{1}{\Lambda^2}(\bar\chi \gamma_\mu \chi) (\bar{e} \gamma^{\mu} e)$, so that all fields involved can be considered massless in the following.

In the limit where the light particles decouple instantaneously from the thermal bath at a temperature $T_0$, it will not receive the entropy released subsequently by annihilating species, so that the final effective degrees of freedom after neutrino decoupling is given by~\cite{Brust:2013xpv}
\begin{align}
 \Delta g_* = g_{LS} \left( \frac{3.38}{g_*(T = T_0) }\right)^{4/3} \ ,
\end{align}
where $g_{LS}$ is the number of degrees of freedom of the light relativistic relic. Translating into an effective number of neutrinos and assuming $T_0 > T_{EW}$ we find the lower bounds of Ref.~\cite{Baumann:2016wac} on the effective number of neutrino, 
\begin{align}
 \Delta \neff (T_0) > 0.027  \Delta g_*  = \left( \frac{106.75}{g_*(T = T_0) }\right)^{4/3} \times \begin{cases} 0.027 \qquad  \textrm{   Scalar} 
 \\ 0.047 \qquad \textrm{     Weyl} 
 \\ 0.054 \qquad \textrm{     Gauge} 
 \\ 0.095 \qquad \textrm{     Dirac} 
 \end{cases} \ .
\end{align}
In particular notice that we have kept $g_*(T = T_0)$ to emphasize that the bound derived in Ref.~\cite{Baumann:2016wac} is only when there are no additional degrees of freedom in the theory beyond the SM ones. This assumption does not hold by definition in our effective theory approach since we do expect new physics to occur around the scale $\Lambda$.\footnote{As an example, the full MSSM has $g_*^{\textrm{MSSM}} = 228.75$, for which the lowest value for $ \Delta \neff$ is actually closer to $ \Delta \neff > 0.01$.} This implies that CMB-S4 experiments will not necessarily rule out every thermally coupled relic, especially in the case of a particularly rich UV sector. Current limits from the Planck experiment~\cite{Aghanim:2018eyx} typically also exclude light relics decoupling below the QCD phase transition at around $100$ MeV. 

Following Ref.~\cite{Baumann:2016wac} we determine the decoupling temperature $T$ by simply comparing the production rate of the dark sector particle $\chi_2$ through the fermion portal operator, $\Gamma_2 (T)$, with the Hubble rate, $H(T)$:
\begin{align}
\label{eq:Neffbound}
 \Gamma_2 (T_R) < H(T_R) = \frac{\pi}{\sqrt{90}} \sqrt{g_{*}(T_R)} \frac{T^2}{M_{pl}} \ ,
\end{align}
where we used the reduced Planck mass $M_{pl} = 2.4 \cdot 10^{18}$ GeV and $g_{*}(T_R)$ is the effective number of relativistic species at the reheating temperature $T$. We obtain the production rate as
\begin{align}
\label{eq:4FGamma2}
 \Gamma_2 \simeq \frac{1}{n^{\textrm{eq}}_{\chi}} \int \frac{d^3p_1}{(2 \pi)^3} \frac{d^3p_2}{(2 \pi)^3} \frac{f_1(p_1)}{2 E_1} \frac{f_2(p_2)}{2 E_2} \mathcal{P}(p_1,p_2) 2s \sigma_{\textrm{CoM}} (s) \ ,
\end{align}
 where we have used the equilibrium density for a Weyl fermion $n^{\textrm{eq}}_{\chi}  = \displaystyle \zeta(3) \frac{3}{4} T^3/\pi^2$ and introduced the thermal distribution functions $f_1, f_2$ as well as a simplified Bose enhancement/Pauli blocking term $ \mathcal{P}(p_1,p_2)$ which, in our four-fermions interaction case are:
\begin{align}
 f_1(E) = f_2(E) = f(E) = \frac{1}{e^{E/T}-1}  \qquad \textrm{and} \qquad \mathcal{P}(p_1,p_2) = (1-f(E_1))(1-f(E_2)) \ .
\end{align}
The main difference with respect to Ref.~\cite{Baumann:2016wac} comes from the dimension-6 nature of the fermion portal operator, which implies that the centre-of-mass production cross-section is of order 
\begin{align}
 \sigma_{\textrm{CoM}} (s) \sim \frac{1}{\pi } \times s \times \frac{g_e^2}{\Lambda^{4}} .
\end{align}
In particular, it is a linear function of the squared centre-of-mass energy $s$ (compared to a constant in the dimension-5 case in Ref.~\cite{Baumann:2016wac}). We then solve using the standard techniques of Ref.~\cite{Gondolo:1990dk} by changing variables from $p_1, p_2$ to $s,~ E_1+E_2,~E_1-E_2$, including the factor $\mathcal{P}$ and solving numerically after extracting all $T$ dependence, to obtain the following order-of-magnitude limits:
\begin{align}
\frac{\Lambda}{\sqrt{g}} ~\gtrsim~ 
\begin{cases} 
\displaystyle 5\times10^3 \text{ GeV}  \qquad &\text{ (Planck)} \\
5 \times 10^{11} \text{ GeV}\times  \left( \frac{T_R}{10^{10} \textrm{ GeV}}\right)^{3/4} \qquad &\text{ (CMB-S4)}  \end{cases} \ ,
\end{align}
where the second limit depends on the reheating temperature $T_R$ since it implies that the light relics were never produced in the first place, while the first limit simply requires it to decouple prior to the QCD phase transition. Notice that the CMB-S4 limit scales as $T_R^{3/4}$ as compared with $T_R^{1/2}$ for the dimension-5 case~\cite{Baumann:2016wac}.

\section{Summary plots and numerical results}
\label{sec:numerics}

We present in this section some illustrative results based on the above formalism. Since the result depends strongly on the choice of effective operator,  we will typically choose particular ratios $g_u : g_d : g_l$ and present the limits in term $\Lambda / \sqrt{g}$, where $\Lambda$ is a mass scale and $g$ represents the overall coefficient. We shall refer to $g$ as an effective coupling though in general it will be a combination of model-dependent factors obtained by a matching calculation to a UV model. For a tree-level UV completion with $\mathcal{O}(1)$ couplings the scale corresponds roughly to the mediator mass. A weaker or stronger coupling could lower or raise respectively this mass scale. The usual caveats then apply regarding the regime of validity of the effective theory~\cite{Contino:2016jqw}.  

\subsection{Vector operator}

\begin{figure}[h!]
	\centering
		\includegraphics[width=0.7\textwidth]{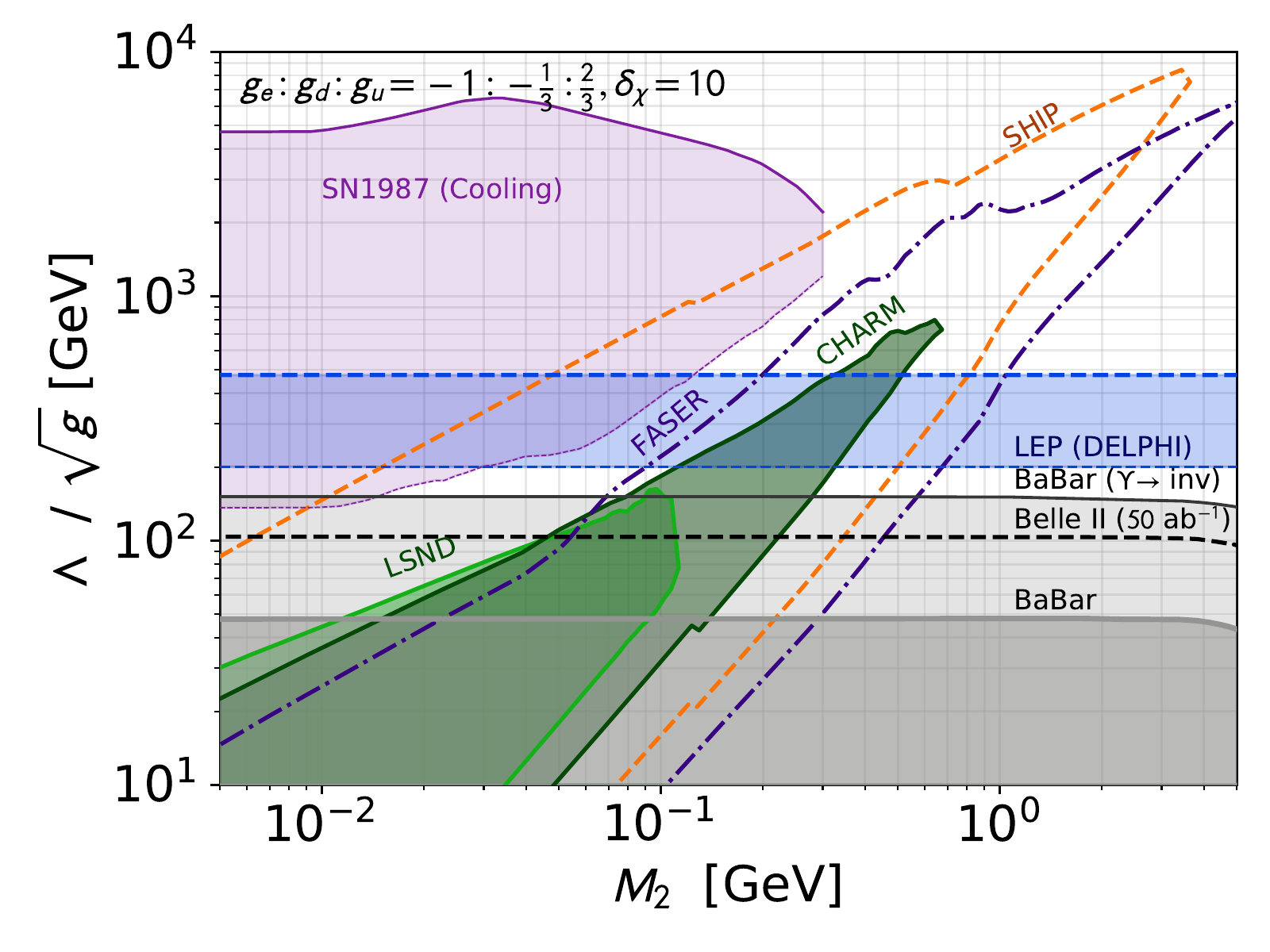}
	\vspace{-0.25cm}
	\caption{Limits and projected sensitivity to the vector operator effective scale $\Lambda/\sqrt{g}$ in the case of effective coupling electromagnetically-aligned as function of $M_2 \gg M_1$ on the x-axis. Grey region indicates coverage from mono-photon at BaBar~\cite{Essig:2013vha}, dashed grey line Belle-II projections~\cite{Essig:2013vha}. The shaded blue region is the mono-$\gamma$ limit from LEP~\cite{Fox:2011fx}. Limits from $\chi_2 \to \chi_1 e^+ e^-$: the green (dark green) regions are the exclusion from LSND~\cite{Darme:2018jmx} (CHARM~\cite{deNiverville:2018dbu}). We show a projection for FASER~\cite{Berlin:2018jbm} in dashed purple and SeaQuest in red (Phase-II defined in~\cite{Berlin:2018pwi}). The 10 events reach by SHiP is shown in dashed orange. The purple region represents the limit from cooling of SN1987A~\cite{DeRocco:2019jti}. The normalised splitting $\delta_\chi$ is defined in Eq.~\ref{eq:dChi}.}
	\label{fig:LimitsEM}
\end{figure}

The first, straightforward case that we consider is the electromagnetically-aligned scenario
\begin{align}
g_u : g_d : g_l = \frac{2}{3} :   -\frac{1}{3} : -1 \, ,
\end{align}
which is typically obtained from an integrated-out dark photon kinetically mixed with the SM photon. The production mechanisms in this case follow closely the ones studied extensively in the literature during the last decade. Due to the off-shell nature of the process, the dark sector production at low mass is dominated by $\eta \rightarrow \gamma \bar{\chi}_2 \chi_1$ (when it is kinematically available) and at high masses by the parton-level production. 

We present in Fig.~\ref{fig:LimitsEM} a summary of the current limits on this scenario as well as some projections for upcoming experiments. Limits from mono-photon signatures at BaBar and LEP, as well as the projection from Belle-II are derived following Sec.~\ref{sec:monoX}. We note that the lower limit from LEP bounds arises due to a breakdown of the effective approach around the LEP centre-of-mass energy; complete models of the dark sector including direct mediator production at LEP would then likely be excluded in this region.

The limits for LSND, CHARM, FASER, SeaQuest and SHiP that are based on the decay of long-lived dark sector states are presented for the saturation case where $M_1 \ll M_2$. The upper bound for each of those experiments can thus be seen as the maximal attainable reach. We have included current limits from LSND recasted from Ref.~\cite{Darme:2018jmx} (equivalent to the one from Ref.~\cite{Izaguirre:2017bqb}) and the limits of CHARM by Ref.~\cite{Tsai:2019mtm}. We show two future experiments as long-term prospects: a naive 10-events projection for SHiP based on the production and detection processes described above (hence not including geometric and detector efficiencies), and projected limits for FASER phase-2 (based on the study of Ref.~\cite{Berlin:2018jbm}).

In all cases, the limits are recasted from a small splitting limit between $M_2$ and $M_1$ to the saturation case $M_1 \ll M_2$ following the procedure presented in Sec.~\ref{sec:decay}. The lower limits for all these experiments are obtained following Sec.~\ref{sec:decay} and therefore combine two main effects: the opening of different decay channels (in particular hadronic) for larger $M_2$ masses, and the average boost factor associated with the various production mechanisms. As discussed before, the presented lower bounds are thus conservative since we can expect that the high-momentum tail of the dark sector particles' distribution should dominate the signal events in this region. Notice also that we did not include secondary production through up-scattering as presented in Ref.~\cite{Jodlowski:2019ycu}. The purple region represents the limits obtained from the cooling of SN1987A, recasted from Ref.~\cite{DeRocco:2019jti} as described in Sec.~\ref{sec:astrophyics}. Note that the lower limit is dashed to represent the significant uncertainty on the trapping regime.

Altogether, the existing set of limits on our fermion portal scenario presents an interesting complementarity, similar to dark photon searches, for example (but with some differences in the phenomenology, as previously discussed). Mono-X and missing energy limits tend to exclude an effective scale independently of the dark sector particles masses, but do not extend beyond $\Lambda$ around a few hundred GeV. Interestingly, and contrary to the situation for a dark photon, the limit from SN1987A directly overlaps with the mono-X limits and extends to several TeV for dark sector masses below $\sim 100$ MeV. Finally, the parameter space coverage of experiments based on decay searches typically extends diagonally, as could be expected from Eq.~\eqref{eq:chi2Gamma} since theses searches are most effective when the long-lived state decays a fixed distance of tens to hundreds of meters from the beam dump.

\begin{figure}[t]
\centering
		\includegraphics[width=0.7\textwidth]{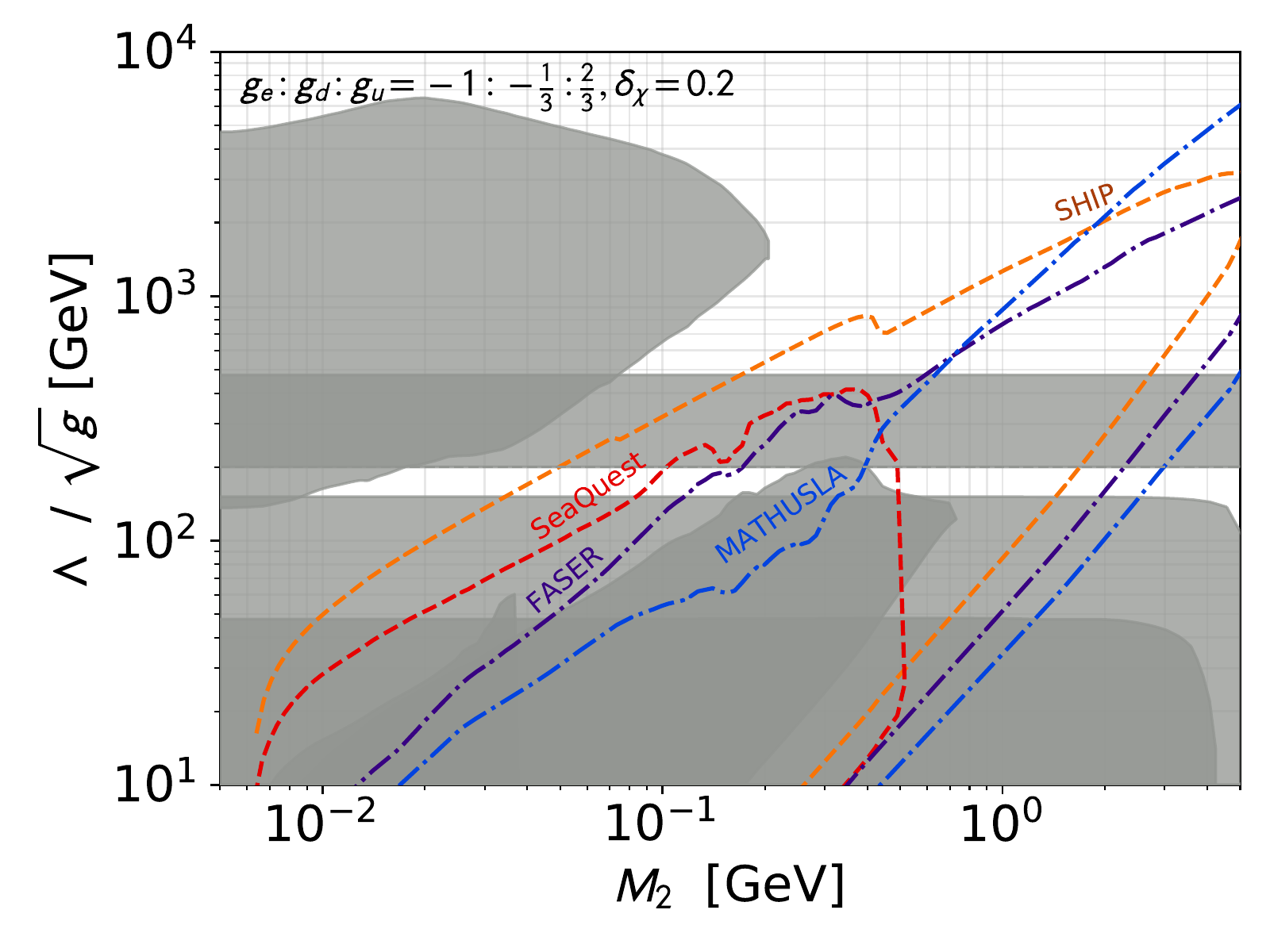}
		\vspace{-0.2cm}
	\caption{Prospective sensitivity at future intensity frontier experiments on the vector operator effective scale $\Lambda/\sqrt{g}$ in the case of electromagnetically-aligned effective couplings as a function of $M_2$ on the x-axis for $\delta_\chi = 0.2$ (defined in Eq.~\ref{eq:dChi}). Grey areas indicate already covered parameter space. We show prospects for FASER~\cite{Berlin:2018jbm} in dashed purple and SeaQuest in red (Phase-II defined in~\cite{Berlin:2018pwi}), MATHUSLA~\cite{Berlin:2018jbm} in blue and SHiP in green (based on our 10 events projections).}
	\label{fig:LimitsVProspective}
\end{figure}

The limits we have shown in Fig.~\ref{fig:LimitsEM}, for electromagnetically-aligned couplings with $\delta_\chi = 10$, emphasise current limits with some representative future projections also displayed (others are omitted for clarity). In the next decade, many new experiments searching for decays of dark sector states have been proposed or are already planned. We show in Fig.~\ref{fig:LimitsVProspective} the projected reach for a selection of future experiments including  FASER~\cite{Feng:2017uoz,Berlin:2018jbm}, SeaQuest (following the Phase-II proposal defined in Ref.~\cite{Berlin:2018pwi}), MATHUSLA~\cite{Lubatti:2019vkf,Berlin:2018jbm} and SHiP~\cite{Anelli:2015pba} (based on our 10 events projections), for electromagnetically-aligned couplings with splitting $\delta_\chi = 0.2$.

\begin{figure}
	\centering
		\includegraphics[width=0.7\textwidth]{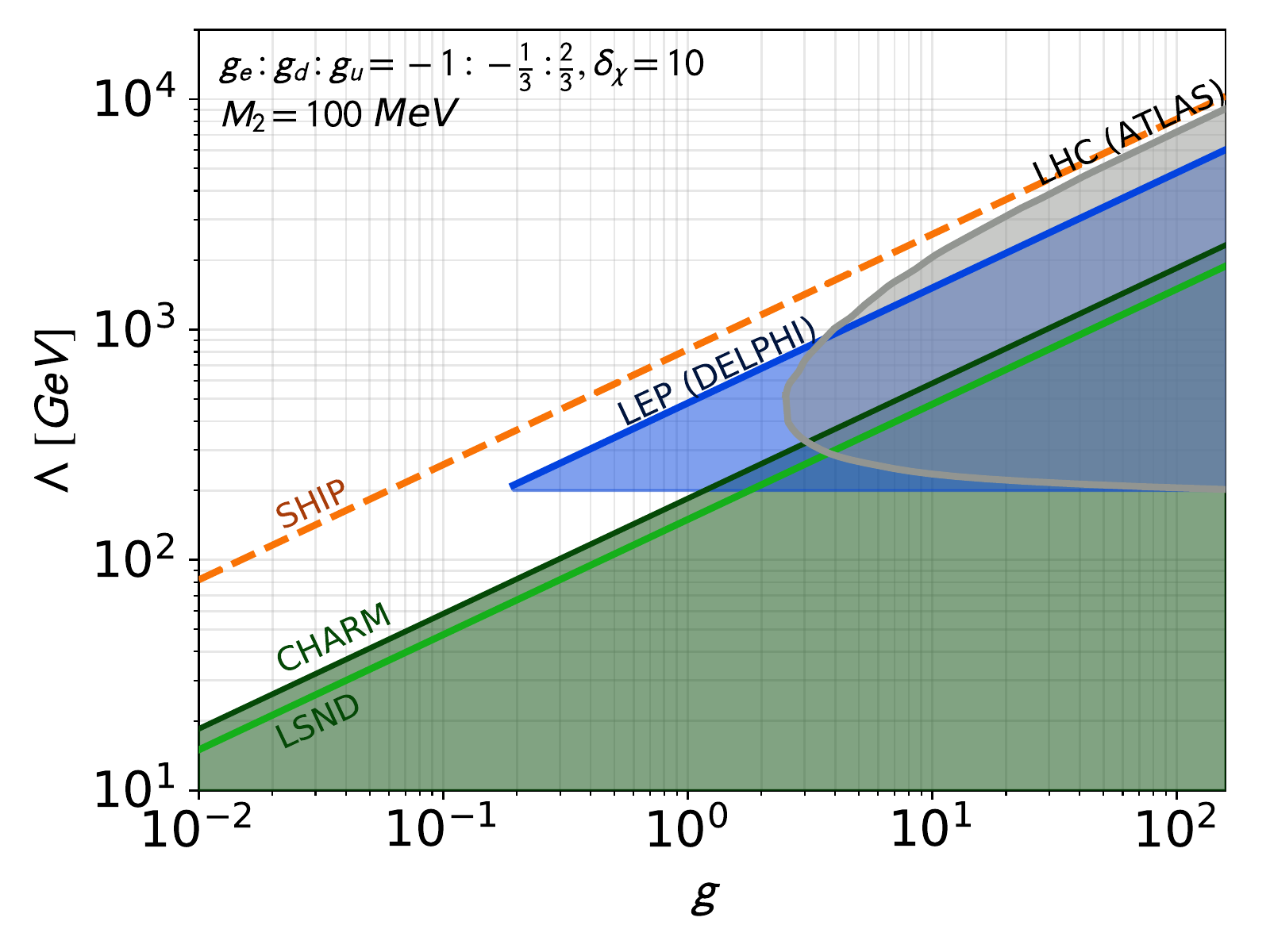}
	\vspace{-0.2cm}
	\caption{Limits and prospective experimental sensitivity when varying the overall scaling $g$ for $M_\chi = 100$ MeV for the vector-current operator as a function of $M_2 \gg M_1$ on the x-axis. Exclusion from LSND~\cite{Darme:2018jmx} (CHARM~\cite{deNiverville:2018dbu}) are shown in green, as well as 10 events reach by SHiP in dashed orange. The shaded blue region is the mono-$\gamma$ limit from LEP~\cite{Fox:2011fx}. The grey region is the exclusion from ATLAS mono-jet~\cite{Aaboud:2017phn}. The normalised splitting $\delta_\chi$ is defined in Eq.~\ref{eq:dChi}.}
	\label{fig:limg}
\end{figure}

\begin{figure}
	\centering
		\includegraphics[width=0.7\textwidth]{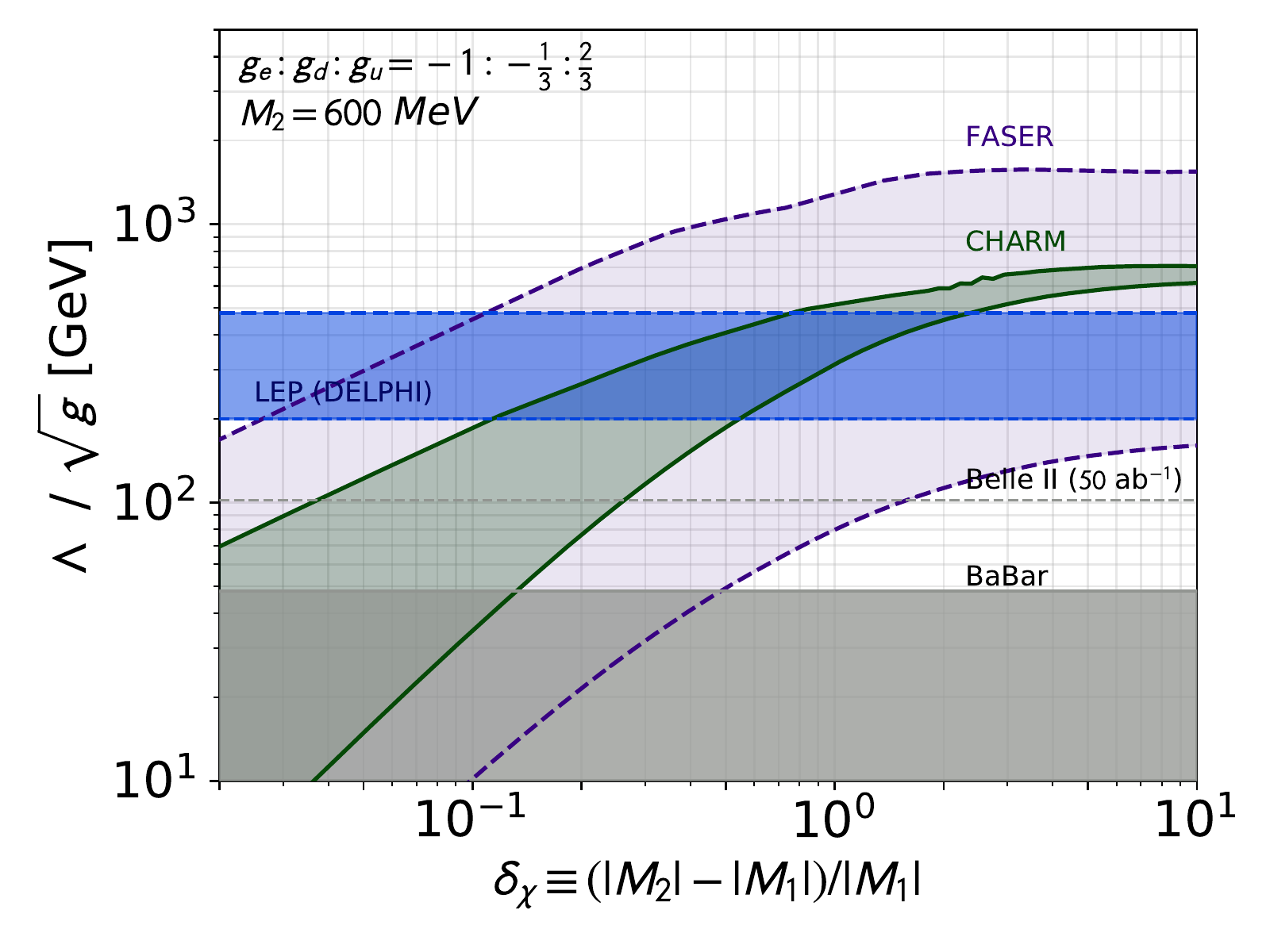}
	\hspace{0.02\textwidth}
		\vspace{-0.3cm}
	\caption{ Limits and projected sensitivity to $\Lambda/\sqrt{g}$ as function of $\delta_\chi$ as defined in Eq.~\ref{eq:dChi}.The shaded blue region is the mono-$\gamma$ limit from LEP~\cite{Fox:2011fx}.  Exclusion from CHARM~\cite{deNiverville:2018dbu} is shown in green, and projection from FASER~\cite{Berlin:2018jbm} in purple.}
	\label{fig:limDelta}
\end{figure}

The above results can vary depending on the effective coupling $g$ or the mass splitting between the two dark sector states. We illustrate this dependence in Fig.~\ref{fig:limg} for the limits on $\Lambda$ as a function of the effective coupling $g$; note that the scaling is not necessarily trivial since mono-photon and mono-jet limits from LEP and ATLAS enter above a certain energy threshold, as described in Sec.~\ref{sec:monoX}. In Fig.~\ref{fig:limDelta} we show how the typical limits depend strongly on the splitting in $\chi_2 \rightarrow \chi_1 e^+ e^-$ decays for the CHARM and FASER phase-2 experiments (based on the limits from Ref.~\cite{Berlin:2018jbm}).

\begin{figure}[t]
\centering
		\includegraphics[width=0.7\textwidth]{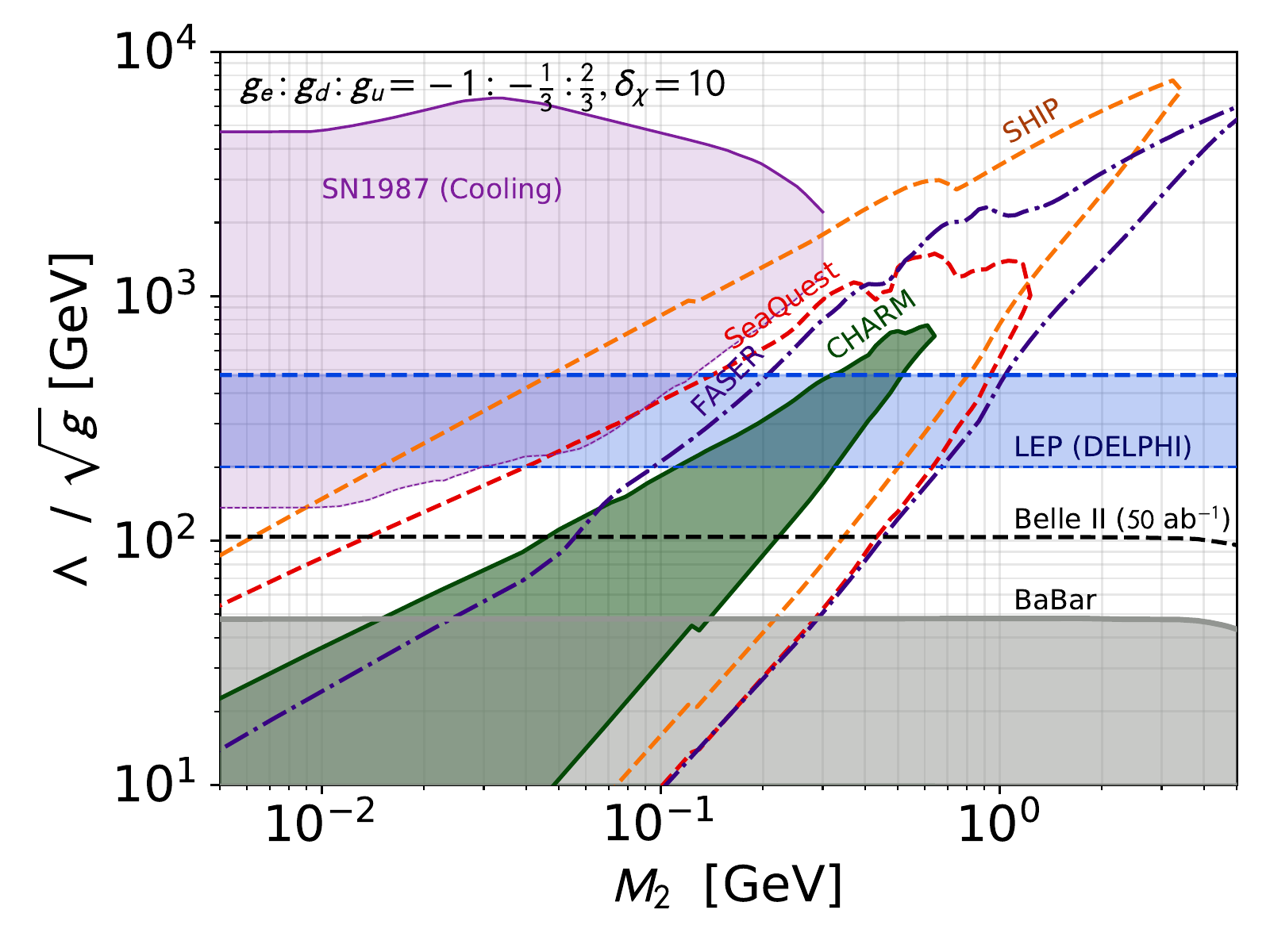}
			\vspace{-0.2cm}
	\caption{Limits and projected sensitivity to the vector operator effective scale $\Lambda/\sqrt{g}$ in the case of protophobic couplings as a function of $M_2 \gg M_1$ on the x-axis. Grey region indicates coverage from mono-photon at BaBar~\cite{Essig:2013vha}, dashed grey line Belle-II projections~\cite{Essig:2013vha}. The shaded blue region is the mono-$\gamma$ limit from LEP~\cite{Fox:2011fx}. Limits from $\chi_2 \to \chi_1 e^+ e^-$: the dark green region is the exclusion from CHARM~\cite{deNiverville:2018dbu}. We show a projection for FASER~\cite{Berlin:2018jbm} in dashed purple and SeaQuest in red (Phase-II defined in~\cite{Berlin:2018pwi}). The 10 events reach by SHiP is shown in dashed orange. The purple region represents the limit from cooling of SN1987A~\cite{DeRocco:2019jti}. The normalised splitting $\delta_\chi$ is defined in Eq.~\ref{eq:dChi}.}
	\label{fig:Limitsphobic}
\end{figure}

As a last vector operator example for the light flavours, we present in Fig.~\ref{fig:Limitsphobic} the limits in the case of proto-phobic couplings,
\begin{align}
g_u : g_d : g_l = -\frac{1}{3} :   \frac{2}{3} : -1 \ .
\end{align}
The main difference with the previous electromagnetically-aligned case is that $\pi^0$ decay production is strongly suppressed, so that searches at the LSND experiment do not lead to a significant limit. On the other hand, other beam dump experiments rely mostly on the $\eta$ meson decay which is only mildly modified, as can be seen in Table~\ref{tab:effcoup}. Fig.~\ref{fig:Limitsphobic} is for $\delta_\chi = 10$.

\begin{figure}[t]
	\centering
		\includegraphics[width=0.7\textwidth]{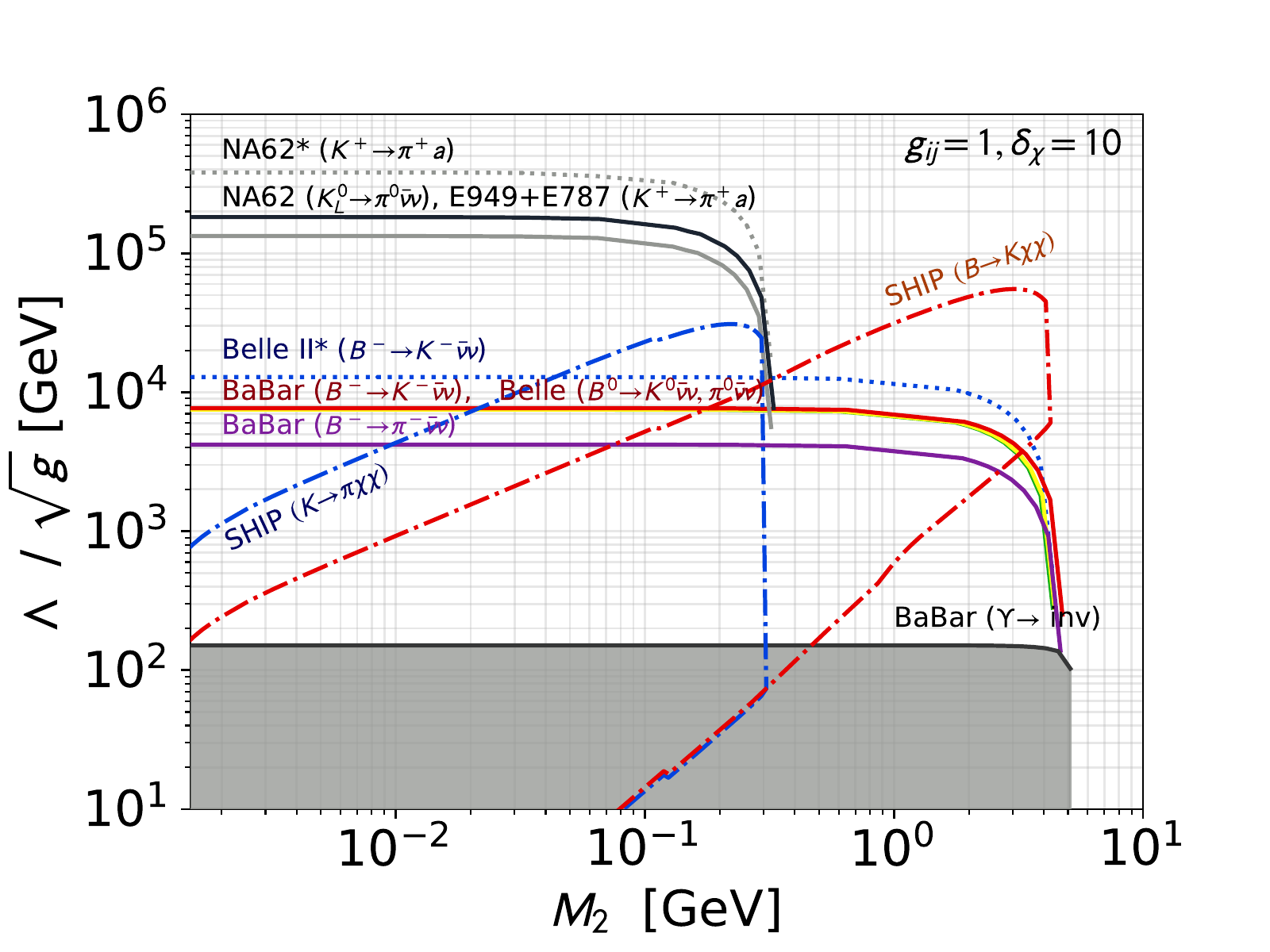}
			\vspace{-0.2cm}
	\caption{Heavy $B$ and $K$ meson limits and projected sensitivity to $\Lambda/\sqrt{g}$ for the vector-current operator as a function of $M_2 \gg M_1$ on the x-axis. Regions outlines by dashed blue (red) lines show the 10 events projection at SHiP for $\chi_2 \to \chi_1 e^+ e^-$ decay processes, produced by  $K \to \pi \chi_1 \chi_2$ ($B \to K \chi_1 \chi_2$). The solid lines denote actual bounds from invisible $B$ and $K$ decays while dotted lines are future projections. The normalised splitting $\delta_\chi$ is defined in Eq.~\ref{eq:dChi}.}
	\label{fig:limHeavymes}
\end{figure}

While most of the limits above are based on the coupling with first generation fermions, effective vector operators for heavy flavours can also be constrained using the same techniques. We present the limits for SHiP based on $K$ and $B$ meson decays in Fig.~\ref{fig:limHeavymes}. The invisible decay bounds are also shown as labelled for BaBar, Belle (II), NA62 and E949/787. We see that the heavier masses of the mesons involved can significantly extend the fermion portal sensitivity to higher effective scales. Notice that the invisible meson decay constraints appear to exclude a large region of the parameter space that SHiP will probe, in particular for the $K$ meson production case, but we recall that we assumed here the same scale suppression and couplings for both production and decay.

\subsection{Axial-vector operator}

\begin{figure}[t!]
		\vspace{-1cm}
	\centering
		\includegraphics[width=0.61\textwidth]{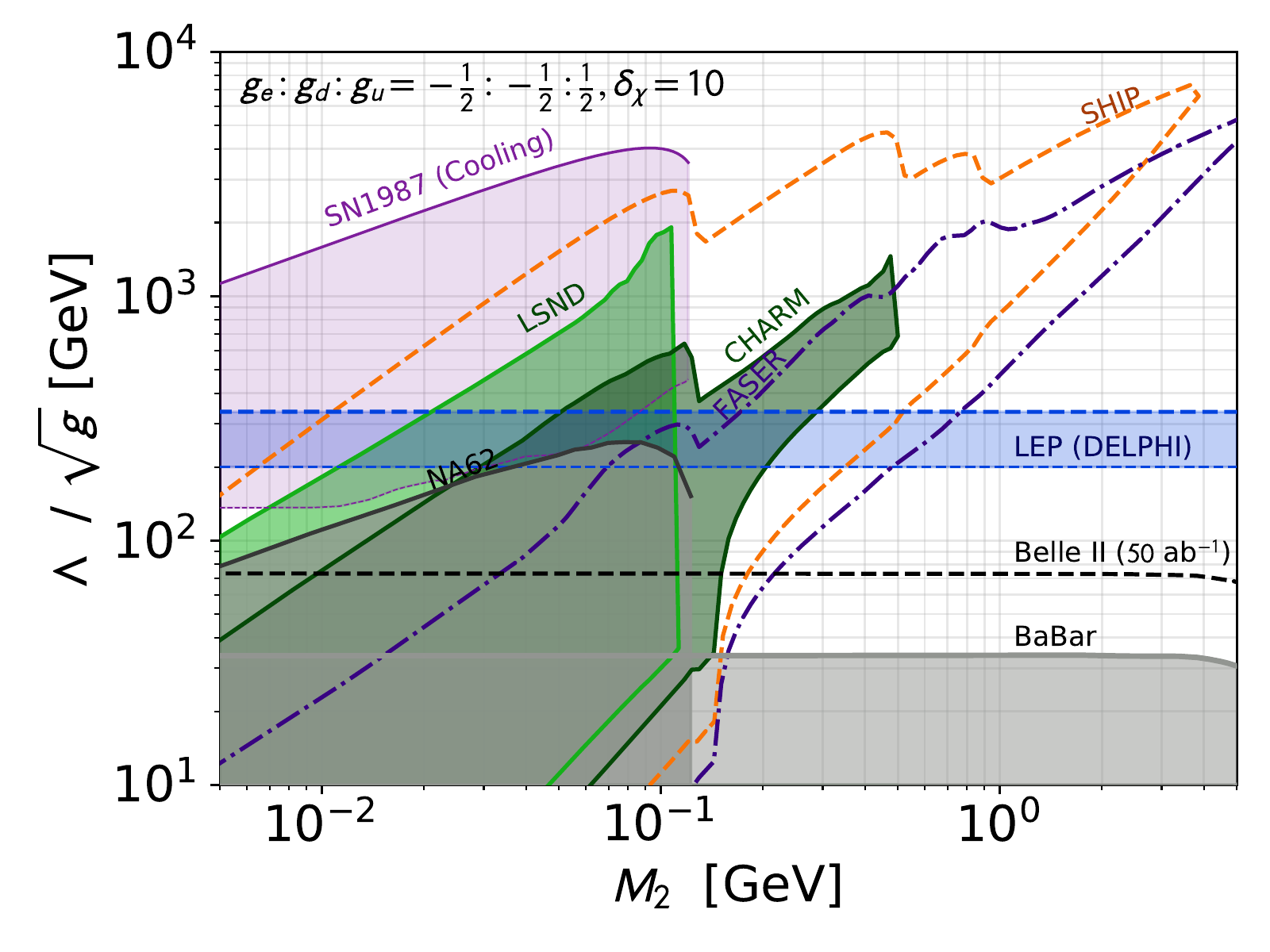}
	\hspace{0.02\textwidth}
	\vspace{-0.35cm}
	\caption{Limits and future sensitivity to the axial-vector operator effective scale $\Lambda/\sqrt{g}$ in the case of $Z$-aligned effective couplings as a function of $M_2 \gg M_1$ on the x-axis. Grey flat regions indicates mono-$\gamma$ limit at BaBar~\cite{Essig:2013vha}, dashed grey line Belle-II projections~\cite{Essig:2013vha}.  The light grey regions at low $M_2$ indicate NA62 $\pi^0 \to $inv limits~\cite{NA62:Kaon2019}. The shaded blue region is the mono-$\gamma$ limit from LEP~\cite{Fox:2011fx}. Limits from $\chi_2 \to \chi_1 e^+ e^-$: the green (dark green) regions are the exclusion from LSND~\cite{Darme:2018jmx} (CHARM~\cite{deNiverville:2018dbu}). We show a projection for FASER~\cite{Berlin:2018jbm} (dashed purple) and SeaQuest in red (Phase-II defined in~\cite{Berlin:2018pwi}). The reach of SHiP for 10 signal events is shown in dashed orange. The purple region represents the $\pi^0 \to \nu \nu$ limit from~\cite{Natale:1990yx} based on cooling of SN1987A. The normalised splitting $\delta_\chi$ is defined in Eq.~\ref{eq:dChi}.}
	\label{fig:LimitsAV}
		\vspace{-0.4cm}
\end{figure}

\begin{figure}[h!]
		\vspace{-0cm}
	\centering
		\includegraphics[width=0.61\textwidth]{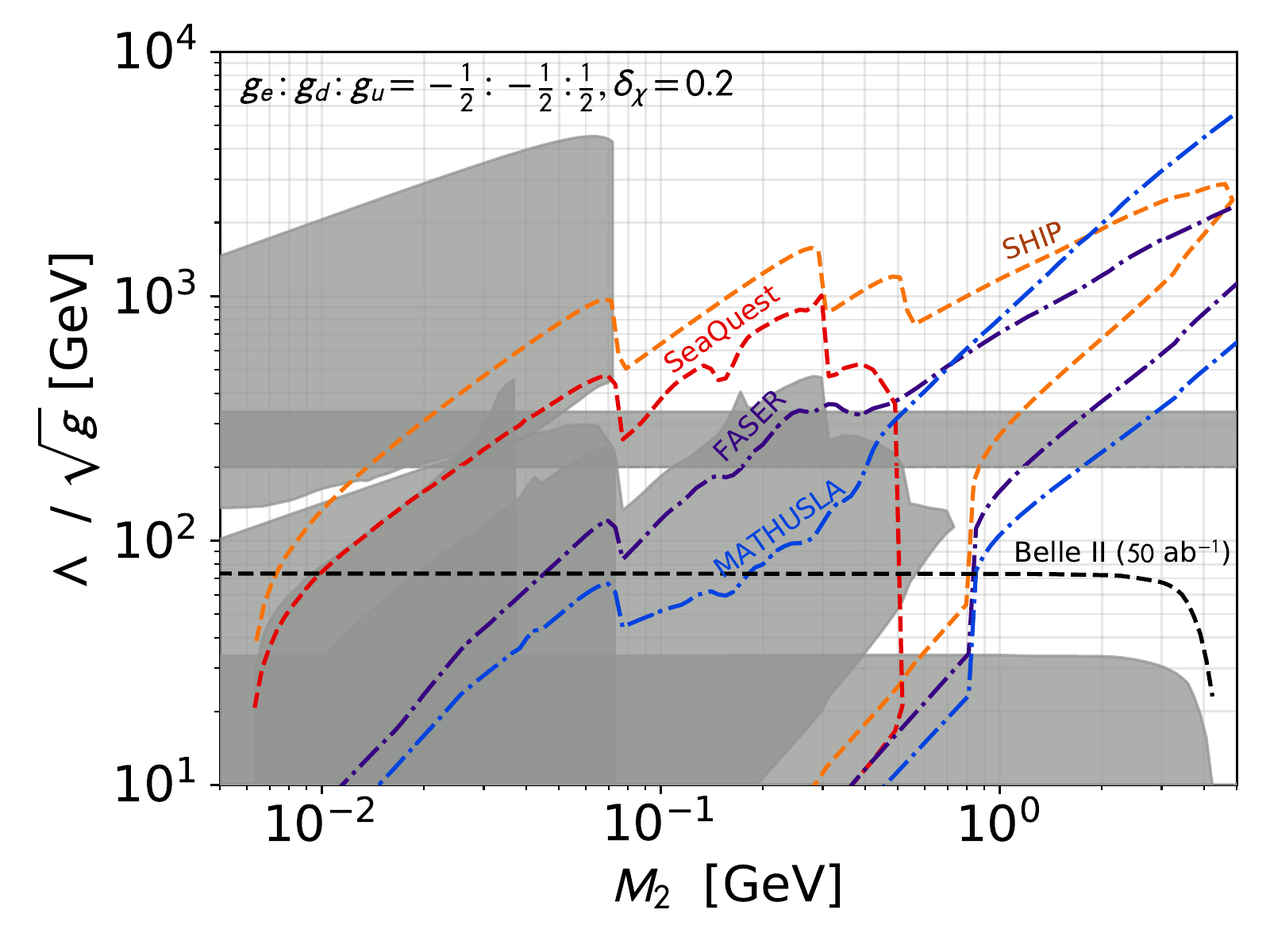}
			\vspace{-0.4cm}
	\caption{Prospective sensitivity to the axial-vector operator effective scale $\Lambda/\sqrt{g}$ from future intensity experiments in the case of $Z$-aligned effective couplings as a function of $M_2$ on the x-axis for $\delta_\chi = 0.2$. Grey areas indicate already covered parameter space. We show prospects for FASER~\cite{Berlin:2018jbm} in dashed purple and SeaQuest in red (Phase-II defined in~\cite{Berlin:2018pwi}), MATHUSLA~\cite{Berlin:2018jbm} in blue and SHiP in green (based on our 10 events projections). The normalised splitting $\delta_\chi$ is defined in Eq.~\ref{eq:dChi}.}
	\label{fig:LimitsAV_prospective}
\end{figure}

As a first example of limits based on the axial-vector operator, we focus on the $Z$-aligned limit with 
\begin{align}
g_u : g_d : g_l = \frac{1}{2} :   -\frac{1}{2} :  -\frac{1}{2}  \ .
\end{align}

We present the result of current limits and a representative selection of future sensitivities, for a splitting $\delta_\chi = 10$ in Fig.~\ref{fig:LimitsAV}. Notice that the two-body meson decay production mechanism strongly enhances the limits at low masses. In particular, the recasting of LSND searches leads to bounds up to the TeV scale in this case. An interesting feature of the lower limits for $\chi_2$ decay is the strong reduction of the limits for $M_2  - M_1 > M_\pi$ due to the opening up of the $\chi_2 \rightarrow \chi_1 \pi^0$ decay channel, as described in Sec.~\ref{sec:decay}. In Fig.~\ref{fig:LimitsAV_prospective} we emphasise the projected reach for future experiments, including FASER~\cite{Berlin:2018jbm}, SeaQuest (following the Phase-II proposal defined in~\cite{Berlin:2018pwi}), MATHUSLA~\cite{Berlin:2018jbm} and SHiP (based on our 10 events projections), again for the $Z$-aligned couplings but with a smaller splitting $\delta_\chi = 0.2$. Shown also shaded in grey are the present exclusions from SN1987A~\cite{Natale:1990yx}, LSND~\cite{Darme:2018jmx}, CHARM~\cite{deNiverville:2018dbu}, LEP~\cite{Fox:2011fx}, BaBar~\cite{Essig:2013vha} and NA62~\cite{NA62:Kaon2019}.

\begin{figure}[t]
	\centering
		\includegraphics[width=0.7\textwidth]{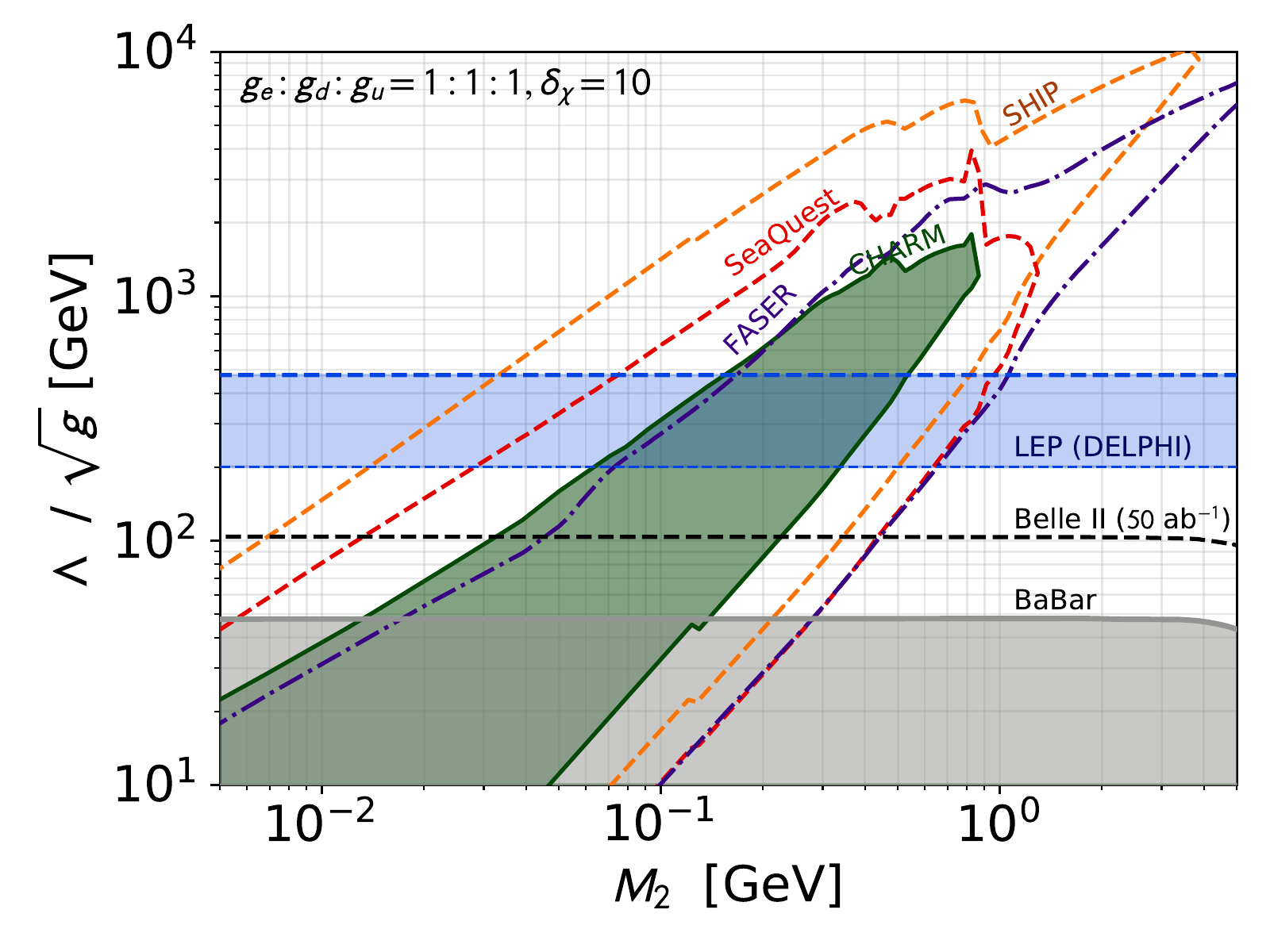}
			\vspace{-0.25cm}
	\caption{Limits and projected sensitivity to the axial-vector operator effective scale $\Lambda/\sqrt{g}$ in the case of a universal effective coupling, which translate into an effective pion-phobic scenario as function of $M_2 \gg M_1$ on the x-axis. Grey flat regions indicates coverage from mono-photon at BaBar~\cite{Essig:2013vha}, dashed grey line Belle-II projections~\cite{Essig:2013vha}.  The light grey regions at low $M_2$ indicate the exclusion from NA62 $\pi^0 \to $inv limits~\cite{NA62:Kaon2019}. The shaded blue region is the mono-$\gamma$ limit from LEP~\cite{Fox:2011fx}. Limits from $\chi_2 \to \chi_1 e^+ e^-$: the dark green region is the exclusion from CHARM~\cite{deNiverville:2018dbu}. We show a projection for FASER~\cite{Berlin:2018jbm} in dashed purple and SeaQuest in red (Phase-II defined in~\cite{Berlin:2018pwi}). The 10 events reach by SHiP is shown in dashed orange. The SN1987A limit is not shown as explained in the text. The normalised splitting $\delta_\chi$ is defined in Eq.~\ref{eq:dChi}.}
	\label{fig:LimitsAV_phobic}
\end{figure}

Finally, in Fig.~\ref{fig:LimitsAV_phobic}, we consider the case of an axial-vector effective operator with ``universal'' couplings
\begin{align}
g_u : g_d : g_l =1 : 1 : 1 \ .
\end{align}
This translates into an effective pion-phobic scenario due to the form of the effective coupling to pions, as shown in Table~\ref{tab:effcoup}. As in the proton-phobic case of the last section, the limit from LSND vanishes. Additionally the heavy state decay $\chi_2$ to a $\pi^0 \chi_1$ is strongly suppressed, extending downwards the limits. Finally, the limits from SN1987A obtained by considering the invisible decay of neutral pions~\cite{Natale:1990yx} no longer apply. A limit from SN1987A from $e^+e^-$ annihilation in the supernova core should still apply, but we have not computed it here. It would likely be similar to the $e^+ e^-$ annihilation constraint for the vector operator shown in e.g. Fig.~\ref{fig:LimitsEM}.

\subsection{Small mass splitting}

\begin{figure}
	\centering
		\includegraphics[width=0.7\textwidth]{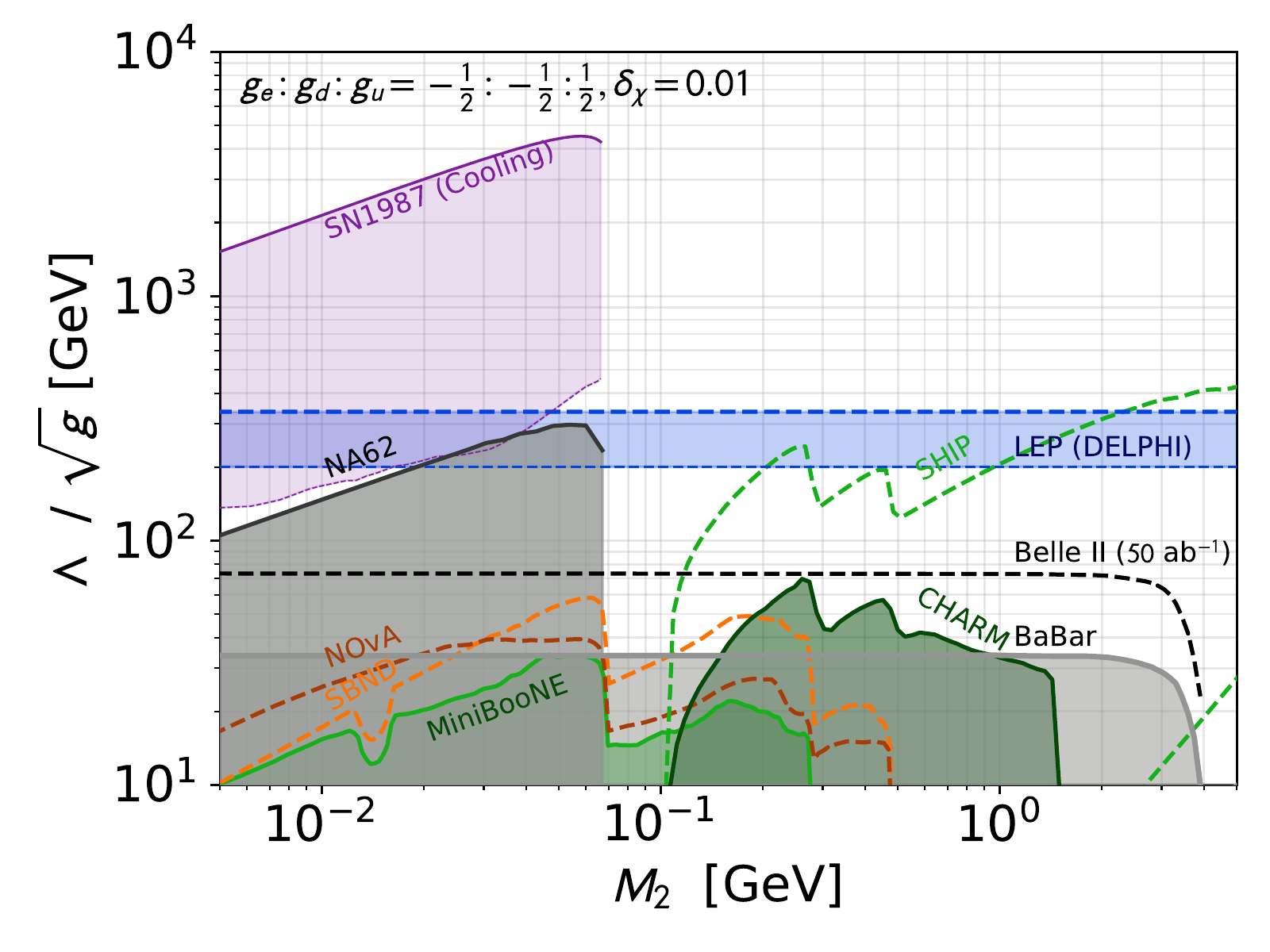}
		\includegraphics[width=0.7\textwidth]{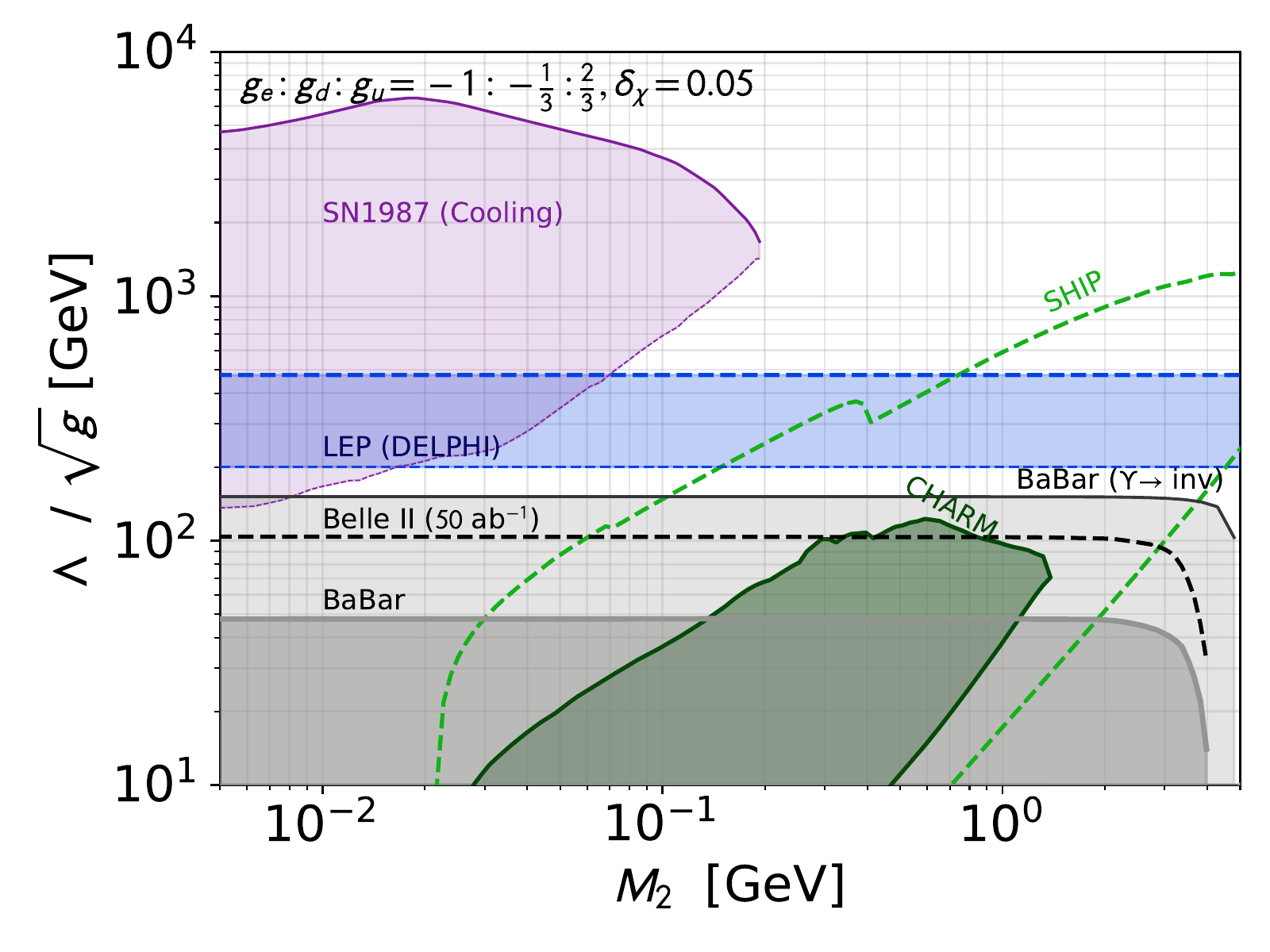}
	\caption{Limits and projected experimental sensitivity for small dark sector mass splitting for (a) an axial-vector current operator with Z-aligned coupling, with $\delta_\chi = 0.01$, and (b) a vector current operator with electromagnetically-aligned coupling for $\delta_\chi = 0.05$.  Grey flat regions indicates coverage from mono-photon at BaBar~\cite{Essig:2013vha}, dashed grey line Belle-II projections~\cite{Essig:2013vha}.  The light grey regions at low $M_2$ indicate the exclusion from NA62 $\pi^0 \to $inv limits~\cite{NA62:Kaon2019}. The shaded blue region is the mono-$\gamma$ limit from LEP~\cite{Fox:2011fx}. Limits from $\chi_2 \to \chi_1 e^+ e^-$: the dark green region is the exclusion from CHARM~\cite{deNiverville:2018dbu}. The 10 events reach by SHiP is shown in dashed orange. Scattering limits at MiniBooNE~\cite{Aguilar-Arevalo:2018wea} are shown in light green, and SBND~\cite{deNiverville:2016rqh} and NO$\nu$A~\cite{deNiverville:2018dbu} in orange and red. The normalised splitting $\delta_\chi$ is defined in Eq.~\ref{eq:dChi}. }
	\label{fig:smallsplit}
\end{figure}

In most of the plots presented above, the limits coming from the decay of the heavier $\chi_2$ dark sector state were important but highly dependent on the lifetime. Since the lifetime of the heavy state scales as $\Delta_\chi^5$ in the limit of small splitting between the dark sector masses, these limits are reduced for small splitting. We illustrate this aspect in Fig.~\ref{fig:smallsplit} on the top and bottom plots for the ($Z$-aligned) axial-vector and (electromagnetically-aligned) vector case with $\delta_\chi = 0.01$ and $0.05$, respectively. In particular, we have represented the limits based on scattering of the lighter state in MiniBooNE (based on Ref.~\cite{Aguilar-Arevalo:2018wea}), SBND (projections from Ref.~\cite{deNiverville:2016rqh}), and NO$\nu$A (from the $3\cdot10^{20}$ protons on target projection of Ref.~\cite{deNiverville:2018dbu}). While these limits are not competitive with the missing energy searches from BaBar, they may become relevant in a more lepton-phobic scenario. Notice additionally that the lower mass thresholds for the $\chi_2 \to \chi_1 e^+ e^-$ decay to be allowed are shifted to higher masses relative to figures where the mass splitting is large.

\subsection{Concrete scenario: GeV scale dark photon}

We end with a practical application of the approach presented above, in terms of the familiar dark photon benchmark as a possible UV completion of the fermion portal. Limits on dark photons decaying invisibly to dark sector fermions with a couplings $g_D$ are currently relatively weak, in the tens of GeV range, with the current best limits arising from LEP~\cite{Curtin:2014cca}. Interestingly, such a heavy dark photon $V$ has a sizeable mixing with the Standard Model Z boson, leading naturally to an axial vector current coupling with the SM fermions. One can go from the effective approach to the simplied model (in the off-shell dark photon limit) using
\begin{align}
    \varepsilon_{\rm lim} = \frac{M_V^2}{\Lambda_{\rm lim}^2 \sqrt{4 \pi \alpha_{\rm em }} g_D}
\end{align}

\begin{figure}[t]
	\centering
		\includegraphics[width=0.7\textwidth]{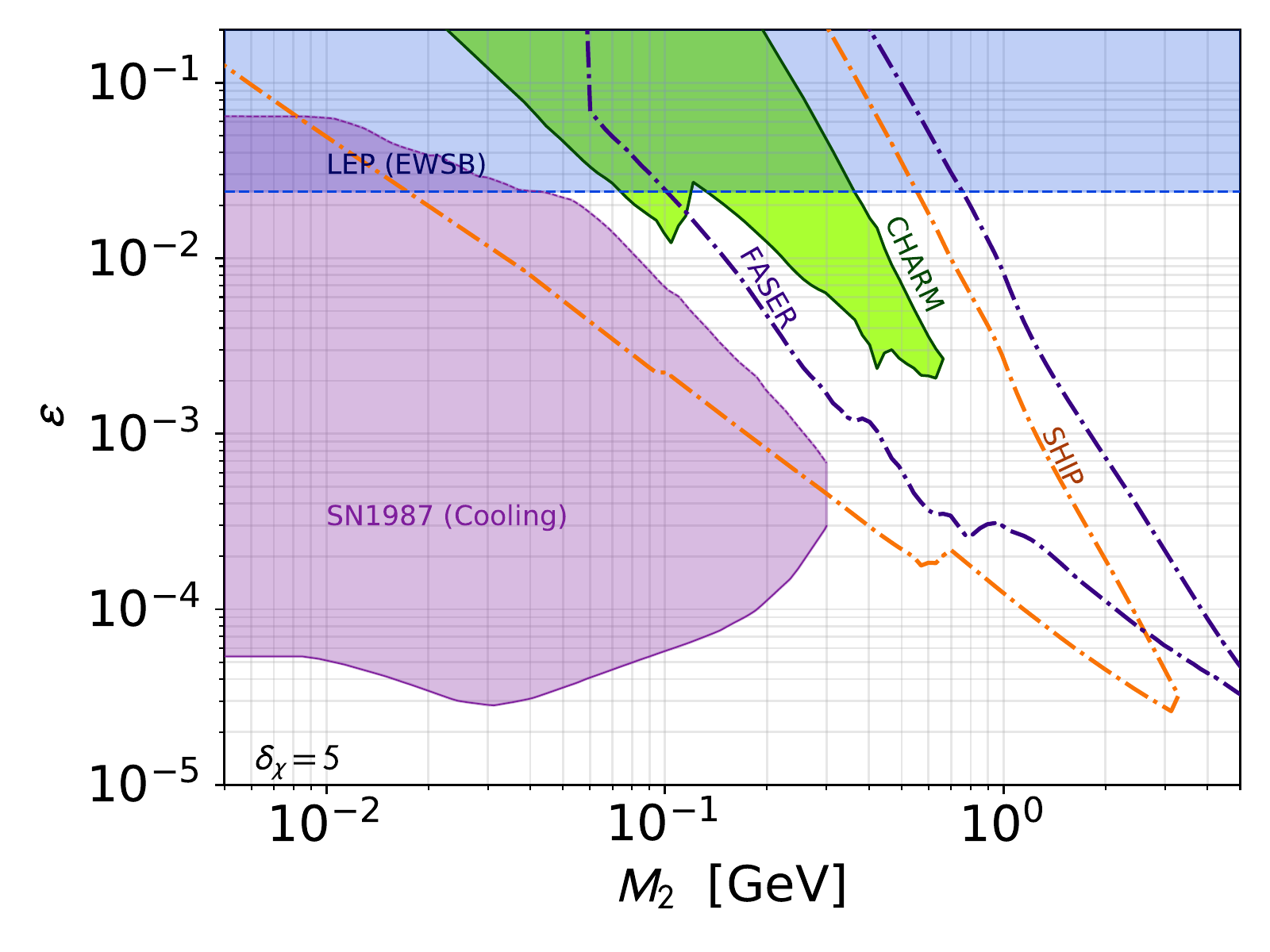}
	\hspace{0.02\textwidth}
	\vspace{-0.5cm}
	\caption{Limits and projected sensitivity to a heavy dark photon mediator coupled to dark sector fermions, with $M_V = 20$ GeV. LEP mono-$\gamma$ bounds are shown in shaded blue. For limits from $\chi_2 \to \chi_1 e^+ e^-$, the dark green region is the exclusion from CHARM~\cite{deNiverville:2018dbu} and the 10 events reach by SHiP is shown in dashed orange. We also show a projection for FASER from~\cite{Berlin:2018jbm}. The normalised splitting $\delta_\chi$ is defined in Eq.~\ref{eq:dChi}.}
	\label{fig:LimitsDP}
\end{figure}

Noting that at first order in the kinetic mixing $\varepsilon$ the axial vector coupling are $\delta^2$-suppressed where $\delta \equiv M_V / M_Z \ll 1 $, we have  
\begin{align}
    \tilde{g}_{u, 11} = - \tilde{g}_{d, 11} = -\tilde{g}_{e, 11} \simeq -\frac{\delta^2 e \varepsilon }{4 c_W^2} \ ,
\end{align}
where $c_W$ is the cosine of the Weinberg angle.
The vector couplings are not significantly modified as long as $\delta^2 \ll 1$:
\begin{align}
    g_{e, 11}=-e\varepsilon = -\frac{3}{2} g_{u, 11}= 3 g_{d, 11} \ .
\end{align}

An important point is that while production of dark sector states can now proceed through either of the operators (including the significant boost observed at low masses in the axial-vector case), the lifetime depends on the sum of the two decay width. In particular, ``mixed'' contributions where production proceeds through the axial-vector operator and decays through the vector one dominates at low masses (when meson decay production for $\pi^0$ and $\eta$ dominate, as seen in Sec.~\ref{sec:production}). We illustrate these effects and summarise the bounds in Fig.~\ref{fig:LimitsDP}, where we show the limits on $\varepsilon$ for $M_V = 20$ GeV. Note that the limit from FASER are conservative in that this experiment will have access to enough energy to produce on-shell a dark photon of such mass.

\section{Conclusion}

Light dark sectors are a class of new physics beyond the SM that present special challenges and opportunities for discovery at the intensity frontier. Since they are neutral under the SM gauge groups, dark sector fields only interact with the SM through so-called portal operators --- gauge singlet combinations of SM fields. Much work has been done on studying the phenomenology of the three lowest-dimensional portal operators: the Higgs, vector, and neutrino portals. Here we have presented a study of the next lowest-dimensional ``fermion'' portal. We focused on the dimension-6 four-fermion operator combination of a fermion portal to a pair of light dark sector fermions. The higher-dimensional nature of this portal can arise naturally from the off-shell limit of one of the renormalisable portal interactions. The scale suppression could moreover explain the weakness of the interaction between visible and dark sectors. Our effective field theory approach has the advantage of encapsulating the phenomenology of light dark sectors interacting with the Standard Model through heavy mediators in full generality.

In our study the typical production mechanisms at intensity frontier experiments present several interesting modifications compared to the standard vector or scalar portals, for instance. Typically, light meson production is subdominant due to the scale suppression arising from the higher-dimensional nature of the portal. An interesting exception is that light pseudo-scalar mesons can have a fully invisible two-body decay in the case of the axial-vector fermion portal, leading to a strong enhancement of the reach of low-energy experiments for such scenarios. Additionally, different phenomenology may be exhibited in bremsstrahlung production of light fermion pairs via an off-shell mediator at electron beam dump experiments. Despite the scale suppression, this may be a relevant mechanism that we will consider in future work. 

Similarly, the detection strategies adopted for the renormalisable portals are modified by the off-shell nature of the fermion portal. In particular, missing energy limits (e.g. mono-photon searches at BaBar or Belle-II) only lead to weak bounds in our case due to the absence of a resonance in the missing mass spectrum. Limits from the scattering or decay of heavy dark sector states can also be adapted to the case of a fermion portal. To place most of our limits on the fermion portal, we simulate the production and decay of light states and use the comprehensive existing analyses for renormalisable portals, such as dark photon and inelastic dark matter models, to estimate experimental efficiencies. Note that in this analysis, we do not consider possible differences in experimental efficiencies due to kinematics. We also do not account for possible renormalisation group running and consequent mixing of fermion portal operators between the various scales involved, which can vary from MeV for decay processes to TeV for parton-level production at the LHC (see e.g. Refs.~\cite{DEramo:2016gos,Arteaga:2018cmw} for the dark matter case). Our framework for applying limits on the fermion portal is available as the public \texttt{DarkEFT} code (described in Appendix~\ref{app:Code}).

We outlined the parameter space coverage of existing and future experiments for both vector and axial-vector operators, considering flavour scenarios such as electromagnetically-aligned or Z-aligned couplings and proto-phobic or pion-phobic couplings. Depending on the nature of the operator, we found that the effective scale for light flavours was typically constrained to lie in the hundreds of GeV to TeV range for effective couplings $g \sim \mathcal{O}(1)$. The combination of missing energy searches typically lead to the requirement that $\Lambda / \sqrt{g} \gtrsim 500$ GeV for dark fermion masses of up to a few GeV. There is a gap in the limits where the effective field theory approach breaks down when considering searches at LEP. This would typically be excluded in a complete model including a kinematically accessible mediator. This gap can also be mostly covered from limits from invisible meson decays when those are available. For smaller dark sector masses up to $m_\chi \sim 100$ MeV, astrophysical constraints from SN1987A cooling play a key role in constraining the effective scale up to several TeV, although the lower ``trapping'' bound has a strong model-dependence as discussed in Section~\ref{sec:astrophyics}. Limits from experiments involving heavy $B$ and $K$ mesons could extend the sensitivity to effective scales in the tens of TeV. While we focused mostly on dark sector fields heavier than several MeV, we also derived an order of magnitude estimate of around $5$ TeV from Planck on the limits on the effective scale from $N_{\rm eff}$ when the dark sector fields behave as relativistic matter.

This paper has focused on fermion portal operators which are either flavour-diagonal in both lepton and quark sectors or flavour-breaking in the quark sector. In particular we do not currently consider flavour-breaking operators in the leptonic sectors and leave to future work limits on the neutrino-based operators. These should typically be generated in the UV along with leptonic ones --- especially when considering the axial-vector fermion portal operator. We also note that the restriction to fermion portal operators to a pair of dark sector fermions, with a $(\bar\chi \Gamma \chi)$ structure in the dark sector, is actually not particularly restrictive for characterising the fermion portal more generally, since most of the phenomenology will depend on the SM part of the higher-dimensional operator. Our results can then be considered as good guidelines for fermion portals to other dark sector combinations (for instance $S \partial_\mu S$ with $S$ a dark scalar).

For future work there are more directions where the bounds can be either refined or improved, some of which were discussed above. One important addition would be to include and simulate the production rates from heavy meson decay and their detection prospects in terms of decay or scattering in high energy frontier experiments. The order of magnitude estimate for SHiP presented here points to limits significantly stronger than those from standard light invisible meson decays. Another refinement of the limits would be to simulate more thoroughly the production and decay of heavy dark sector states through the fermion portal; in particular near short lifetime limits, where our conservative estimates could be improved since the high-energy tail of the spectrum will dominate the expected events. Other portal operators could also be investigated. 

In this work we have taken a step towards a systematic study of the Standard Model portal operators through which dark sectors necessarily interact. As the next decade begins, the intensity frontier of particle physics could enable a thorough exploration of the universe's dark sectors.

\bigskip
\noindent \textbf{Acknowledgments}
\medskip

LD thanks J. Wagner for helpful discussions. SARE thanks G. Marques-Tavares for useful discussions. LD was supported in part by the National Science Center (NCN) research grant No.~2015-18-A-ST2-00748 during the course of this work. SARE is supported in part by the U.S. Department of Energy under Contract No. DE-AC02-76SF00515, and by the Swiss National Science Foundation, SNF project number P400P2\_186678. TY is supported by a Branco Weiss Society in Science Fellowship and partially supported by STFC consolidated grant ST/P000681/1. SARE thanks the Galileo Galilei Institute (GGI) for Theoretical Physics for its hospitality and support.

\appendix

\section*{Appendix}

\addcontentsline{toc}{section}{Appendices}

\section{Meson decay amplitudes\label{app:mesdecaycalc}}
In this appendix we detail the calculations and references used to obtain the effective coupling and decay rates presented in Sections~\ref{sec:lightmesdecays} and \ref{sec:heavymes}.

\subsection*{Light meson two-body decays}
In order to find the various mesonic decay amplitude, we need to determine the matrix element for the corresponding interpolating current. In the case where the meson directly decays to two dark sector particles, this implies determining the decay constant defined such that
\begin{align}
\bra{0}  \sum_{q \in u,d,s}  \frac{\tilde{g}_q}{\Lambda^2}  (\bar{q} \gamma^{\mu} \gamma^{5} q)  \ket{M(p)}  ~\equiv~ i \frac{f_M}{\Lambda^2}  p^\mu 
\end{align}
for the case of a pseudo-scalar meson $M$ and as
\begin{align}
\bra{0}  \sum_{q \in u,d,s}  \frac{g_q}{\Lambda^2}  (\bar{q} \gamma^{\mu} q)  \ket{V(\varepsilon_\lambda)}  ~\equiv~ i \frac{M_V f_V}{\Lambda^2}  \varepsilon^\mu_\lambda
\end{align}
for the case of a vector meson $V$ with polarisation vector $\varepsilon_\lambda$. The amplitude then follows straightforwardly. For $\rho, \omega$, we use directly the summary from~\cite{Straub:2015ica}, which accounts for the mixing effects between vector mesons. The decay constants associated to each quarks are:
\begin{align}
    f^{(u)}_\rho = 222 \mev &&& f^{(u)}_\omega = 192 \mev \nn \\ 
    f^{(d)}_\rho = 210 \mev &&& f^{(d)}_\omega = 201 \mev  \\
    f_\phi = 233 \mev \nn \ .
\end{align}
We can therefore straightforwardly project our operators on the interpolating current for the vector mesons
\begin{align*}
    \mathcal{A}^\mu_{\omega / \rho} = \frac{1}{\sqrt{2}}\left( \bar{u} \gamma^{\mu} u 
    \pm  \bar{d} \gamma^{\mu} d \right) \qquad , \qquad  \mathcal{A}^\mu_\phi =  \bar{s} \gamma^{\mu} s 
\end{align*}
to obtain the effective couplings presented in Table~\ref{tab:effcoup}.

For the direct decay of a pseudo-scalar meson $M\rightarrow \chi \chi $ in the case of axial-vector current, we follow the approach for neutrino decays presented in~\cite{Herczeg:1981xa}, and extend it to the $\eta$ and $\eta^\prime$ case using~\cite{Escribano:2005qq} to account for the mixing effects. In more details, we introduce the currents $\Azero, \Aoct$ as:
\begin{align}
\Azero &= \frac{1}{\sqrt{3}} \left[ \bar{u} \gamma^{\mu} \gamma^{5} u + \bar{d} \gamma^{\mu} \gamma^{5} d +  \bar{s} \gamma^{\mu} \gamma^{5} s \right] \nonumber \\
\Aoct &= \frac{1}{\sqrt{6}} \left[ \bar{u} \gamma^{\mu} \gamma^{5} u + \bar{d} \gamma^{\mu} \gamma^{5} d  - 2  \bar{s} \gamma^{\mu} \gamma^{5} s \right] 
\end{align}
And the corresponding decay constant $f_8$ and $f_0$ defined as
\begin{align}
\bra{0}  \mathcal{A}_a^\mu   \ket{M(p)}  ~\equiv~ i f_a  p^\mu \ .
\end{align}
In the two angle mixing scheme (see e.g~\cite{Escribano:2005qq}), the $\eta$ and $\etap$ are then represented by:
\begin{align}
  \ket{\eta} &= \cos \theta_8  \Aoct  \ket{0} - \sin \theta_0 \Azero  \ket{0} \\
    \ket{\etap \ } &= \sin \theta_8  \Aoct  \ket{0} + \cos \theta_0 \Azero  \ket{0} \ ,
\end{align}
where $\theta_8 \sim -22^{\circ}$ and  $\theta_8 \sim -9^{\circ}$.
Finally, we can project our set of operators on the $ \Azero,\Aoct$ basis to obtain (using the usual shorthand notation for $\cos$ and $\sin$ as $c,s$):
\begin{align}
    f_\eta &= f_8  c_8 \frac{g_u + g_d - 2 g_s}{\sqrt{6}} -  f_0  s_0 \frac{g_u + g_d + g_s}{\sqrt{3}} \\
    f_{\eta^\prime} &= f_8  s_8 \frac{g_u + g_d - 2 g_s}{\sqrt{6}} +  f_0  c_0 \frac{g_u + g_d + g_s}{\sqrt{3}} \ .
\end{align}
Using the fitted values $f_8 = 1.28 f_\pi, f_0 = 1.2 f_\pi$ from~\cite{Escribano:2005qq}, we obtain the results of Table~\ref{tab:effcoup}.

\begin{figure}[t]
	\centering
	\subfloat[]{%
		\label{fig:RadDecayDiag_a}
		\includegraphics[width=0.3\textwidth]{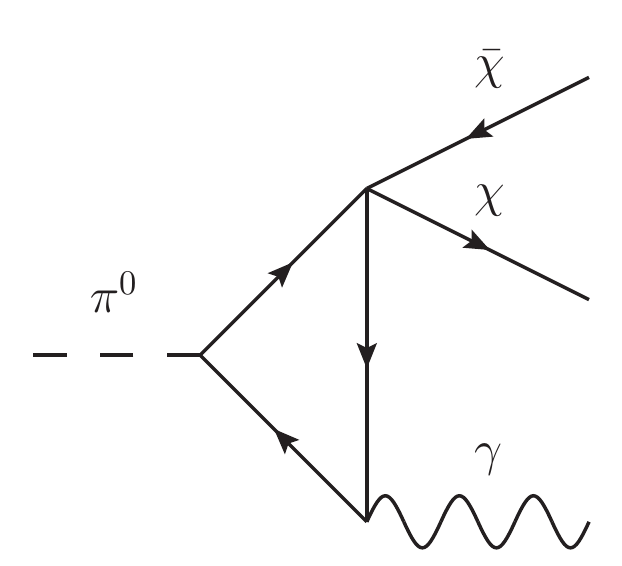}
	}%
	\hspace{0.02\textwidth}
	\subfloat[]{%
		\label{fig:RadDecayDiag_b}
		\includegraphics[width=0.3\textwidth]{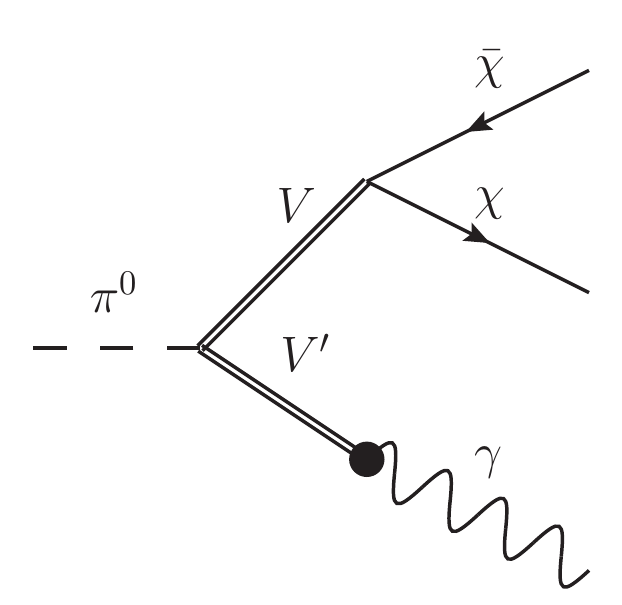}
	}%
	\hspace{0.02\textwidth}
	\caption{(a) Triangle diagram approach to the radiative $\pi^0$ decay. (b) VMD approach radiative to the $\pi^0$ decay}
	
\end{figure}

\subsection*{Light meson three-body decays}

In the case of an ``associated'' radiative decay $M \rightarrow X \gamma$. We obtained the results from the main text using two distinct procedures. The first approach relies on the Vector Meson Dominance (VMD) advocated in~\cite{Fujiwara:1984mp} which has already been used extensively for the case of new dark vector particles searches (see e.g.~\cite{Tulin:2014tya,Ilten:2018crw}). We define the $U(3)$ generators for the relevant mesons as:
\begin{align}
    T_{\pi^0} &= \frac{1}{2} \textrm{diag} (1,-1,0) &&& T_{\rho} &= \frac{1}{2} \textrm{diag} (1,-1,0) \nonumber \\
    T_{\eta} &= \frac{1}{\sqrt{6}} \textrm{diag} (1,1,-1) &&& T_{\omega} &= \frac{1}{2} \textrm{diag} (1,1,0) \\
    T_{\etap} &= \frac{1}{2 \sqrt{3}} \textrm{diag} (1,1,2) &&& T_{\phi} &= \frac{1}{\sqrt{2}} \textrm{diag} (0,0,1) \nonumber \ ,
\end{align}
where we have used the same simplified approach with a single $\eta-\etap$ mixing angle $\theta$ with $\cos \theta \sim \sqrt{6}/3$ and $\sin \theta \sim -1/3$ as in~\cite{Tulin:2014tya,Ilten:2018crw} (see also~\cite{Batell:2014yra}), based on~\cite{Feldmann:1999uf}. Furthermore, we can define the electromagnetic and dark coupling matrices as:
\begin{align*}
Q &= \frac{e}{3} \textrm{diag} (2,-1,-1) \\
Q_D &= \textrm{diag} (g_u,g_d,g_s) \ .
\end{align*}
Using the Feynman rules following~\cite{Fujiwara:1984mp} and defining the coupling $\grpp \sim 6.1$, we first consider the amplitude for the process $V \rightarrow \chi \bar{\chi}$, where $V$ is a vector meson of polarisation vector $\epsilon^V$:
\begin{align}
    \mathcal{A}_{V \rightarrow \chi \bar{\chi}} = \frac{M_V^2}{\grpp \Lambda^2} 2 \Tr (T_V Q_D) \ \epsilon_\mu^V \bar{u}\gamma^\mu v ~\equiv~ M_V f_V^{\rm eff} \epsilon_\mu^V \bar{u} \gamma^\mu v \ .
\end{align}
This leads to effective coupling constants within $ \sim 10 \%$ of the complete results from~\cite{Straub:2015ica} used in Table~\ref{tab:effcoup} (this result can also be recovered by taking the heavy dark vector limit based on the expressions from~\cite{Tulin:2014tya,Ilten:2018crw}). Based on the calculation of the $\pi^0 \rightarrow \gamma \gamma$ amplitude from~\cite{Fujiwara:1984mp}, we can now estimate the amplitude corresponding to the three-body decay of a pseudo-scalar meson $P (p^\pi) \rightarrow \gamma (q) \bar{\chi_1} \chi_2$ from Fig.~\ref{fig:RadDecayDiag_a} as:
\begin{align}
    | \mathcal{A}_{P \rightarrow \gamma \chi \bar{\chi}} |  = & \left( 4 |\grop| \times \frac{1}{\Lambda^2 \grpp^2} \times \sum_{V = \rho, \omega, \phi}  2 \Tr \left[Q T_V T_P \right] \Tr \left[Q_D T_V \right] \right)\\
    & \times \varepsilon^{\mu \nu \rho \sigma} q_\mu \varepsilon_\nu (p^\pi_\sigma-q_\sigma) (\bar u(\chi_1) \gamma_\rho v(\chi_2)) \nonumber 
\end{align}
where we have used $\grop = - \frac{3 \grpp^2}{ 8 \pi^2 F_\pi}$, with $F_\pi \simeq 93$ MeV. Note that we have neglected the momentum dependence in the vector propagator so that it simply amounts to inserting a $1/M_V^2$ factor; we briefly discuss this approximation at the end of this Appendix. We can deduce the effective coupling $g_P$  from the first line of this equation. By comparing with the coupling in the $\pi^0 \rightarrow \gamma \gamma$ case, we obtain:
\begin{align}
    g_P = 12 \! \sum_{V = \rho, \omega, \phi}  \Tr \left[Q T_V T_P \right] \Tr \left[Q_D T_V \right] \ .
\end{align}
This expression agrees with the one of Table~\ref{tab:effcoup} (based on the second approach below) at $10 \%$ level for $g_u,g_d$ and $30 \%$ level for $g_s$, all the observed discrepancies are related to the single angle scheme, and replacing the $T_P$ of $\eta$ and $\etap$ by the two mixing angles scheme which we will use below leads to perfect agreement.

The second approach builds directly on the effective amplitude for the radiative decay of a $\pi^0$ into a photon with polarisation $\varepsilon^\lambda$ and a dark vector boson $V$ of polarisation $\tilde{\varepsilon}^\kappa$ :
\begin{align}
\mathcal{A}_{P \gamma V} \supset \frac{e g_{Q}}{4 \pi^2 F_\pi}  \varepsilon^{\mu \nu \rho \sigma} q_\mu \varepsilon_\nu \tilde{\varepsilon}^\kappa (p^P_\sigma-q_\sigma) \ ,
\end{align}
where the effective coupling is estimated from the triangular diagram presented in Fig.~\ref{fig:RadDecayDiag_b} as (see~\cite{Feldmann:1999uf,Bernstein:2011bx,Tulin:2014tya})
\begin{align}
\label{eq:effcoupPgam}
    g_{Q} = 6 \Tr [Q Q_D T_P] \ ,
\end{align}
and where the $T_P$ matrices arises according to the quark content of the interpolating currents of the pseudo-scalar mesons. We obtain the amplitude in our case for a vector-current interaction by replacing the dark photon polarisation vector by the dark vector current $(\bar\chi_1 \gamma_\rho \chi_2) / \Lambda^2$ (which is of course similar to integrating out the dark photon): 
\begin{align}
\mathcal{A}_{P \gamma \chi \bar\chi} \supset \frac{e g_{Q}}{4 \pi^2 F_\pi \Lambda^2}  \varepsilon^{\mu \nu \rho \sigma} q_\mu \varepsilon_\nu (\bar u(\chi_1) \gamma_\rho v(\chi_2)) (p^P_\sigma-q_\sigma) \ .
\end{align}

While the matrices $T_P$ have been described above, for the case of $\eta$ and $\etap$ meson, we can go beyond the single mixing angle approach presented above using~\cite{Escribano:2005qq}. Defining 
\begin{align*}
  T_0 &= \frac{1}{\sqrt{3}}\textrm{diag} (1,1,1) \\
    T_8 &= \frac{1}{\sqrt{6}}\textrm{diag} (1,1,-2)\ ,
\end{align*}
we obtain
\begin{align}
  T_\eta &= \frac{f_\pi}{\sqrt{2} \Delta_f} (c_0 f_0 T_8 - s_8 f_8 T_0) \\
   T_{\eta^\prime} &= \frac{f_\pi}{\sqrt{2} \Delta_f} (- s_0 f_0 T_8 + c_8 f_8 T_0) \ ,
\end{align}
where $\Delta_f =  f_0 f_8 ( c_0 c_8 + s_0 s_8) $
We can then use directly these matrices in Eq.~\ref{eq:effcoupPgam} to recover the results presented in Table~\ref{tab:effcoup}.

Let us close this section by a comment regarding the Vector-Meson-Dominance approach. One of the main prediction of this approach, already noticed in~\cite{Tulin:2014tya}, is the presence of the propagator for the vector meson, leading in particular to resonances in the case of dark photon production from $\etap$ decay. In our approach, the vector meson can decay directly to a pair of dark sector fields. This implies, firstly, that the resonance will be automatically integrated over, limiting its effect and, secondly, for $\etap$ decays where these effects are relevant, the main production channel is in fact the direct decay $\rho, \omega \rightarrow \chi \bar{\chi}$ which dominates over the associated $\etap \rightarrow \gamma  \chi \bar{\chi}$.

\subsection*{Heavy meson three-body decays}
\label{app:heavymesdecaycalc}
The amplitude for the three-body decay of a heavy pseudo-scalar meson $P_1$ to another meson $P_2$ and two dark sector fermions $\chi_3, \chi_4$, proceeding via a four-fermion operator with a vector coupling $g_{12}$, can be written as
\begin{equation}
\mathcal{M}(P_1 \to P_2 \chi_3 \chi_4) = -i \frac{g_{12}}{\Lambda^2} \bar{u}(p_3)\gamma^\mu v(p_{4}) f_+(0) (p_{1} + p_{2})_\mu \, ,
\end{equation}
where we take the hadronic form factor $f_+(q^2)$ in the limit $q^2 \to 0$. This approximation is valid for light dark sector fermions and since the form factor only varies by $\mathcal{O}(1)$ factors we shall make this assumption in our calculation~\cite{Bjorkeroth:2018dzu}.

Finally the three-body differential decay width, averaging over spin states and integrating over angles, is then given by 
\begin{equation}
d\Gamma = \frac{1}{(2\pi)^3}\frac{1}{32 m_1^3} |\mathcal{M}|^2 dm_{23}^2 dm_{34}^2 \, ,
\end{equation}
where $m_{ij}^2 = (p_i + p_j)^2$. Note that $m_{23}^2 + m_{34}^2 + m_{24}^2 = \sum_{i=1}^{4} m_i^2$. The decay width is obtained by integrating over the phase space range
\begin{align}
(m_{23}^2)_\text{min} &= (m_2 + m_3)^2 \quad , \quad (m_{23}^2)_\text{max}= (m_1 - m_4)^2 \, , \\
(m_{34}^2)_\text{min} &= (E_3^* + E_4^*)^2 - \left(\sqrt{{E_3^*}^2 - m_3^2} + \sqrt{{E_4^*}^2 - m_4^2} \right)^2 \, , \\
(m_{34}^2)_\text{max} &= (E_3^* + E_4^*)^2 - \left(\sqrt{{E_3^*}^2 - m_3^2} - \sqrt{{E_4^*}^2 - m_4^2} \right)^2 \, , 
\end{align} 
with 
\begin{align}
E_3^* = \frac{m_{23}^2 - m_2^2 + m_3^2}{2 m_{23}} \, , \\
E_4^* = \frac{-m_{23}^2 - m_4^2 + m_1^2}{2 m_{23}} \, .
\end{align}

\section{Available limits and \texttt{DarkEFT} companion code} \label{app:Code}

In order to simplify the use of the results presented in this paper, we have released a companion code \texttt{DarkEFT}, written in python and available at: \href{https://github.com/Luc-Darme/DarkEFT}{https://github.com/Luc-Darme/DarkEFT}.
Its main features are:
\begin{itemize}
    \item A database of relevant analysis and limits, along with the relevant references and a small description.
    \item Amplitudes for various relevant production and decay mechanisms for dark sector states within the effective field theory presented above.
    \item A set of tools to recast the stored limits to the fermion portal case.
\end{itemize}

Importing the main python module of the code can done as

\begin{lstlisting}[frame=single] 
import LimitsList as lim
\end{lstlisting}

DarkEFT allows to very simply recast existing limit for any choice of the effective couplings. As an example, recasting the limits from the MiniBooNE collaboration from light dark matter scattering presented in~\cite{Aguilar-Arevalo:2018wea}, for a electromagnetically-aligned vector operator and for a splitting of $25\%$ between $\chi_1$ and $\chi_2$ can be done by

\begin{lstlisting}[frame=single] 
geffem={"gu11":2/3.,"gd11":-1/3.,"gl11":-1.} 
xi_full,Lim_full= lim.miniboone_scattering.recast(0.25,geffem,"V")
\end{lstlisting}

The full details and possibility of the code are presented directly in the Readme file. The current sets of implemented limits are presented in Table~\ref{tab:explist}.


\begin{table}[t]
	\begin{center}
		\begin{tabular}{|c|c|c|}
				\hline
				\rule{0pt}{14pt}Experiment & Search & Ref.  for recasting \\
				\hline
				\hline
				\rule{0pt}{14pt}
			    MiniBooNE & $e^-$ scattering &\cite{Aguilar-Arevalo:2018wea}, Fig. 24  \\[0.1em]
			    NO$\nu$A & $e^-$ scattering &\cite{deNiverville:2018dbu}, Fig. 2 \\[0.1em]
			    SBND & $e^-$ scattering& \cite{deNiverville:2016rqh}, Fig. 9b \\[0.1em]	
			    SHiP & $e^-$ scattering &\cite{deNiverville:2016rqh}, Fig 24a \\
			    \hline
			    \rule{0pt}{14pt}
			    MATHUSLA & $\chi_2 \rightarrow \chi_1 e^+ e^-$ decay & \cite{Berlin:2018jbm}, Fig. 7 \\[0.1em]
			    FASER & $\chi_2 \rightarrow \chi_1 e^+ e^-$ decay & \cite{Berlin:2018jbm}, Fig. 7  \\[0.1em]
			    SeaQuest & $\chi_2 \rightarrow \chi_1 e^+ e^-$ decay & \cite{Berlin:2018pwi}, Fig. 12 \\[0.1em]
			    LSND & $\chi_2 \rightarrow \chi_1 e^+ e^-$ decay & \cite{Darme:2018jmx} Fig. 5a or \cite{Izaguirre:2017bqb} Fig. 6 \\[0.1em]
			    CHARM & $\chi_2 \rightarrow \chi_1 e^+ e^-$ decay & \cite{Tsai:2019mtm} Fig. 1e \\
			    \hline
			    \rule{0pt}{14pt}
			    SHiP & $\chi_2 \rightarrow \chi_1 e^+ e^-$ decay & 10 events line \\
			    \hline
			    \rule{0pt}{14pt}
			    BaBar &  mono-photon & \cite{Essig:2013vha} Fig. 4 \\[0.1em]
			    Belle-II &  mono-photon & \cite{Essig:2013vha} Fig. 4 \\[0.1em]
			    LEP &  mono-photon & \cite{Fox:2011fx} Fig. 2 \\[0.1em]
			    ATLAS & mono-jet & \cite{Aaboud:2017phn} and \cite{Belyaev:2018pqr} Fig. 7
			    \\
			    \hline
			    NA62 & Inv. $\pi^0, K$ meson decay & \cite{Fantechi:2014hqa} \\[0.1em]
			    BES & Inv. $J/\Psi$ meson decay & \cite{Ablikim:2007ek} \\[0.1em]
			    BaBar & Inv. $\Upsilon, B$ meson decay & \cite{Aubert:2009ae,Aubert:2004ws,delAmoSanchez:2010bk} \\[0.1em]
			    Belle (II) & Inv. $B$ meson decay & \cite{Grygier:2017tzo,Abe:2010gxa} \\[0.1em]
			    E949/787 & Inv. $K$ meson decay & \cite{Adler:2008zza} \\[0.1em]
			    \hline
			    \rule{0pt}{14pt}
			    SN1987A & cooling & \cite{DeRocco:2019jti}
			    \\[0.1em]
		        SN1987A & cooling, $\pi^0 \to \nu \nu$ & \cite{Natale:1990yx} \\
			    \hline
			\end{tabular}
		\caption{List of experimental searches currently implemented, along with some details about the process. Note that for SHiP decay limits, the 10-event limits are obtained directly using the tools described in this paper}
		\label{tab:explist} 
	\end{center}
\end{table}

Notice that while most of the limits are obtained from recasting existing works, the tools presented in this article can also be used to obtain naive estimate for various setups (not including the detection and geometric efficiency). for instance, limits from long-lived state at SHiP are in fact a 10 events line obtained in this way.

The code also allows to recast a large sets of experiments and print directly the results to files. For instance the following code

\begin{lstlisting}[frame=single] 
ExperimentsList=np.array(["lsnd_decay","charm_decay", \
                          "babar_monogam","belle2_monogam"])
geffZal={"gu11":1/2.,"gd11":-1/2.,"gd22":-1/2.,"gl11":-1/2,"gl22":-1/2}
Lim,LabelLimit = lim.GetLimits(ExperimentsList,10,geffZal,"AV",True)
\end{lstlisting}
creates a list of analysis to be recasted, loads Z-aligned couplings for an axial-vector effective operator and prints the resulting recasting into files. The output variable \texttt{Lim} contains a python dictionary of array $(M_2,\Lambda)$ with keys given in \texttt{ExperimentsList}.

We have finally enclosed with the distributed code various example files along with some plotting routines, including those for generating all the plots in this paper.

\bibliographystyle{utphys}
\bibliography{biblio}

\end{document}